# How Life Works: Darwinian Evolution of Proteins

J. C. Phillips

Dept. of Physics and Astronomy, Rutgers University, Piscataway, N. J., 08854

Abstract

We review the development of thermodynamic protein **hydropathic** scaling theory, starting from backgrounds in mathematics and statistical mechanics, and leading to biomedical applications. Darwinian evolution has organized each protein family in different ways, but dynamical **hydropathic** scaling theory is both simple and effective in providing readily transferable dynamical insights for many proteins represented in the uncounted amino acid sequences, as well as the 90 thousand static structures contained in the online Protein Data Base.

Critical point theory is general, and recently it has proved to be the most effective way of describing protein networks that have evolved towards nearly perfect functionality in given environments (self-organized criticality). Darwinian evolutionary patterns are governed by common dynamical **hydropathic** scaling principles, which can be quantified using scales that have been developed bioinformatically by studying thousands of static PDB structures. The most effective dynamical scales involve hydropathic globular sculpting interactions averaged over length scales centered on domain dimensions.

A central feature of dynamical **hydropathic** scaling theory is the characteristic domain length associated with a given protein's functionality. Evolution has functioned in such a way that the minimal critical length scale established so far is about nine amino acids, but in some cases it is much larger. Some ingenuity is needed to find this primary length scale, as shown by the examples discussed here. Often a survey of the Darwinian evolution of a protein's sequences suggests a means of determining the critical length scale. The evolution of Coronavirus is an interesting application; it identifies critical mutations.

## 1. Introduction: Protein molecular complexity

Before we discuss the mathematical and statistical basis for protein scaling, we should first consider the complexity of the problem. A typical protein chain may contain 300 sites, each potentially occupied by one of 20 amino acids chosen by evolution. The resulting number of possible amino acid sequences is $300^{20} \sim 10^{46}$, a number far larger than is normally encountered in physics, even in astrophysics. We shall later see biomedically important examples where a single site amino acid mutation is enough to change success into failure. Is it possible to explain such sensitivity using fundamental principles? Perhaps, although every protein family is different.

Several ambitious statistical studies of protein sequences are available. A broad evolutionary 2001 study of 29 proteomes for representatives from all three kingdoms: eukaryotes, prokaryotes, and archaebacterial, showed that proteins have evolved to be longer in eukaryotes, with more signaling heptad transmembrane helices in eukaryotes [1]. These are the signaling proteins which are the basis for about half of modern drugs. This very well-studied field is now classical, while research on most other proteins has just begun. A sophisticated group wish list (> 30 authors) of evolutionary problems associated with sequence evolution reported in 2012 that "current efforts in interdisciplinary protein modeling are in their infancy" [2].

Simulations usually rely on optimized force fields, and do not attempt to construct general principles for all protein structure-function relations. Here we also study specific protein families, and our methods succeed in connecting Darwinian evolution of sequences to function. Newtonian all-atom models are often unable to connect structural dynamics to function [3,4]. Evolutionary improvements are not easily recognized, as structures are seldom available for most species, and backbone structural differences are nearly always too small to explain Darwinian evolutionary progression. For example, the chicken and human peptide backbone coordinates of lysozyme *c* (Hen Egg White) are indistinguishable, even at highest attainable X-ray resolution in hundreds of PDB structures [5]. An historical note: science has grown rapidly in the last 60 years, and the science curriculum has struggled to keep pace. Here we often use the word "classical" to indicate topics that are old enough to have gained wide acceptance, and "modern" to describe the revolutionary changes that have taken place, often since ~ 2000. This review includes many modern ideas, which have enabled simple, powerful tools to organize molecular biology in new and often very accurate directions.

## 2. Mathematical tools

Phase transitions are central to understanding protein functions and evolution, and they occur also in the very large and rapidly growing scientific literature. Mathematical and scientific literature, as measured by numbers of papers published each year, is growing super-linearly, with the number of specialized journals proliferating in the 21$^{st}$ century. Citations are widely used as a measure of the impact of research efforts, if not the absolute importance of their content. This growth has stimulated more than 700 papers analyzing the literature scientometrically. The most spectacular scientometric study so far surveyed nearly all the 20$^{th}$ century literature, consisting of 25 million papers and 600 million citations, with $10^6 - 10^7$ more entries than typical data bases, and unique in the history of epistemology [6]. This study identified a citation transition, which occurred around 1960, and which is the earliest example of the cultural effects of globalization [7].

We can compare the development of different mathematical tools by using carefully chosen key words to search the Web of Science [8]. For example, although analysis is also a branch of mathematics, "analysis" is used in far too many non-mathematical contexts to be useful in a key word search. Useful mathematical words are algebra, geometry and topology; these yield surprising phase transitions over the last 30 years (Fig. 1). It is plausible that the glasnost emigration of Russian computer scientists around 1990 facilitated development of many biological data bases and contributed to the success of the human genome project. The fastest growing field, topology, is well suited to discussing protein network structure and Darwinian evolution. One might have supposed that geometry, the oldest of the mathematical disciplines, would have stagnated by now, but modern geometry has evolved into differential geometry, a tool whose concepts are well suited to discussing the functions of sculpted globular proteins. The two-year 1989-1991 jump of special functions, a subclass of analysis, by a factor of > 20, is far larger than any of the general jumps in Fig. 1. Other subclasses also show large 1989-1991 jumps, for instance, "differential geometry", a factor of 9, and "network topology", a factor > 20.

## 3. Statistical Mechanics

In his enormously popular Dublin 1943 lectures and book, "What Is Life?", Erwin Schrodinger proposed that we could progress in answering this question by using statistical mechanics and partition functions, but not quantum mechanics and his wave equation. He described an "aperiodic crystal" (today we would call it a glass) which could carry genetic information, a description credited by Francis Crick and James D. Watson with having inspired their discovery of the double helical structure of DNA [9,10]. Schrodinger arrived at this picture from thermodynamic theories concerning protein stability and information content. He observed that proteins are not only exponentially complex, but also must be near thermodynamic equilibrium, as they function nearly reversibly. Thus one can say that Schrodinger may have been the first theorist to conjecture that protein functionality could be usefully described thermodynamically.

Given what we know today, Schrodinger might have started more simply with the isothermal curves of the van der Waals equation of state (Fig. 2). Apart from providing a quantitative description of the liquid-gas transition of many molecular liquids, this equation exhibits spinodal phase separation topped by a critical point on the critical temperature isothermal. The critical features of this model are common to many systems at or very near thermodynamic equilibrium. Protein scientists often focus only on repulsive contact (or near contact) short-range interactions in static structural models, such as salt bridges, or orientations of helical segments. The van der Waals equation results from the combination of both short-range repulsive and long range (allometric) attractive interactions, and these are balanced at the critical point. Thus the short-range structural interactions are only half of the story of protein Darwinian evolution and functionality.

There is more to be learned from the balance of attractive and repulsive interactions. The rates of chemical reactions are determined not by energy minima and ground state configurations, but by saddle point (excited state) configurations. Little is known about these, but one can expect quite generally that chemical trends in protein-protein interactions can be usefully studied over longer length scales, where they are dominated by many weak attractive interactions.

Specifically one can imagine that protein functionality can involve transitions between two end states, called "open" and "closed" by Karplus in his protein example [3]. Then if one knows from general considerations some quantitative properties of any protein sequence, one may be able to recognize extrema and critical points and quantify balanced short- and long-range interactions in their neighborhood, without using elaborate Newtonian simulations. The comparative advantages of this more abstract thermodynamic method are its simplicity and universality, which makes it transferable. Such a method can easily treat Darwinian evolutionary protein sequences by following their nearly critical properties, and may also facilitate identifying essential features in any protein and relating them to its functionality.

The importance of critical points in statistical mechanics has led to many studies of their properties in toy models, especially Ising models , which are lattice models with only nearest neighbor interactions. Ising himself showed his models for dimensionality d = 1 do not contain a phase transition (a general property of all d = 1 models), and was unable to persuade himself that a transition existed for d = 2. L. Onsager studied critical point neighborhoods for a two-dimensional Ising toy model (1944), which he had solved exactly [11]. For d ≥ 4 the critical exponents of correlation properties are integers or half-integers (mean-field theory, used by L. Landau in his book on statistical mechanics). The most difficult case, d = 3, was solved using the renormalization group methods of particle theory. Its critical exponents are irrational sums of power series [12].

The existence of irrational exponents for d = 2 and 3 toy model phase transitions is suggestive, because empirical power law fits (which appear to be linear on log-log plots) are common in engineering discussions of nearly optimized systems. Power laws describe self-similarity (a power of a power is a power), and self-similarity is an appealing way of fitting together proteins over a wide range of length scales. Mandelbrot discussed geometrical self-similarity in terms of fractal exponents and power-law iterations (Mandelbrot sets) [13]. Per Bak connected these ideas to science in [14], and to evolution in [15].

## 4. Dynamical Scaling and Criticality

The general ideas discussed in 2012 in [2] and similar subsequent discussions of evolutionary dynamics and function [16,17] were unable to make contact with statistical mechanics and critical points. General ideas are all very well, but can they produce tangible biophysical and biomedically relevant results [18]? The positive answer to this challenge came from an unexpected direction, bioinformatic thermodynamic scales. The Kelvin temperature scale is central to entropy and general statistical mechanics, and T = 0 on the Kelvin scale corresponds to -273.16 C. Protein globular shapes are determined by competing hydrophobic forces (pushing segments inwards towards the globular cores) and hydrophilic forces (pulling segments outwards towards the globular water interface). Moreover, the leading physicochemical properties determining protein mutations are hydrophobicity, secondary structure propensity and charge [19].

To quantify these effects in the classic period of biophysics (before 2000), no less than 127 hydropathicity scales were proposed. Each hydropathicity scale had its merits, but few attempts were made to compare their accuracies, or applicability to properties other than those used in their definitions [20]. Meanwhile, the number and accuracy of PDB static structures had grown enormously, creating the opportunity to re-examine the early geometrical definitions based on average neighboring hydropathicity volumes [21] or average surface areas [22] in Voronoi partitions [23] of proteins into amino-acid centered units with van der Waals radii. Note that the combination of Stokes' theorem and modern differential geometry suggests that there should be a close connection between the volume and surface definitions. The surface one in particular emphasizes hydrogen bonding to the water films which have shaped globular proteins in evolution, much as rocky coastlines are shaped by the pressures of tidal water waves [13,24].

With the stage set, in 2007 Brazilian bioinformaticists Moret and Zebende (MZ) built the interdisciplinary bridge connecting proteins to statistical mechanics and critical points that everyone else had missed [25]. They evaluated solvent-exposed surface areas (SASA) of amino acids in > 5000 high-resolution (< 2A) protein segments, and fixed their attention on the central amino acid in each segment. The lengths of their small segments L = 2N + 1 varied from 3 to 45, but the interesting range ran from 9 to 35. Across this range they found linear behavior on a log-log plot for each of the 20 amino acids (aa):

$$\log SASA(L) \sim const \ - \Psi(aa) \ const.\log L \quad (\ 9 \leq L \leq 35) \quad (3)$$

Here $\Psi(aa)$ is recognizable as a Mandelbrot fractal. It arises because the longer segments fold back on themselves, occluding the SASA of the central aa. The most surprising aspect of this folded occlusion is that it is nearly universal on average, and almost independent of the individual protein fold. Thus this striking universal result transcends and compresses thousands of individual protein folding simulations. The 20 MZ fractals based on > 5000 protein segments have effectively compressed a huge amount of three-dimensional structural data into 20 one-dimensional parameters. It was always possible these parameters could have existed in principle, but no one had thought such a simple and sweeping discovery was possible.

How much information is contained in these 20 fractals? A lower limit is the number of different hierarchical orders from hydrophilic to hydrophobic for 20 parameters. This is 20!, which is $\sim 10^{46}$; we encountered this astronomical number before, in Sec. 1, as the number of possible aminol acid sequences in a 300 amino acid protein. This means that these fractals represent a potential magical wand that can analyze protein complexity on realistically large length scales, involving not only domains, but also inter-domain interactions. A second question is how accurate are their numerical values? The curves they displayed are based on about 250 data points per protein, and are likely to yield values of $\Psi(aa)$ accurate to about 1%.

It is plausible that the MZ fractals exist because Darwinian protein evolution has brought average protein SASA near critical points in each of the 20 aa subspaces. Remarkably, these subspaces span the same length frame $9 \leq L \leq 35$, independently of whether the aa is hydrophilic (usually near the globular surface) or hydrophobic (usually buried in the globular interior). One cannot "prove" with mathematical rigor these connections, but one can test (prove) them in the context of the Darwinian evolution and functionality of many protein families and thousands of individual proteins. Comparing the results with those obtained with the classical hydropathicity scales enables us to estimate the relative merits of various hydropathicity scales.

When an {X(aa)} scale is available, it can be shifted linearly to a new {X′(aa)} scale

$$\{X'(aa)\} = a\{X(aa)\} + b \tag{1}$$

without altering its relative values. Given the two constants a and b, one can arrange all scales to have the same average value and difference of largest and smallest values

$$<X'(aa)> = <\{X(aa)\}>, \quad \{X'(aal)\} - \{X'(aas)\} = \{X(aal)\} - \{X(aas)\} \tag{2}$$

Moret and Zebende compared their fractal scale hierarchically to seven classical hydropathicity scales, the closest of course being [22], which utilized only the SASA of a single aa, averaged over a few entire protein structures available in 1985. Paradoxically, the MZ Mandelbrot fractals based on short segment log-log plots may be more significant, and not just because of critical points in configuration space. Both protein dynamics (on a scale of ms) and Darwinian protein evolution (on scales as short as thousands of years) are difficult to understand in the context of their $\sim 10^{46}$ configurational complexity (Levinthal's paradox) [26]. At first, it was not obvious why the MZ success occurred, but retrospectively we can see how fractal segmental character implicitly includes Darwinian evolutionary optimization through exchange of modular building blocks [27]. Amino acid hydromodularity is apparently the best-documented example of effective parameter space compression [28].

Although [22] is closest to MZ, it is convenient to compare results obtained from the MZ hydropathicity scale (which implicitly describes second-order phase transitions), with those obtained from first-order protein unfolding measured by enthalpy changes from water to air (1982 KD Ψ scale; this is also the most popular of > 100 pre-2000 Ψ scales) [29]. The differences between results obtained by the two hydropathicity scales should reflect improvements in accuracy, which results from balancing a few strong short-range interactions against many weak long-range interactions. The overall correlation of the KD and MZ hydropathicity scales is about 85%. This is large enough to show they measure the same water-protein interactions. At the same time, it assures us that the 1% accuracy of the MZ hydropathicity scale is sufficient to enable us to separate first- and second-order thermodynamic effects.

In a few cases hydropathic (inside/outside) Ψ shaping may not be the most important factor, in which case we can turn to secondary structure (inside/outside) propensities of α helices and β strands [30]. Hydrogen bonding is longitudinal for α helices, and transverse for β strands, so we are not surprised to find that the (inside/outside) differences are small for α helices, and larger for β strands which can bind outside. Protein binding can involve the β strand exposed propensities for some cases. Another β strand scale was constructed from the core sequence of amyloid β [31], and we will return to this later. It is designed to emphasize aggregative "Hot Spot" propensities for binding on a length scale of 7 aa, and it will be referred to as βHS.

At first it may seem complex to have two hydropathicity scales Ψ and two β strand scales. Still, by comparing not only the scales with each other (through their Pearson correlation functions **r**), but also their performances on protein regions known to be structurally important, we will gain insights into molecular binding otherwise unobtainable. The best part is that comparative calculations with these hydropathicity scales are transparent and extremely easy, and can be implemented using only EXCEL spreadsheets, and their included software subprograms.

## 5.   Critical Length Scales

In normal liquids critical opalescence couples long wave length light waves to long wave length density fluctuations. Near a critical point could long wave length ice-like film waves couple to long wave length solvent-exposed protein area fluctuations, thus explaining the origin of the MZ fractals? This is a difficult question. In the apparently simpler system of quenched (non-equilibrium) glasses there has been much theoretical discussion of the possibility of a diverging length scale at the glass transition [32], especially connected with long-range stress fields [33]. To discuss phase diagrams and physical properties of network glasses, one must start from specifics of their chemical bonds [34-36], as well as both the local and extensive topological properties, which include stress percolation.

Here we argue that biomedically important results are obtainable by judiciously combining specific length scales W = 2M + 1 with one or more of the $\Phi$ = {hydropathic $\Psi/\beta$ strand scales}, denoted generically by $\Phi$. Given $\Phi$(aa), we calculate the modular average

$$\Phi(aa,W) = \text{average}(\Phi(aa - M), \Phi(aa + M)) \qquad (4)$$

which is a rectangular window from aa -M to aa +M . It is possible to iterate this process, for instance one can easily see that

$$\Phi(aa,W,W) = \text{average}(\Phi(aa - M,W), \Phi(aa + M,W)) \qquad (5)$$

gives a triangular window extending from aa -2M to aa +2M In practice for most cases rectangularly smoothed $\Phi$(aa,W) appear to give best results. By using $\Phi$ hydropathicity scales we "dress" modular building blocks and enable protein interactions to appear to be short range with cutoffs.

Given that $\Phi$(aa,W) is a good variable, how do we determine W? Experience with many examples suggests that all protein families are different, because their functions differ. We reflect some of these differences by choosing an optimized W, and others appear through comparisons between different components $\Phi$. Because all four $\Phi$ scales have general meanings, comparison of $\Phi$(aa,W) profiles often produces easily interpreted results. Far from being complex, the tools associated with $\Phi$(aa,W) are powerful aides for exploring protein complexity. They automatically incorporate universal aspects of globular Darwinian evolution.

## 6. Hinges and Pivots

Given a protein hydropathic profile $\Phi$(aa,W), one notices immediately that it has two kinds of extrema, hydrophobic maxima and hydrophilic minima. It is natural to suppose that the maxima act as pivots or pinning points for the conformational motion that is functionally significant, while the minima act as hinges. This language does not specify the conformational motion in Euclidean space that is functionally significant, as it jumps directly from the universal sequence geometries of Φ to function. What happens if we attempt to go elastometrically only between sequence and structure, by using the isotropic vibrational amplitudes of individual amino acids measured in structural studies?

The picture of differential aqueous sculpting of globular protein surfaces near a critical point [15] can be compared to elastometric treatments of hinge-bending conformational transition pathways [37-39]. A recent example is [40], which studied first-order open/closed conformational transitions. These are similar to the Ψ(KD) hydropathicity scale, which depends on first-order enthalpy changes between folded (in water) and open (in air). If there is a single mode that contributes significantly to the open/closed transition, it might have functional significance. However, the results showed only a wide range of overlap between conformational changes and single modes. It is unlikely that such changes or modes have functional significance. Although muscle contraction is mediated by a myosin cross-bridge which exists in two (open/closed) conformations, these were not explained by known conformers of myosin [41]. Instead there is an iterated hand-over hand motion of myosin along actin filaments [42].

## 7. Variance, Correlations and Level Sets

One of EXCEL's convenient software tools is variance, which can be used to quantify trends in extremal hydropathic (phobic-philic) widths as functions of both the choice of Φ (for instance, ΨKD verses ΨMZ) and W. Informally variance is known as "mean of square minus square of mean", or

$$\mathrm{Var}(\Phi(aa,W)) = \Sigma \, ((\Phi(aa,W) - <\Phi(aa,W)>)^2 = \Sigma \, \Phi(aa,W)^2 - n(<\Phi(aa,W)>)^2 \qquad (6)$$

where the sum is taken over n consecutive amino acid sites. In the context of the MZ Ψ scale, the variance measures the hydropathic roughness of the globular surface of the n sites of the protein chain segment. The local or global roughness can affect dynamical functions, which should occur neither too fast nor too slowly, in order to synchronize with other protein motions [43]. Mixtures of rougher granules have lower packing densities, and the granular knobs can jam kinetics [44].

Variance is a useful quantity in studying protein Darwinian evolution and dynamics because it combines extremal hydropathic pivots and hydrophilic hinges on an equal footing. One might suppose that such a simple function would have been used for long times in biology. In fact its bioinformatic importance was first realized only in 1911 by R. Fisher (then a student), but its publication centenary came only in 2018. Fisher used it to describe "The Correlation between Relatives on the Supposition of Mendelian Inheritance". It conveniently represents the random combinations of parent genes.

Pearson correlations themselves are normalized cross variances between two functions, for instance two sequences of the same protein from different species or strains X and Y ($-1 \leq r \leq 1$)

$$r = <(X(aa)\} - <X(aa)> (Y(aa) - <Y(aa) >)/(<(X(aa))^2> - <X(aa)^2>^{1/2} >) (<(Y(aa))^2> - <Y(aa)^2>^{1/2} \quad (7)$$

Level sets were developed to track the motions of continuum interfaces [45] – applied here to protein globular surfaces. Mathematically oriented readers will find "simple" explanations of their background and comparative computer science advantages online, for instance, under "Level Set Methods: An initial value formulation". Practical applications of level sets have emphasized image analysis [46,47], and have gradually evolved to include Voronoi partitioning, just as has been used for deriving protein hydropathicity scales since 1978 [48,49]. We expect, of course, that hydrophobic pivots move most slowly, while hydrophilic hinges move fastest. When there are two or more level pivots or hinges, it is likely that this is not accidental (nothing in proteins is), and we can test this assumption by comparing profiles with different hydropathicity scales, the KD $\Psi$ and MZ $\Psi$ scales, for example.

## 8.  Darwinian Evolution and Mutated Aggregation: Hen Egg White

We begin our tour of the protein world with lysozyme $c$ (aka Hen Egg White, or HEW), which was for some time the most studied protein: the PDB contains more than 200 human and 400 chicken HEW structures. HEW is also present in many other species, not only in the 400 million year old chicken sequence, but in most other vertebrates, almost unchanged in its peptide backbone structure. The backbone structure is exceptionally stable, with human and chicken $C_\alpha$ positions superposable to 1.5A°, while the aa sequence mutates from chicken to human with 60% aa conservation [10], well above the 40% minimum usually necessary for fold conservation [50].

HEW is a comparatively small 148 aa protein, which has a nearly centrosymmetric tripartite α helices (1- 56 and 104-148) and β strands (57-103) secondary structure. During its long career, HEW has performed at least three functions, as an enzyme, an antibiotic, and an amyloidosis suppressor. The relative importance of these functions has changed from species to species, and it seems likely that these changes are reflected in the amino acid mutations that have maintained the centrosymmetric structure.

Because we are most interested in the Darwinian evolution of the properties of HEW from chickens to humans, we plot the roughness or variance R as $R_S(W)/R_{Human}(W)$ in Fig. 3, using the fractal MZ $\Psi$ scale, for a range of species S. There is a broad peak, together with a narrow peak, both centered on W = 69, roughly half the protein length. The peak also occurs at W = 69 with the KD $\Psi$ scale, but its amplitude is only ~ 60 % as large, so the Darwinian roughness evolution is better described as thermodynamically second order. Similar narrow-broad peak structures have been observed in critical opalescence spectra, where the broad peak, associated with phonons, is called the Mountain peak [51]. Note that this "universal" peak applies to the terrestrial species, but not to zebrafish.

The structure-function relations giving rise to these peaks can now be profiled with W = 69, as shown in Fig. 4 for the extreme terrestrial cases of chicken and human. Note that normalizing $R_S(W)$ by $R_{Human}(W)$ in Fig. 3 is natural, because human structures have evolved to be closest to critical and smoother. This means that in most cases the critical limiting behavior of ideal functionality is nearly reached with the human sequence, described partly by $R_{Human}(W)$. It is obvious from Fig. 4 that on the W = 69 length scale, which is approximately half the HEW length, the largest effect of Darwinian evolution has been to stiffen the stiff central β strands by making them less hydrophilic.

While the results shown in Fig. 4 dramatically confirm the previously hidden content of HEW amino acid sequences, what is it? The characteristic length scale for amyloidosis is W = 40, because this is the length of the Aβ fragment (also discussed in Section 13) responsible for forming amyloid fibrils. On this length scale one can examine the effects of mutations on HEW aggregation rates, and a detailed discussion shows consistent shifts [50]. Presumably amyloid suppression is a key function for advanced species with larger neuronal networks which must be stable for longer lifetimes.

One can also discuss mutated aggregation rates in a similarly centrosymmetric protein, measured in [19], and analyzed there using W = 5 only (no other values of W were considered). The 98 aa α/β protein acylphosphatase (AcP) resembles HEW (αβα) in that its sequence is still nearly Ψ centrosymmetric, but it is more complex with five regions instead of three (β<1-19>, α<20-32>, β<33-54>, α<55-70>, β<71-98>). A parameterized W= 5 method for studying HEW mutations was applied to AcP, with disappointing results: increases in mutated aggregation rates were expected only in the central region, and found only in the N- and C- terminal wings [19].

When the S/human roughness ratios are plotted for Acyl-1, the results are much more complex than in Fig.3 for HEW. The difference is attributed to its richer α/β structure [50]. The overall scale, as measured by chick/human, is about 30 % enhancement, which is about 10 times smaller than that shown for HEW in Fig. 3. Nevertheless, three features were easily identified. Chick/Human gave peaks at W = 43 and W = 25, as well as a human peak at W = 13. These three values of W probably reflect interactions with three other (unknown) proteins. One can assume that mutations of the human sequence tend to "undo" Darwinian evolutionary improvements and cause mutated human profiles to regress towards chicken profiles. The three profile difference patches then agree well with the patterns of mutated aggregation rates [50].

The aggregation of globular proteins, such as well-studied lysozyme *c* (Hen Egg White), may involve unfolding, and is thus more complex than that of Amyloidβ, a known product of A4 fragmentation. The smallest lysozyme amyloid nucleus is 55-63 (9 aa) GIFQINSRY, called K peptide [52,53]. K peptide is the strongest amyloid former of nine related small (< 9 aa) peptides over a pH range from 2 to 9. Profiles for entire human lysozyme in Fig. 4 show that the 9 aa K peptide nucleus is located at the center of centrosymmetric α-β-α lysozyme. The 69 aa wide central β region 57-103 is hydropathically level, so its β strands are nonamphiphilic [50].

9.  **Level Set Synchronization: the Case of HPV Vaccine**

Synchronized motions of actin cell skeleton proteins guide cell surface and interface deformations, a protein realization of Stokes' theorem that also involves criticality [54,55]. Most proteins have > 300 aa, and their functionality at the molecular level also involves large-scale conformational motions which are optimized by synchronization. The first example of molecular level set criticality is likely to come as a surprise, as it concerns a single amino acid mutation of a 505 amino acid protein, which alters the self-assembly rate of protein complexes of HPV cervical cancer vaccine by a factor of $10^3$ [56].

The long road that led to cervical cancer vaccines began in 1976 when Harald zur Hausen published the Nobel hypothesis that human papilloma virus (HPV) plays an important role in the cause of cervical cancer. HPV is a large capsid protein, but it was found that only the 505 aa L1 part was needed to make a good vaccine that conformationally self-assembled into morphologically correct virus-like particles (VLPs). L1 from HPV 16, taken from lesions that had not progressed to cancer, self-assembled $10^3$ times faster than the HPV 16 L1P that researchers everywhere had been using; the old strain L1P had been isolated from a cancer, which differed from L1 by only a single amino acid mutation D202H [56]. The huge increase in self-assembly rate could well be due to conformational synchronization, but this is not easily quantified using Newtonian methods.

Given the lower bound of L = 9 in (3), and the remarkable properties of L1, its profile Ψ (aa,9) with the fractal scale (3) was examined near the 202 mutation site [57]. The striking feature is the presence of two almost level L1 hydrophobic peaks in the region between 191 and 231, shown enlarged in Fig. 5. The narrow peak α is centered near 202, the mutated site distinguishing L1 from L1P. The level condition is satisfied to within 1% by L1, but by only 5% by L1P and by two other singly mutated strains recently added to the PDB. Note that no mutations were found in the stabilizing broad peak. Note also the deep hydrophilic minimum near 215, which functions as a plastic hinge accelerating self-assembly.

This example also brings out the advantages of scaling with Φ(aa,W). The excellent agreement shown in Fig. 5 disappears when W is reduced to 7, below the cutoff in (3). It also disappears when ΨMZ is replaced by ΨKD. In other words, the pre-2000 efforts involved in constructing 127 different Ψ scales were involved in a good direction, but proteins are so complex that success was possible only bioinformatically after the PDB structures became numerous and more accurate [57].

In addition to the types of HPV that cause cancer, there are "milder" types that cause only warts (self-limited growth). It might appear that the differences between these two types, which occur on a cellular level, could not be analyzed on a molecular level. However, there are many self-similar aspects to proteins and cells, so one should look at the differences between the HPV16 (cancer) and HPV 6 (warts) profiles (Fig. 6). There are large differences in the amide (N) – terminal region, far from the plasticity hinge (which is almost unchanged) seen in Fig. 5. Thus the main self-assembly function is unchanged, but the N-terminal region changes can account for the sometimes serious side effects. These small differences are much reduced with the classic KD hydropathicity scale, which is suited to some large open-closed transitions, but not small conformational changes [57].

## 10. Level Set Synchronization: the Case of the False Aspirin

Hippocrates (~ 400 BC), the father of modern medicine, mentioned the miracle drug aspirin as a powder made from the bark and leaves of the willow tree [58]. By 2000 chemists knew the molecular mechanisms of aspirin (acetylsalicylic acid) [59,60]. Aspirin and other non-steroid anti-inflammatory drugs (NSAIDs) inhibit the activity of cyclooxygenase (COX-1) which leads to the formation of prostaglandins (PGs) that cause inflammation, swelling, pain and fever. However, by inhibiting this key enzyme in PG synthesis, the aspirin-like drugs also prevent the production of physiologically important PGs which protect the stomach mucosa from damage by hydrochloric acid, maintain kidney function and aggregate platelets when required. This conclusion provided a unifying explanation for the therapeutic actions and shared side effects of the aspirin-like drugs, which when overused, cause ulcers.

A unique example of a natural enzyme twin was discovered in the early 1990's, COX-2. It has nearly the same length (~ 600 amino acids) as COX-1, but is encoded by a different gene! BLAST sequence comparison of COX-1 and COX-2 showed 60-65% sequence identity of COX-1 and COX-2, which should be enough to prevent elucidation of their functional differences from structural data alone [60], even when augmented by Newtonian simulations (which have so far not been attempted).

COX-2 inhibitors became attractive when it was realized that they could have the same anti-inflammatory, anti-pyretic, and analgesic activities as nonselective inhibitors NSAIDs, with little or none of the gastrointestinal side-effects [60]. An inhibitory COX-2 drug passed through all its tests, and was duly marketed for several years. Then it was discovered that COX-2 inhibitors caused the appearance of significant cardiovascular toxicity associated with chronic use (2–4% of patients after 3 years).

What causes the difference between COX-1q and COX-2 inhibitors? This difference remains unexplained, although there are few enzymes of lipid biochemistry for which there is such a wealth of structural and functional information [61]. However, this structural information is limited to COX complexed with inhibitors, which are known to strain the wild structures [62]. In other words, here (as often) the ~ 600 amino acid structural differences are both small and complex. Enzyme inhibition occurs in transition states (saddle points in an energy landscape), which are conformationally different from the ground state minima of a bound complex. Moreover, as we shall see, there are indications that many large protein interactions occur near cell membranes, which orient the conformational distortions in the transition states.

Because of the large sequence identity of COX-1 and COX-2, it may appear difficult to analyze their differences quantitatively. However, their similarity can be turned to advantage by comparing their overall roughnesses or sequence variances (see (6)). The R(W) differences are small for small W, but increase with increasing W, finally peaking at W = 79 [63]. We can interpret this large value of W as an average spacing between the centers of modules that are involved in critical large-scale conformational changes that distinguish COX-1 from COX-2. A close inspection of the sequences aligned with BLAST suggests splitting each type into three domains, A: 1-110, B:110-290, and C: 290-end, but this separation is more obvious from the hydropathic profiles.

The general result from comparing the MZ9 and KD9 hydropathicity profiles is that the MZ9 scale is much more accurate, and the central domain B:110-290 is most scale-sensitive. There are many quantitative differences between the MZ and KD profiles [64], which could have a direct bearing on self-assembly kinetics. The greater precision of the MZ hydropathicity scale is expected, because all the differences must be conformational (large-scale and small). The differences could be explored with mutagenesis studies of self-assembly kinetics in the type-sensitive regions.

## 11. Globins and the Correlated Darwinian Evolution of Metabolism

Myoglobin (Mb) and neuroglobin (Ngb) are two globin-based proteins responsible for oxide-binding in tissue. As their names would suggest, Mb is typically found in muscle tissue where it stores oxygen [65], while Ngb is concentrated in the nervous system. While globins have typically been thought to have respiratory purposes, recent studies of Ngb have revealed that they also can have a number of other physiological functions [66]. These physiological functions are closely related to the structures of the oxygen channels of the proteins. Further analysis of how the oxygen channels differ across species of Ngb and how they differ from the oxygen channel in Mb helps understanding these physiological functions.

At the center of each globin is the heme unit, which consists of Fe bound to the center of porphyrin, a large planar ring composed of 4 smaller C- linked rings. The Mb structure (~ 150 amino acids) was the first protein structure to be determined (in the 1950's), and discussion of its structure-function relations continued well into the $21^{st}$ century. Perhaps the most interesting question is the nature of the oxygen channels which control $O_2$ and CO passage. These channels contain a sequence of pockets, observed in Xe-doped samples. Herculean Newtonian simulations have shown that the $O_2$ and CO ligands have a choice of two channels, from opposite sides of the porphyrin [67,68]. The protein chain wraps the porphyrin macro-ring, and passage through the two closest pockets is controlled by two Histidine (pKa = 7, neutral) gates. The spacing of these gates is 29 amino acids in Mb, while in Ngb, the spacing is 32 amino acids. This spacing is nearly constant across all terrestrial species, although its amino acid composition varies.

Given the key role played by the 29 amino acid His gate spacing, it is natural to profile both human Mb and Ngb using W = 29 [69], as shown in Fig. 7. There are some similarities between the two hydropathic profiles, but there are also striking differences, which are directly connected to differences in function (see Fig. caption). A detailed discussion [69] shows that the principal Euclidean features of the different oxygen channels of Mb and Ngb correspond quite accurately to the fine structure of these profiles.

A striking feature of Fig. 7 is the approximate three-fold symmetry of the human Ngb profile, with three nearly level and equally wide hydrophobic pivots, and two nearly level hydrophilic hinges. This is somewhat similar to lysozyme (Fig. 4), but here the symmetry is even more pronounced. Is this accidental? Fortunately we have Ngb sequences for many species, enabling us to study its Darwinian evolution from cold-blooded aquatic species across the Darwinian evolutionary panorama up to humans. The simplest case is tropical and temperate freshwater fish, shown in Fig. 8. In tropical zebrafish, the Ngb profile is very nearly tripartite level, while in temperate fish the structure is stabilized by hydrophobic wing peaks.

Because Ngb performs the most demanding metabolic functions – supplying oxygen for neural and retinal signaling – its remarkably balanced hydropathic profiles could be expected from its Darwinian evolution. Mb shows one spectacular feature in its Darwinian evolution from chickens to humans. Mice are favorite meals for predators, yet they flourish because of their ability to escape rapidly (a favorite theme for cartoonists!). To do so, their muscles must be supplied with oxygen rapidly. Comparing mice with chickens, we see (Fig. 9) a striking difference, associated with the appearance in mice of a hydroneutral peak associated with the apex (center) of the 29 amino acid chain connecting the two His gates.

The critical role played here by the His gates is an intrinsic part of the fractal MZ scale. The stability of histidine makes it the central and most conserved element of many catalytic triads [70], the most studied examples being Serine-Histidine-Aspartate (chymotrypsin) and Cysteine-Histidine-Aspartate. Catalytic triads form a charge-relay network (central His has pKa = 7), and are excellent examples of convergent Darwinian evolution [71].

The standard scale for mutation rates (BLOSUM 62, used in BLAST) exhibits a deep hydroneutral minimum in mutation rates near its center [64]. With the MZ scale this minimum is associated with alanine (A), glycine (G), the smallest amino acid, and histidine (H). Table I of [25] shows that none of the older scales places all three of these amino acids at its center. In terms of RMS deviations from the average value of each scale, the off-center differences are 7 times larger for the second-best KD hydropathicity scale, and much larger for all other scales. Note that these differences are not much larger than the quoted error bars in the classical KD work; the modern fractal MZ hydropathicity scale, based on an L = 20 centered range of solvent accessible areas, is more accurate in principle, because it benefits from its post-2000 bioinformatic survey of more than 5000 segmental structures.

Another way of looking at the hydroneutral amino acid group between the hydrophobic and hydrophilic groups is topological. The hydrophobic amino acids are concentrated on the protein inside (minimum distance from globular center), while the hydrophilic ones are similarly on the onside (maximum distance from globular center). Between these two extrema there are saddle points, where the hydroneutral amino acids should be concentrated. The saddle points are ideal for catalyzing reactions, as they can be tipped either towards the inside or the outside.

Hemoglobin is the classic protein oligomer. It colors red blood cells, and it transports oxygen from the lungs throughout the body. It is a dimer of dimers based on Mb-like monomers. Whereas Mb and Ngb ligand interactions involve charge, the Hb tetramers absorb and release oxygen primarily through interactions between strain fields localized near hemes and extended strain fields associated with dimer interfacial misfit [72]. Strain fields are the microscopic mechanism for long-range ("allosteric") protein interactions, because charge interactions are screened by the large dielectric constant (~ 100) of water. Hydropathically driven strain fields were anticipated by Hopfield's model of linear distributed energies [73] Linearity implicitly supports allosteric hydropathic waves. Allosteric aspects of globin functionality, especially the Christian Bohr (father of Niels Bohr) cooperative oxidation of tetrameric Hgb, have been the subject of 6500 papers [74].

Simply by comparing the **hydropathic** profiles of the Hb dimers ($\alpha,\beta$) with the Mb and Ngb profiles, we see immediately that Hb strain field interactions are synchronized by more level extrema, much like the Ngb interactions, and qualitatively different from the Mb interactions, which are dominated by a single hydrophilic hinge (Figs. 7,9,10). Strain fields - their observation and theoretical quantification – are one of the most difficult areas of condensed matter physics (for example, microscopic theories of cuprate high temperature superconductors [75]). When two fields interact, their spectra often exhibit (in the atomic case, Fano) antiresonant interference dips following resonant peaks [76,77]. An antiresonance is observed in Hb ($\alpha,\beta$) correlation (see equation (7)) spectra [73]. The importance of strain fields in the globins suggests amusing similarities between globin functionality and superconductivity (strain-field generated Cooper pairs).

## 12. Self-Organized Criticality and Color Vision

Next to the neural network itself, three-color vision (with depth perception!) is perhaps the most complex living function. Rhodopsin is the primary color opsin, and comparison of various species shows human rhodopsin has the smallest hydropathic MZ variance, that is, it is the smoothest. This suggests that its function is probably best preserved for long times by recoil involving small rotations with little wear and tear. The relative smoothness of human rhodopsin increases with increasing W [78]. One might suppose that other color opsins are also smoothest in humans, and this is true for most species. However, cats are nocturnal predators, who rely on their red opsins, and their red opsins are the smoothest. Dogs, who hunt in packs, have inferior vision [79].

All retinal frameworks have a rod-and-cone structure. There is a broad Darwinian evolutionary difference between vertebrate and invertebrate rhodopsin roughness profiles $\Re(W)$. The small-scale cellular structure of compound invertebrate eyes is less susceptible to cumulative amplification of lateral instabilities (inter-rod slippage), which are suppressed in large single-cell vertebrate retinas by rougher rhodopsin. This enables insects to utilize smoother rhodopsin, which probably enhances their visual temporal resolution (200 images/sec in bees, compared to 30 images/sec in humans), as their optical responses are less slowed by multiplied inter-rod constraints. Invertebrates have compound eyes with 2 primary cells and ~ 10 secondary pigment cells per retina, compared to ~ $10^5$ rod cells/retina in mammals [80]. This difference corresponds to a change in lateral molecular correlation lengths (average retinal size) of a factor of 100.

The smoother invertebrate roughness profiles $\Re(W)$ for W < 25 reflect a trade-off between long-range smoothness for W > 25, which is ineffective with only 10 pigment cells/retina, in favor of smoother photoreceptor-membrane interactions in their ommatidia. It is striking that the retinal cellular morphological differences (single/compound lens) between vertebrates and invertebrates are qualitatively obvious in their rhodopsin molecular roughness profiles $\Re(W)$. In principle the dominance of $\Re(W)$ for W > 25 in large mammalian retina reflects the nearly scale-free property at large lengths that characterizes self-organized criticality in more recently evolved species.

Empirically it is known that color discrimination (or quantization) can be described by three 8, 8 and 7 bytes/color channel, and that this discrimination is almost perfectly multiplicative, leading to a rough estimation of the number of distinguishable colors in the optimal color space as $3.2.10^6$ (22 bytes) [81]. Retinal organization in primates, which have a complex visual behavioral repertoire, appears relatively simple. However, quantitative connections between primate vision, retinal electronic connectivity (or neural network) and color quantization have so far surprisingly not emerged from studies of physiological models [82].

Balance between competing channels is characteristic of many dynamical critically self-organized networks [83], and is independent of chemical details. The networks are nearly optimized near self-organized criticality, and their dynamics follows another extremal principle, maximum rate of entropy production [84], in this case, fastest download and reset utilize the same number of degrees of freedom as activation. The separation of visual signals into three channels will be most efficient if those channels are nearly equally informative, and this is found to be the case [85]. This implies that the primate 20 aa opsin sequence is also nearly optimally designed to interact with all three pigment download paths, and that these paths can be primarily influenced by opsin aa hydrophobicities.

To count the number of independent optoelectronic channels accurately, one should make allowance for amino acid hydroredundancy. Even if the MZ hydrophobicities are not well separated, two aa may still make distinctive contributions if their strongly $\pi$ polarizable ring contents are different. Combining these two criteria, hydroredundancy occurs in only three cases: for one-ring Tyrosine ($\psi(Y) = 0.222$) and Phenylalanine ($\psi(F) = 0.218$); aliphatic chain (no rings) Isoleucine ($\psi(I) = 0.222$) and Methonine ($\psi(M) = 0.221$), and Alanine ($\psi(A) = 0.157$) and Glycine ($\psi(G) = 0.156$). Thus there are effectively N = 17 remaining functionally independent aa available, with a byte content of 4.12. The number of possible combinations N! of these channels is given by Stirling's formula ($\log N! = N\log N - N$).

The visual system is optimized to monitor moving objects, and this action involves resetting the retina to its dark state after each observation, in order to be ready for the next observation. This two-step ON/OFF process with nearly optimal two-step serial uploading and downloading must resolve 2(4.12) = 8.24 ~ 8 bytes/tricolor channel, which is in very good agreement with the empirical CIELAB color difference formula (8,8,7) [81]. Note that this two-step reversing model is easily countable, while two-step chemical mechanisms (phosphorylation, etc.) are not. The two co-exist, but in terms of shaped globular amino acid configuration space, the mechanical model is more fundamental [86]. The interface of protein structural biology, protein biophysics, molecular Darwinian evolution, and molecular population genetics forms the foundations for a mechanistic understanding of many aspects of protein dynamics [2].

## 13. Platelet Aggregation Inhibitors

Hemodialysis is used to treat patients with total or partial kidney failure, which results in protein aggregation. Several blood-based biomarkers have been tested to monitor partial kidney failure [87, 88], a globulin and an enzyme. We can compare their hydropathically profiled Darwinian evolutions [89]. β-2 microglobulin (β2m) is a small (119 aa) single-domain protein, characterized by a seven-stranded β-sandwich fold typical of the immunoglobulin domain family. We see in Fig. 11 how β2m has evolved from chicken to human, with large amino acid changes in the partially conserved (17-116) region, Iden., 48% , Posit 67% (BLAST), with enough similarity that the fold is still conserved. The feature that has evolved most is the leveling of the two central hydrophobic peaks. This small protein assists in removing molecular refuse from kidneys [20,21], and it can best do this by synchronizing its C- and N- halves. This is more important in long-lived humans, so on going from chickens to humans, β2m has evolved level hydrophobic peaks. Similar leveling of hydrophobic peaks differentiates the Darwinian evolution of myoglobin (154 aa, larger because it is wrapped around the heme). Compare especially the leveling of Mgb hydroprofiles of tropical and temperate freshwater fish, Fig. 8.

The 190 aa enzyme prostaglandin-H2 D-isomerase (P2GDS) has evolved much more rapidly than β2m, and the human and chicken folds are different [63]. The results of recent Darwinian evolution from rat to human (conserved fold, Iden., 69% , Posit 80%) are shown in Fig. 12. Both the hydrophobic peaks and the hydrophilic dips are much more level in the human profile.

## 14. Evolution of Flu Glycoproteins

Influenza is an RNA virus with a high mutation rate, nearly one mutation per replication [90]. It is also highly infectious, extending across an effective population size of donor-recipient pairs estimated to be approximately 100-200 contributing members [91]. This high mutation/transmission rate facilitates rapid selection for viruses that succeed in evading both antibody-mediated immunity and vaccines, and may involve ~3-5% substitutions in the antibody-recognized regions. Traditionally viral evolution (sometimes called antigenic drift and shift) has been represented by phylogenic trees [92], but the utility of this representation has been questioned on fundamental grounds [93,94]. Trees are dimensionally limited, as their effective dimension d is d = 1+ δ, where δ << 1 is the average rate of formation of new branches.

Epidemics occur when a new evasive strain cluster emerges in a minimal mutation space with dimension d $\geq$ 2 [93].

The public data base for flu strains amino acid sequences contains hundreds of thousands of strains, is probably the most extensive in protein science, and so should be highly suited to analysis by scaling theory. The ultimate goal is an effective and timely determination of suitable target strains for vaccine development. Because hundreds of millions of flu vaccines are produced annually, this goal challenges most methods based on general considerations only [93].

There are several types of flu viruses, and the current vaccine cocktail treats the most common three. The oldest type A/H1N1 was responsible for both the deadly 1918 pandemic and the unexpectedly harmless 2009 "swine flu" pandemic. The H1N1 type may be as much as 500 years old, and it appears to have stabilized, with $\delta \ll 1$. The success of traditional methods based on phylogenic trees for H1N1 left virologists unprepared for vaccine failure (effectiveness reduced from 50% to 15%, with higher morbidity and mortality) associated with the 2013-4 A/H3N2 strains, which are "young" (only 50 years old) [95].

Influenza virus contains two highly variable envelope glycoproteins, hemagglutinin (HA) and neuraminidase (NA). The structure and properties of HA, which is responsible for binding the virus to the cell that is being infected, change significantly when the virus is transmitted from avian or swine species to humans. HA is used to make vaccines, and its variations can be measured antigenically (through time-consuming ferret blood responses). Scaling theory quantifies the simpler problem of the much smaller human individual evolutionary amino acid mutational changes in NA, which cleaves sialic acid groups and is required for influenza virus replication. These two glycoproteins combine to form spikes on the nearly spherical viral surface.

When one studies the panoramic evolution of NA variances, a striking trend emerges (Fig. 13). The strains have not evolved randomly, as is often assumed. Instead, the N1 strains became overall smoother as they evolved, except for occasional outbreaks [96]. This smoothing enabled the strains to evade antibodies generated by vaccines, but they also made the virus less virulent. The long-term benefits of sustained vaccination programs shown in Fig. 13 are large and are not easily recognized in short-term studies [97]. Meanwhile, what was happening to H1?

The HA1,3 story has turned out be too complex for treatment by scaling theory [98], and is discussed in the following paragraphs. The NA2 story has turned out to be quite interesting, as it explains why H3N2 mutates so rapidly that it escapes phylogenic predictions, and why it is more virulent than H1N1. We can compare the hydroprofiles of NA1 [96] and NA2 [95]. There is a deep hydrophilic minimum of N2 near 335, which is associated with a wide disordered 37 amino acid region (no α helices or β strands). The three deepest hydrophilic minima of N1 are less deep, and are associated with shorter disorder: (150), 17; (220), 15; and (390) 13 – less than half as large disordered regions (PDB 4B7N). The deep hydrophilic minimum of N2 creates a large, exposed and elastically soft surface region, which is both easily mutated and can flexibly enhance cell penetration. The interactions of the two glycoproteins HA and NA are strongly coupled, and it is N2 that "drives" the dangerous H3N2 strains.

Although it is outside the scope of this review, Deem et al, using heuristic methods [99,100] (inspired by modern statistical mechanics [101]) on the large sequential data base appear to have made the greatest progress, starting with a seminal H3N2 paper in 2006, when greatest interest was focused on H1N1. They found that simply compressing the H3 sequence mutations into a 2-dimensional principal component space was enough to yield a reliable picture that identifies new strain clusters, without involving the slow and low-resolution ferret antigenicity measurements that are used to monitor vaccine safety. They also showed that vaccine effectiveness is most easily predicted from mutations of nearly conserved sites on the HA heads, which form distinct subsets called epitopes. Here the binding function of HA switches from one epitope to another, and the most effective vaccines follow switches. Fig. 14 is a recent d = 2 map showing punctuated creation of H3N2 strain clusters in timely 6 week periods [102].

## 15. Secondary Thermodynamic Scales

Protein globular shapes are determined by competing hydrophobic forces (pushing segments inwards towards the globular cores) and hydrophilic forces (pushing segments outwards towards the globular water interface). Differential (fold-conserving) changes in globular structures which involve hydropathic extrema are essentially topological, and correspond to inside/outside structural features. Differential functional properties are usually determined by outside (surface) mutations, as the inside amino acids are more densely packed, and not easily rearranged while conserving the fold.

The leading physicochemical properties determining protein mutations are hydrophobicity, secondary structure propensity and charge [19]. Secondary structure consists of α helices (like Myoblobin) and β strands (amyloids), which for our purposes can be regarded as longitudinal and transverse H-bond networks. The inside/outside bifurcation has been developed for of α helix and β strands in two ways: bioinformatically by surveying > 2000 structures, labelled FTI [103], and by a Hot Spot model based on central mutations of a central β amyloid 7-mer segment (labelled HS) [104]. FTI found, as one would expect, that the inside/outside propensities were small for α helices and large for β strands.

Comparison of these inside/outside scales shows Pearson correlations r ~ 0.9. This means that small differences between the scales can have large functional consequences. Some examples are discussed in [105]. A common example of an all-β protein is Tenascin, an extracellular matrix glycoprotein containing 14 repeats and > 2000 amino acids. It has many functions, including guidance of migrating neurons as well as axons during development of neural networks.

## 16. Holy Grail of Cancer

Vincent DeVita was instrumental in developing combination chemotherapy programs that ultimately led to an effective regimen of curative chemotherapy for Hodgkin's disease and diffuse large cell lymphomas. Along with colleagues at the NCI, he developed the four-drug combination, known by the acronym MOPP, which increased the cure rate for patients with advanced Hodgkin's disease from nearly zero to over 70%. This was the first cure for a solid tumor. The MOPP drugs are simple and inexpensive, and a dramatic demonstration of the power of combinational chemotherapy [106].

The MOPP drugs separately are less effective. Synergistic cancer drugs are often described in terms of checkpoints in the native immune system. There are >20 different ligand–receptor molecular checkpoint interactions between T cells and antigen-presenting cells [107]. Drug treatments of most of these reactions separately are ineffective, and often depend on the individual set of cancer mutations. There are > 10 million amino acid mutations associated with cancer, so extending combinational chemotherapy to the molecular level is a challenging problem, not so much for computers, as for building an adequately large DNA data base to be treated computationally.

If a simple, cost-effective cancer biomarker for early cancer detection (ECD) could be developed, it would immediately enable many cancers to be treated effectively using known and also cost-effective methods. It would also make building a combinational chemotherapy checkpoint data base much easier, so DeVita has called such a biomarker the **Holy Grail of Cancer** [106]. At present the leading candidate for such a marker appears to be the 15-mer epitopes selected from 400 aa p53 and mucin in the only large-scale (50,000 patients, hence cost-effective) clinical studies reported so far [108]. The discovered epitopes are more sensitive for ECD than p53 itself, which is more sensitive than any other protein.

Small antigenic regions (epitopes, as small as 10 aa) interact with small antibody regions (paratopes). Small peptide epitope sequences are printed cost-effectively on microarrays. Due to their miniature format they allow for the multiplex analysis of several thousands of peptides at the same time while requiring a minimal sample volume [109]. The most famous and best-studied (~ 100,000 articles) protein is the "cancer suppressor" p53 (~ 400 aa). 15-mer p53 epitope scans [108] revealed a bifurcation in the autoantibody pool between early cancer and tumor populations, which is the critical test for ECD. The tumor epitope lies near the center of p53, while the most sensitive early cancer epitopes lie in the N- and C- terminal wings. This central/wing pattern occurred in the mutations of HEW and AcP (Section 8), and is predicted correctly only by using the best MZ hydropathicity scale.

Can we use scaling to confirm the (early cancer)/tumor||wing/central parallel dichotomy epitope sensitivity structure for p53? The answer is both No and Yes. Hydropathic $\Psi$ scaling does not identify any of the epitopes discovered in [108]'s epitopic scan, but scaling with secondary (Sec. 15) exposed $\beta$ strand propensities [103] is successful [110]. Most proteins are hydropathically compacted into globules, but the tumor suppressor p53 (393 aa) forms a flexible, tetrameric four-armed starfish [111], quite distinct from the globular structures which most proteins (even when oligomerized) form. Among all proteins p53 is much more

hydrophilic than average, and it is also elastically much softer, with about half its structure dominated by β strands, while the remainder (especially the N-terminal quarter) is disordered.

The most interesting ECD example of the fit to the experimental 15-mer overlapping epitopes numbered XY, containing amino acids $5(10X + Y) - 4$ to $5(10X + Y) + 10$, is XY = 09,10. The profile in Fig. 15 shows that overlapping 15-mer epitopes -9 and -10 were successful because they share a common 7-mer. The profile -9,-10 obtained with the FTI βexp scale is much more useful than that with ψ hydropathicity scales [110]. Such comparisons are themselves simple and easily applied "checkpoints" for testing the theory.

A subtle point in the scanning profiles is the weakness of the XY = 34 central epitope sensitivity in early cancer stages, even though it is the only dominant epitope in the presence of tumors [108]. Initially the immunogenic interactions may be weaker and longer ranged, so that -34 does not initially form β strand bonds to antibody paratopes, whereas -9,-10 does. At later cancer stages, the interactions could be stronger and shorter ranged, with entangled interactions involving nonlinear amino acid interactions. Such nonlinear interactions are not easily connected to DNA personalized mutations, whereas such connections may well be possible with the weaker interactions discovered in ECD. Not only are drugs addressing weaker interactions much more likely to be effective, but they could also be much more easily identified.

The superior biomarker sensitivity of p53 epitopes to p53 itself is understandable, considering that the autoantibody paratopes probably form families themselves (so far unknown). The sensitivity of the individual p53 epitopes is ~ 30%, but with better statistics and correlations with dynamically monitored individual p53 mutations it could be higher. In any case, mucin is another whole protein cancer biomarker that can be used as a source of autoantibody detective epitopes, which can probe different autoantibody families [112,113]. The structure of mucin (~1250 aa) is dominated by 20 aa repeats, whose length increases from 16 in mouse to 30 -90 in humans. The ~ 1000 aa repeat domains are more stable in the center and more variable in the periphery, so [112,113] found that the most sensitive mucin epitopes consisted of three center repeats; presumably the two outside repeats buttress the autoantibody binding center repeat.

Although the mucin structure is disordered, it can still be profitably analyzed with scaling theory, using both hydropathic Ψ and transverse β scales [114]. The differences between the two scale profiles are small, but the sensitivity of the epitopes is enhanced by using three repeats from the domain center from the DNA binding domain [113]. Here it is the long-range stabilizing elastic interactions that enhance sensitivity, and not the short –range glycosylation ones [112], probably because glycosylation obstructs β strand hydrogen bond interactions. The detailed analysis of the differences shown in Fig. 16 can be extended to other mucin epitopes studied in [113], with the conclusion that the largest scaling differences correlate well with largest diagnostic sensitivity differences [114].

The direct approach to treating cancer is supported by the Developmental Therapeutics Program (DTP) of the NCI, which maintains the infrastructure and expertise for the operation of cell-free and cell-based high-, medium- and low-throughput assays. The DTP functional genomics laboratory provides molecular analyses including gene expression microarrays, exon arrays, microRNA arrays, multiplexing gene assays, plus others as tools to determine the role of selected genes in the mechanism(s) of drug action and cellular responses to stressors [115].

The direct approach to a problem that involves identifying the $> 5$ key mutations (from a pool of $> 10$ million) that have affected $> 20$ checkpoints in a way dependent nonlinearly on individual DNA's (containing $> 20$ million nucleotides) is advantageous in one respect: it will keep researchers busy for an indefinite period (probably longer than individual careers). However, suppose we reverse the problem, having already identified a small number of epitopic signals in the "linear response" regime of ECD (not the nonlinear, and fully entangled regime of diagnosed tumors). Then, already "knowing the answer", we can work backwards through a data base of millions of patients, to find patterns in the individual DNA mutations that correlate with epitopic patterns (which could grow to include dozens of individual epitopes). That would be a new chapter in the history of science and medicine.

## 17. Recycling Amino Acids with Level Sets: Ubiquitin and Ubiquitinase

Manufacturing amino acids is expensive, and all eukaryotic cells recycle amino acids of damaged proteins. Ubiquitin, discovered less than 50 years ago, tags thousands of diseased proteins for destruction. It is small (only 76 amino acids), and is found unchanged in mammals, birds, fish, and even worms; even the yeast and human ubiquitins differ only at three amino acids. Such conservation indicates that eukaryotic ubiquitin is perfect. Key features of its functionality are identified [116] using critical point thermodynamic and hydropathic scaling theory. These include synchronized pivots and hinges, a stabilizing central pivot, and Fano interference between first- and second-order elements of correlated long-range (allosteric) globular surface shape transitions. Comparison with its closest relative, 76 amino acid Nedd8, shows that the latter lacks all these features. A cracked elastic network model is proposed for the common target shared by many diseased proteins. Ubiquitin's three-peak profile is very well suited to bonding both ends to stain cracks of diseased proteins. The third peak may stabilize bending of ubiquitin when binding to a bent crack.

Ubiquitin tags diseased proteins and initiates an enzyme conjugation cascade, which has three stages. The first-stage enzyme Ubal (El) has evolved only modestly from slime mold to humans, and is >14 times larger than Ub. Critical point thermodynamic and hydropathic scaling theory to connects Ubal (El) evolution from yeast and slime mold to fruit flies and humans to subtle changes in its amino acid sequences [117]. As shown in Fig. 18, the four highest human hydrophobic pivots are nearly level, while the slime mold pivots are tilted by ~ 15%. The choice of W* = 55 is consistent with the size of ubiquitin. With > 1000 amino acid sites, such a large value of W* again emphasizes the importance of W as a tuning element for focusing on large-scale domain interactions.

Structural studies identified four Uba1-E1 building blocks: first, the adenylation domains composed of two motifs (labelled IAD (1-169) and AAD (404-594), for ''inactive'' and ''active'' adenylation domain, respectively), the latter of which binds ATP and Ub; second, the catalytic cysteine half-domains, which contain the E1 active site cysteine (CC (169-268) and CCD (594-860) inserted into each of the adenylation domains; third, a four-helix bundle 4HB (268-356) that represents a second insertion in the IAD; and fourth, the C-terminal ubiquitin-fold domain (UFD (926-1024)), which recruits specific E2s. How do these structural and functional domains compare with our one-dimensional hydropathic profiles? Fig. 19 shows an excellent match, with the domain boundaries associated either with pivots or one amphiphilic side of a hinge. While Cys itself is the most hydrophobic amino acid, the pivots of both catalytic cysteine half-domains are level at hydroneutral in Fig. 19.

## 18. Unfolded Proteins in Human and Cow Milk

The Darwinian evolution of milk from glandular skin secretions is a fascinating story, with many twists and turns from fish to preserving water at and after birth on dry land [118,119]. Eventually Darwinian evolution led to the mammalian micelle solution, which involves an increase in calcium and protein concentrations in water by up to 100- and 1000-fold, respectively. Such an abrupt change seemed unlikely to 19th century opponents of Darwinian evolution [120], but today we recognize that Darwinian selection probably occurs by exchange of modular genetic fragments, sometimes recognizable as both static and dynamic structural domains. Human milk today is probably close to a critical point which balances many factors; here we focus on the recent evolution of the three caseins that stabilize the micelle against precipitation of $CaPO_4$, while still making it available as needed, along with the proteins, fats and sugars in micelles [121].

Micelle structures of bovine milk are well-studied, while there are no similar structural studies of human milk. The simplest micelle nanocluster model is based primarily on observations made by solution X-ray and neutron scattering under conditions that preserved the native environment of the micelle so that the possibility of artefacts is minimized. These determine the radius, mass and average spacing of the calcium phosphate nanoclusters. The model establishes a clear relationship between the structure and its biological (primarily nutritional) functions Overall the kappa caseins stabilize the micelle nanocluster surface, while the alpha and beta caseins are classic examples of unfolded network proteins. These often adopt the extended poly-L-proline type 2 (PP-II) conformation, where the backbone is fully exposed to water which forms a stabilizing hydration network around it. In this backbone neutral Pro and Gly residues are the most abundant; they straighten and stiffen the network [122]. The scattering studies see a micelle interior "gel" or network of alpha and beta caseins covered by kappa caseins separating calcium phosphate nanoclusters.

By comparing hydropathically the sequences of different species, especially bovine and human, we can test the nature of interactions between unfolded alpha and beta casein proteins. Caseins differ from globular proteins in that in the former, disulphide bridges are mostly intramolecular whereas in the latter, they are mostly intermolecular. Thus one supposes that nanocluster structures will be best hydroprofiled with the long-range MZ scale.

The β-casein fraction of human milk comprises more than 85% of total casein, whereas the β-casein fraction of bovine milk is much lower (33% of the total casein) [122]. In bovine milk the αS1 fraction nearly matches the β-casein fraction. An important difference between human and bovine milk is the degree of phosphorylation of β-casein. Bovine β-casein usually occurs in fully phosphorylated forms, whereas human β-casein is found in multiphosphorylated forms having 0 to 5 phosphate groups per molecule [123]. Dephosphorylation of bovine β casein using acid phosphatase resulted in similar physicochemical functionality to human casein as well as increased emulsion stability. Thus **Darwinian** evolution has rebalanced the α and β casein fractions, presumably because in humans β casein supports flexible phosphorylation.

We can quantify casein rebalancing by comparing bovine and human β casein Pro- and Gly-rich regions and their $\Psi(aa,W^*)$ hydropathic profiles [121]. The simplest comparison uses BLAST to compare the two amino acid sequences site-by-site. BLAST gives 66%, which shows that they share a common fold, but it seems too low. Does using the single site correlations of hydropathic scale values $\Psi(aa,1)$ improve matters? The results are somewhat better than, but still similar to BLAST positives for the KD scale (72%) and the MZ scale (69%). Can better correlations be achieved by restricting the region to the polar tracts that are rich in Pro and Gly {(P,Q) rich}? We define these regions as sites from 40 to the C terminal, and again the improvement is small (KD, 74%, MZ, 70%). There is also a small increase in the lengths of the polar tracts from bovine to human.

By now the reader already knows that these disappointing results are caused by the limitation of single site correlations. There was a dramatic improvement when [121] used the best value of $W^* = 29$ to plot $\Psi(aa,W^*)$ hydropathic profiles (Figs. 20,21). These optimized hydropathic profiles $\Psi(aa,W^*)$ correlations are 92% (KD) and 89% (MZ), which is a dramatic improvement over BLAST (~ 60%), without or with restriction to polar tracts. Of course, there is still an intrinsic difference between human and bovine caseins, and it could be about 10%, with a 2% uncertainty. The BLAST values imply a much larger (physiologically unacceptable) difference of about 40%.

The hydropathic wave profiles with the MZ scale (Fig. 20) and the KD scale (Fig. 21) reveal important beta functions. The deep hydrophilic minimum near 31 is associated with the Pi9 phosphated serine-rich cluster. The deeper the minimum, the larger the water density near the Pi9 phosphate cluster. The anionic SerP clusters may be charged, and $Ca_2PO_4$ is cationic. Water is highly polarizable, and it dielectrically screens these charges, lowering the total energy. The KD human hydrophilic minimum is relatively deeper for human compared to bovine for the KD scale. Both figures show a large hydrophobic peak near 75, which is structurally critical for stabilizing the intertwined mesh, as discussed below. This peak is hydrophobically bonded to a peak near 180, and for strongest bonding the two peak heights should be equal. The difference for the KD scale is 1/3 smaller than  for the MZ scale, which again suggests that the KD scale is yielding slightly better results, presumably because the fibril binding interactions are more nearly short-range than long-range. The $\Psi(aa,W^*)$ hydropathic domain profiles for both scales are much more informative than the $\Psi(aa,1)$ profiles (not shown), whose correlations are similar to BLAST positives.

## 19.  Lens Transparency and Cataracts

Granted that neural networks are nature's most amazing invention, what is in second place? One candidate is the human eye, and specifically its transparent lens, which is also strong and flexible. In 1802, philosopher William Paley called the eye a miracle of "design". Charles Darwin wrote in his Origin of Species (1863), that the evolution of the eye by natural selection at first glance seemed "absurd in the highest possible degree". Today, after more than 150 years, most biologists and even biophysicists are unable to analyze such Darwinian protein evolution. The analysis of the Darwinian evolution of the lens is straightforward with hydropathic scaling, which reveals amusing details [124].

Lenses are made of crystallins, which exhibit order similar to spheroidal glasses. There are many crystallins, but the dominant crystallins with the largest molecular weight are the αA and αB crystallins, which contain less than 200 amino acids. The even smaller crystallins may function to support edges and corners of the transparent bulk structure of the lens. Transparency requires a non-absorptive region in the optical spectrum which has been well studied in inorganic solids, where it is known that disorder is energetically unfavorable. The hydropathic scale of the important disorder should be a typical membrane thickness, which leads immediately to the center of the MZ range (Secs. 4,5) at W = 21. This value is confirmed by Darwinian evolutionary variance ratios of the αA and αB crystallins.[124].

From chicken to human the αA crystallins evolve more than the αB crystallins, and most striking is the appearance of synchronization (level sets) of the hydrophobic extrema of αA crystallins (Fig. 22). One might suppose that these subtle Darwinian evolutionary changes at the level of individual crystallin proteins could not be related easily to chemistry at the cell level, but such connections are visible. Healthy lenses are characterized by arrays of nearly parallel cell fibers, which are occasionally cross-branched, as seen by staining. The most unique biochemical characteristic of the eye lens fiber cell plasma membrane is its extremely high cholesterol content. The cholesterol forms lipid bilayers, which increase the fiber cell plasma membrane's resistance to oxygen permeation, thus helping to maintain lens transparency and protecting against cataract formation. These layers are hydrophobic, so that although water fills most of the lens volume, there is an effective oxygen barrier at the fiber membrane interface.

The parallel fibrous structure of healthy lenses is absent in cataracts. Comparing the chicken and human α(A,B) hydropathic profiles in Fig. 22, we can see how Darwinian evolution may have strengthened the fiber lens structure for longer lived humans, especially for αA. There the c-terminal hydrophobic peak in $\Psi(aa,21)$ has increased from 151 in chicken to 158 in human αA crystalline (hydroneutral is 155). One can go further in modeling by supposing that the fiber structure contains α spirals, in which chains are formed by connecting the hydrophobic n- and c-terminal peaks. The spiral is stabilized by the central hydrophobic peak, with the two hydrophilic minima flexibly promoting alternating outside extrema of the spiral structure.

Raman microspectroscopic images show the short-range-order of the fiber cytoplasm remains largely intact in cataracts. At the individual crystallin level the subtle chicken-human Darwinian evolutionary shifts in αA crystallins hydropathic profiles are consistent with improved human fiber stabilization and cataract avoidance. Cataracts appear in rats after only two years. How is this reflected in αA rat/human hydropathic profiles? These are compared in Fig. 23, which shows dramatic differences. The most obvious one is the deletion from the human sequence of 23 amino acids 64-86. The remaining sites are almost unchanged, except for the Ser165Cys replacement, which levels the human c terminal hydrophobic peak. Compared to Fig. 22 for human/chick, the rat profile is much less stable, especially with regard to the amphiphilic trend of the hydrophobic peaks, and the very deep hydrophilic minimum at 83. This should also contribute to forming cataracts, and the very deep hydrophilic rat minimum may be related to its need for dark vision.

## 20. Giant Hubs in Protein - Protein Interaction Networks

Protein–protein interaction network (PPI) studies show that proteins with a large number of interactions (large degrees k) are rare, with the probability P of such degrees following a power law (the P(k) distribution is self-similar: one fractal). Like other characteristics of hubs, the connection between evolutionary conservation and high degree of nodes in PPI networks has been controversial [125]. It appears that these controversies are inevitable when one tries to find universal rules for a large number of proteins using site-by site (BLAST) measures of evolutionary distance.

One can instead use thermodynamic scaling theory (20 fractals) to explore the fine Darwinian evolutionary details of a few proteins which are not merely hubs (k > 10), but are giant hubs (k > 50) [126].  Tyrosine kinase Syk (635 aa) and proto-oncogene tyrosine protein kinase Src (539 aa) are giant hub signaling enzymes that can transfer a phosphate group from ATP to a protein in a membrane. These signaling proteins are well described on their human Web-based Uniprot homepages.  Syk has 7 molecular functions and 57 biological processes (Uniprot), while a recent search [127] listed Syk in its Table 2 as the most connected protein localized in both the cytoplasm and cell membrane (CMP) with 105 interactions.  The corresponding numbers for Src, a mitochondrion CMP, are 114 interactions on the Uniprot list, and 208 interactions by the keyword search methods of [127].   Standard methods based on aa sequence conservation and structural similarities enable division of these large protein into three domains and two transition regions. The changes in their domain structures with Darwinian evolution are small and not easily quantified, yet small changes are often medically significant.

Domain coverage refers to the percentage of the residues in a protein that belong to a domain, thus excluding residues in loops or other structures outside a domain. Domain coverage provides the best genome-wide predictor of organismal complexity [128].  The PPI domain coverage per protein appears to increase with the complexity of the organisms so that proteins in complex organisms contain more domains and perform more specialized functions.

To go further than coarse proteomic observations, one can analyze the detailed **Darwinian** evolution of the domain structures of specific proteins, such as the giant hubs Syk and Src [126]. The results for Syk in Fig. 24 show a close correspondence between the three-dimensional structure and the one-dimensional hydroprofile with W = 21 (membrane length, appropriate to these proteins, which have strong interactions at membranes). Next we turn to Src; as Fig. 25 shows, the overall domain structures and profiles are similar to Syk, but with one striking difference in the new N-terminal U region [129]. This initially hydrophilic region is responsible for binding to targets, and almost all the differences between chicken and human occur here.

The differences shown in Fig. 25 could arise merely from replacing a few hydrophilic amino acids with hydrophobic ones. Instead they involve a purely hydroneutral septad appears almost miraculously in human Src: 25HGAGGGA31. (The unbiased probability that such a segment, with 7 out of 7 aa accidentally being A, G or H, is roughly $\sim 10^{-5}$.) The stability of histidine makes it the central and most conserved element of many catalytic triads, the most studied examples being Serine-Histidine-Aspartate (chymotrypsin) and Cysteine-Histidine-Aspartate. Catalytic triads form a charge-relay network (central His has pKa = 7), and are excellent examples of convergent Darwinian evolution. What is most remarkable about the unique Src heptad is that it contains only one catalytic His, while adding hydroneutral Ala and Gly. The three amino acids A,G, and H lie near the centers of approximate classical hydropathicity scales, while they are centered almost exactly by the thermodynamically ultraprecise MZ scale (Sec. 11).

Is that all? No, there's more. We can look at the effect mammal domestication has had on this hydroneutral cluster. Of course, chicken is primitive, but surely mammals are similar? [126] contains a detailed discussion of a similar, but more complex, hydroneutral cluster in Syk. This hydropathic profile of this cluster is much like that of the human cluster in dogs, cats and horses, but not in pigs and cows. What is impressive here is the great detail available in sequential analysis of the structural features of these complex signaling proteins using hydropathic profiling.

## 21. Cytoskeletons and Microtubules

The cytoskeleton is a complex, dynamic network of interlinking protein filaments present in the cytoplasm of all cells, including bacteria and archaea. It extends from the cell nucleus to the cell membrane and is composed of similar proteins in the various organisms. In eukaryotes, it is involves three main components, microfilaments, intermediate filaments and microtubules, and these are all capable of rapid growth or disassembly dependent on the cell's requirements. In eukaryotes there are also motors that are absent from prokaryotes.

A multitude of functions can be performed by the cytoskeleton. Its primary function is to give the cell its shape and mechanical resistance to deformation, and through association with extracellular connective tissue and other cells it stabilizes entire tissues. The cytoskeleton can also contract, thereby deforming the cell and the cell's environment and allowing cells to migrate. A large-scale example of an action performed by the cytoskeleton is muscle contraction.

When the cytoskeleton was first introduced, it was thought to be an uninteresting gel-like substance that helped organelles stay near the cell center. Roger Penrose helped to show that microtubules vibrate within neurons in the brain, suggesting that brain waves come from deeper microtubule vibrations [128]. Penrose argues further that resonance conductance through tubulins and microtubules could be a quantum effect, but as we shall see, elastic water waves (while classical) are elastically coherent, explain these data, and quantify Darwinian evolution as well. The phases of quantum waves are unlikely to be stable at living temperatures. The problem of explaining the thermal stabilities of such protected "edge" or "surface" waves has been solved topologically for simple examples [129].

Cytoskeletons are self-organized networks based on polymerized proteins: actin, tubulin, and driven by motor proteins, such as myosin, kinesin and dynein. Actin is a polymerized monomer that forms elastically self-organized networks [39]. It is so stable that to obtain substantial Darwinian evolutionary changes, one should compare human with algae (BLAST 84% identities, 92% positives) [130]. The algae/human ratio of the variances of $\Psi(aa, W^*)$ is shown in Fig. 26. It contains the first actin surprise (in fact, protein functions are all different, so there are usually surprises, especially with extremely stable, "nearly perfect" proteins like actin and ubiquitin). Most often $W^*$ is the value of W which maximizes a variance ratio $V_r$, but here $W^*$ corresponds to a maximum in the derivative $dV_r/dW$. What has happened? Actin performs two functions, stabilizing cell shapes after small second-order distortions while functioning, and polymerizing to push membranes forward during cell growth, which is thermodynamically first order. Apparently both functions are optimized by balancing small W values for W < 21 (important for polymerization), against large W values of W (important for stabilizing large cell shapes). This double balance eliminates the maximum in $V_r$, and leaves instead a maximum in $dV_r/dW$ at W* = 21.

Actin double balance is tested by using it to profile hydropathically with $\Psi(aa,W^* = 21)$ the actin Darwinian evolutionary changes from algae to human (Fig. 27). Of course, with high site conservation ~ 90%, these changes are small, but they are strikingly consistent with leveling of hydrophobic extrema (elastic pivots). First there is the large hydrophobic peak near 150, which is unchanged and presumably stabilizes large cell shapes. In humans Darwinian evolution has leveled hydropathic extrema in the central region 180-280. The deep hydrophilic hinge near 110 is also deeper in human. Finally the extreme ends of the N and C terminals are both more hydrophilic in human actin, facilitating the construction of large filaments and filamentary bands.

Like actin, tubulin has evolved very little, so we look for a reference species which is the best starting point for measuring Darwinian eukaryotic evolution. Yeast studies have led to the discovery of genes involved in fundamental mechanisms of transcription, translation, DNA replication, cell cycle control, and signal transduction. However, since the divergence of the two species approximately 350 million years ago, fission yeast appears to have evolved less rapidly than budding yeast, so that it retains more characteristics of the common ancient yeast ancestor, causing it to share more features with metazoan cells.

At the molecular level tubulin is a heterodimer; the polymerizing α and β tubulin monomers are similar in length and secondary structures, as well as having indistinguishable phylogenetic trees [131]. While there appeared to be little difference between α and β monomers in early static structure studies, more recent atomistic simulations of rearrangements upon hydrolysis of template pig (99% human sequence) structures extracted from ~ 100 mostly human structures have shown large differences in elasticity [132]. Given that hydrolysis is thermodynamically first-order, we expect to see these differences especially clearly by comparing the hydropathic KD profiles of α (Fig. 28) and β (Fig. 29). Fig. 4 of [132] exhibits large hydrolysis displacements for α tubulin, and small displacements for β tubulin. Moreover "Remodeling of longitudinal dimer contacts is coupled to {long range} conformational changes in α-tubulin … The observed changes at the interdimer contact, around the E-site, are accompanied by internal rearrangements of the tubulin dimer involving the intermediate and C-terminal domains of α-tubulin …upon hydrolysis the α-tubulin intermediate domain within the microtubule undergoes a shift similar to that reported for the straight to bent transition".

Although calculations were always done with both the hydropathic first-order KD scale and the second-order MZ scale, the figures shown in this Report usually involve only the MZ scale, because it is more accurate and better suited to most proteins' functionality. The case of α and β tubulin is exceptional, because the KD results had higher resolution of domain structure (Figs. 28, 30). This observation confirms the [132] conclusions regarding the large hydrolysis shifts of α tubulin. Other conclusions of [132] are matched by the rich detail of level sets (synchronized extrema) shown in Figs. 28-30. It is gratifying and surprising that simple hydropathic profiles agree so well with such sophisticated atomistic simulations of rearrangements upon hydrolysis. Is it surprising that hydropathic scaling identifies the domain structure of α tubulin without any template input? Surely this is another example of natural magic [79], as anticipated by Francis Bacon (1545) and included in the charter of London's Royal Society (1660). It can be traced back to the critical hydropathic shaping of protein segments (Sec. 4) [25], which in this case has been affected mainly by first-order hydropathic interactions.

## 22. Motor Proteins

Movement along microtubules is based on the action of motor proteins that utilize energy derived from ATP hydrolysis to produce force and movement [133]. Members of two large families of motor proteins—the kinesins and the dyneins—are responsible for powering the variety of movements in which microtubules participate. The third large family is the myosins, which bind to actin. Thermodynamic scaling theory yields new insights into the evolution of these motor proteins, whose mechanisms for transporting proteins from the cell center (organelle) to the cell surface membrane are well understood overall.

An example [134] of the remaining mysteries is dynein, a large protein with three distinct parts: a platform composed of 6 rings (300 aa each), paired antiparallel coiled coil stalks (100 aa each) which connect the platform to the stalk heads, which bind to microtubules. Long-range two-way communication is necessary for coordination between the catalytic cycle of ATP hydrolysis and dynein's track-binding affinities The long range (cooperative) metabolic interactions of hemoglobin occur between hemes separated by 100 amino acids, and were long considered mysterious, but have recently been explained in terms of coupling of linear strain fields. Even more mysterious is the transfer of ATP energy to the microtubule-binding domain, which is structurally separated by about 20 nm and 1500 amino acids from the ATP nucleotide-binding sites in the rings. Thus, long-range two-way communication is necessary for coordination between the catalytic cycle of ATP hydrolysis and dynein's track-binding affinities.

Because dynein is so large, and the connection across 1500 amino acids has been simulated mechanically with a coarse-grained model based on available structural data [135]. Such simplified models contain only a very small fraction of the internal degrees of freedom that are thermally active and likely to disrupt coherent energy transfer using short-range contact forces only. Water waves are an attractive alternative, as tsunamis have travelled thousands of miles across the Pacific Ocean. The general theory of tsunamis has been known for decades [136]. The parallel is close: the oceanic water waves are shallow, while the water film that has shaped proteins is thin. The long range of shaped water waves is a feature consistent with the 20 fractals in the MZ scale [137].

Dynein is large and complex, and its different parts exhibit interesting evolutionary improvements, for instance, in ring-ring coupling [134]. The evolution of the hydromechanical interactions of the shaft and shaft head are shown in Fig. 31. Cargos in tubulin could be moved by rocking the shaft head between the two stronger hydrophobic peaks A and B, with hydrophobic peak C serving as a pivot. This corresponds to the pre- and post-stroke conformations observed with optical tweezers [138,139]. Note that some of the motor features are merely mechanical. A model showing small tilts of flexible C1 (connected to A) and C2 (connected to B) stalks observed by polarized total internal reflection fluorescence microscopy is also consistent with this rocking model [140].

The growth of the protein data base has led to an expansion of the list of members of the myosin superfamily, generated in humans by 40 known or predicted myosin genes [141]. Myosin-1 contains ~ 1940 amino acid side groups and necessarily involves long range interactions. While they differ in many details, motor proteins contain motor heads and lever arms anchored to thick filaments [142]. Uniprot P12882 identifies the myosin-1 motor sites 86 -782, and the coiled coil lever sites 843-1939. The C terminal rod-like tail sequence is highly repetitive, showing cycles of a 28-residue repeat pattern. For myosin the optimized sliding window length W* should be close to 28. At the same time, actin is a key part of the cytoskeleton, and supports myosin, with W* = 21 (typical for membrane proteins) [130]. In tubulin, it was found that a compromise value W* = 25 gave good results for the evolution of tubulin [131], and that value is used in [142] as well.

From yeast to human there is an interesting shift in synchronization of hydrophobic and hydrophilic extremes from the N-terminal myosin-1 region (1-420) to the central region (350-650). As shown in Figs. 32 and 33, in yeast the N-terminal extrema are synchronized, while in eukaryotes the extrema of the central region are synchronized. The central region synchronization improves from fruit fly to round worm to human (Fig. 33).

How accurate are the second order MZ profiles for Myosin-1? According to BLAST, the chicken sequence P02565 registers 96% positives with the human sequence P12882. The important central regions are compared in Fig. 34. The two hydrophilic extrema are both level, and the human extrema are both 1.5 (0.5)% more hydrophilic than the chicken extrema. This accuracy of 0.5% can be compared to the BLAST positives difference of 4%. Myosin-2 human is similar to Myosin-1 chicken, and thus is less evolved than Myosin-1 human.

## 23. Level Set Synchronization: Coronavirus Darwinian Evolution

Coronavirus 2019 has evolved to be much more dangerous than CoV2003. It differs greatly from other airborne viruses (like flu) because its infections are much more likely to be fatal. It is also much more contagious, with a median 5 day asymptomatic incubation phase that can extend to 14 days. Coronaviruses are large, roughly spherical particles with unique surface projections called spikes (S). The S protein is composed of S1 and S2 subunits: the S1 subunit forms the head of the spike, and has the receptor binding domain, while the S2 subunit forms the stem which anchors the spike in the viral envelope. There are two natural questions, why has the coronavirus 2019 been so successful, and what is its source? These questions can be discussed by comparing coronavirus 2019 to coronavirus 2003, which has been much studied.

According to BLAST site-by site (W = 1) sequence comparisons, the largest and most striking difference between CoV19 (human) and CoV03, as well as coronaviruses of many species, is the insertion of a 4 amino acid PRRA sequence near the S1/S2 cleavage site, Fig. 1C of [143]. Experiments showed that this insertion facilitates spike reassembly after cleavage and is presumably the main factor responsible for the high severity of CoV19 infections. What is the cause of the asymptomatic incubation phase? The spikes contain > 1200 amino acids, and BLAST shows that ~ 300 of these are mutated. Can only a few of these 300 mutations, far from the two cleavage sites, also be contributing to the extremely strong viral interaction of CoV-2? The extreme effectiveness of CoV19 suggests self-organized criticality [144], but how do we identify these possibly critical distal and so far unobserved sites?

The spikes are long and stick out into water, so they appear to be ideally suited to hydropathic thermodynamic scaling [145]. As usual, we choose W = 35 to maximize the hydropathic shape differences between CoV-1 and CoV-2, as measured by their variance ratio. The two cleavage sites S1/S2 and $S_{2'}$ [2] of CoV-1 have moved lower (hydrophilically, further outside) in CoV-2 (Fig. 35), consistent with the very accurate MZ scale. When a cleavage segment is further outside, there is more space for cleavage and reassembly, which will occur more rapidly. The insertion PRRA in CoV-2 was identified [143] with BLAST (W = 1) as unique to CoV-2, but with BLAST alone one cannot show that this change has made CoV-2 more dangerous.

The hydropathic results shown in Fig. 35 again exhibit level sets (synchronized extrema) for CoV-2 but not CoV-1, just as in many other protein profiles in this review (see Fig. 34, etc.). Viruses must act rapidly before being destroyed by antibodies, and they could do this through synchronized motion by leveling their hydrophilic (outside) extrema. As shown in Fig. 35, such a leveling of minima 1-3 occurs in CoV-2, while it is absent from CoV-1. The change in minimum 2 is especially striking: it is caused by a cluster of four critical mutations from CoV-1 to CoV-2. These synchronized minima provide a natural explanation for the occurrence of the asymptomatic incubation phase that has made CoV-2 so dangerous. One could discard either the PRRA insert, or some or all of these four mutations, and possibly improve a vaccine based on spikes.

The abstract of [145] predicted that, because critical synchronization is so easily disrupted, a "very successful" vaccine based on spikes was possible. At this writing, this prediction appears to have been confirmed, by the report of spike-based vaccines that have been more than 90% successful in large scale trials. By comparison, flu vaccines usually achieve 20-50% success.

## 24. Kinesin Motors and the Evolution of Intelligence

Intelligence is often discussed in terms of neural networks in the cerebral cortex, whose evolution has presumably been influenced by Darwinian selection. There are more than 14 kinesin motor families, and the mechanics of the hand-over-hand kinesin step dragging cargo along tubulin have been well studied. There are three kinesin families that are very widespread among species: Kif 1, 5, and 14 [146] (see their Fig. 1). According to Uniprot, Kif1 provides "anterograde axonal transport of synaptic vesicle precursors", Kif5 is "required for slow axonal transport of neurofilament proteins", and "during late neurogenesis, Kif14 regulates the cerebellar, cerebral cortex and olfactory bulb development through regulation of apoptosis, cell proliferation and cell division".

The evolution of Kif5 from Chicken to Human is shown in Fig. 36, with W = 9. The Chicken profile contains a single large hydrophobic peak near its C-end at 196.4, with a human peak at 194.1. Human Kif5 has a second hydrophobic peak near its N-end at 191.1. The two nearly level human hydrophobic edges provide dynamical balance for Human Kif5 acting across the neural network, which is absent in Chicken Kif5.

The effects of evolution on Kif14 are illustrated in Fig. 37, which compares human and chicken profiles [147]. The chicken level set has 3 peaks which grow to 5 peaks in human. Alternatively, one can find the deviation from the mean for the five highest hydrophobic peaks. This is smallest in humans, and it increases for earlier species [147]. In humans Kif5 exhibits only two level strongly hydrophobic peaks, near the N and C terminals. Note that Kif5 is not localized in the prefrontal cortex. Altogether the qualitative differences here are good circumstantial evidence that Kif14 is important to building and refining neural networks in the prefrontal cortex.

## 25. Schrodinger's Dream – Historical Perspective

In his enormously popular Dublin 1943 lectures and book, "What Is Life?", Erwin Schrodinger proposed that we could progress in answering this question by using thermodynamics. A few years later, high energy physicists gave us hydrogen bombs, while solid state physicists gave us transistors.  Pauling discovered the alpha helix in proteins and remarked that "as the methods of chemistry are applied to physiological problems it will be found that the significance of the hydrogen bond is greater than that of any other structural feature".  Over the next 50 years high energy physicists produced nuclear power and the LHC, while solid state physicists gave us solar power, the Internet, and billions of fantastic gadgets.  Molecular biologists joined molecular solid state physicists to generate an enormous data base of protein structures and functions, which have formed a platform for miraculous medical treatments.   The DNA sequences of many species are known, and from them the amino acid sequences of many proteins: the 21[st] century is truly the genomic era [148].

The historical path from Schrodinger's classical discussion to modern thermodynamic hydropathic scaling involves many modern technical tools from mathematics and the statistical physics of phase transitions.  It also benefitted from intuitive insights, and here special mention should be made of Walter Kauzmann's work in the 1950's [149].  Kauzmann emphasized that proteins function reversibly on long time scales (ms), and behave like a deeply supercooled liquid with very high viscosity.  Darwinian evolution has achieved Kauzmann's network qualities by going beyond classical glassy "funnels" [26] and approaching fractal critical points. Kauzmann anticipated the modern viewpoint through his emphasis on the key role played by hydropathic forces in shaping protein globules.

Thermodynamic scaling may enable us to realize Schrodinger's dreams, and refine new medical platforms. Specifically it exploits the protein data base to describe the connections between amino acid sequences and protein functions. For chemists still thinking in 20[th] century terms of isolated amino acids the uses of hydropathic scaling to describe protein functions in terms of phase transitions [144] still seems to be little more than conjectural.  They assert that "the notion of 'hydrophobicity scale' is more qualitative than quantitative"[150].   It is most fitting that Schrodinger's dreams have been realized only in the 21[st] century with the discovery by Brazilian physicists of 20 universal amino-acid specific fractals in solvent-accessible surface areas of large protein segments [25].

Recently there have been several sophisticated efforts to extend the classical discussions of protein folding [26] to include energy landscapes and evolution [151], but their applications have been limited to miniproteins (< 50 aa) [152], such as the villin headpiece (35aa, noted for ultrafast folding). With thermodynamic scaling the Darwinian evolution of villin sequences (827 aa) from chickens to mice to humans has now been analyzed [153]. It turns out that the villin 35 aa headpiece is part of a 100 aa fast amphiphilic C-terminal cascade that has become stronger in humans. While miniproteins exhibit instructive dynamical properties, this example illustrates the advantages associated with thermodynamic scaling, which can analyze (Sec. 22) the dynamics even of very large motor proteins like dynein (several thousand aa). The present method frequently finds functionally significant differences between first- and second-order effects, which are inaccessible to dynamical simulations.

In the past many biologists have insisted on using only site-by-site (W = 1) comparisons (similar to BLAST) to search for evidence of positive evolution, with consistently discouraging results [154]. Dynamical scaling is a more general concept, attractive to many physicists [144]. By now thermodynamic domain scaling, which uses evolution to fix W self-consistently, has been successful for many proteins, as seen in this review.

*Postscript.* Simple laboratory toy models of sub-critical self-organized in/out globular shaping exhibit a variety of long-range interactions between water surface droplets, analogous to protein amino acid very long-range interactions [155]. Statistical methods alone applied to genome intron/extron analysis failed to identify ordering of protein sequences [156]. This illustrates why the combination of topological, geometrical and structural data in scaling theory is needed to connect protein sequences to function and biomedical applications. Students will find concise, self-contained introductions to fractals and many of the classical tools of statistical mechanics used here in [157,158].

J. C. Phillips is an applied theoretical physicist, who has studied semiconductors (basic to the opto-electronic revolution), network glasses (Gorilla glass), and protein sequences (since 2007). He is an NAS (USA) member.

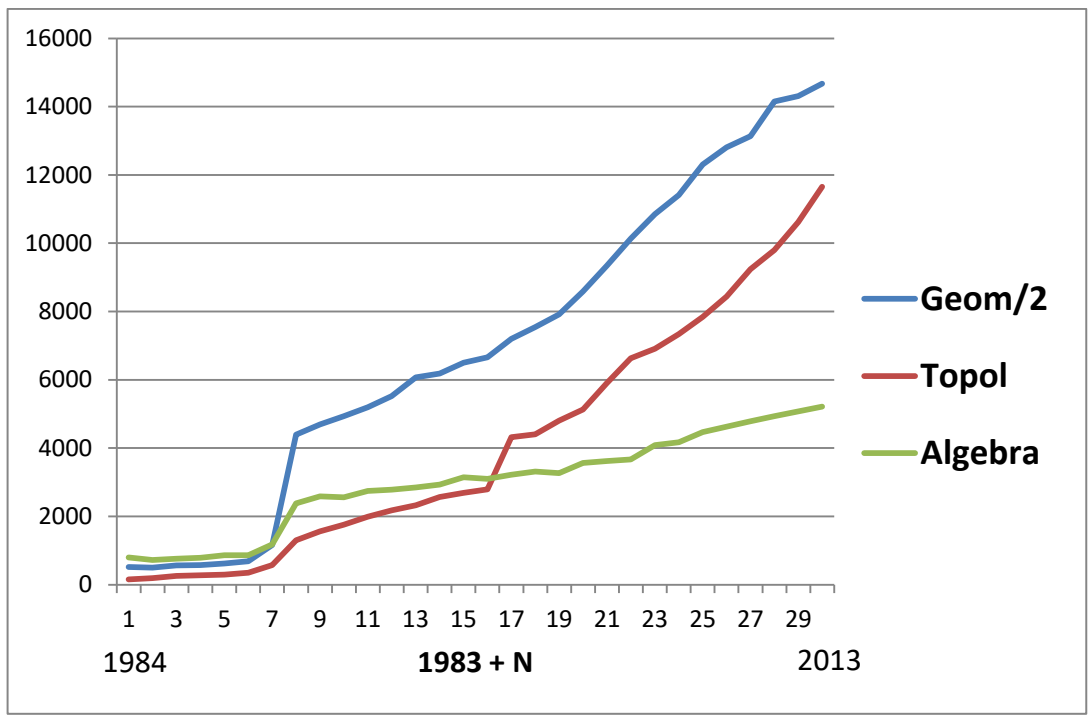

Fig. 1  The annual numbers of papers on each of three mathematical disciplines, with the geometry number divided by 2.  The entire geometry profile covers 330,000 papers.  Note not only the abrupt increases around 1990 associated with glasnost, but also the rapid increase in topology, which crosses over algebra with the advent of the Internet in 2000.  Topology is also the key mathematical tool  used  in  thermodynamic scaling.

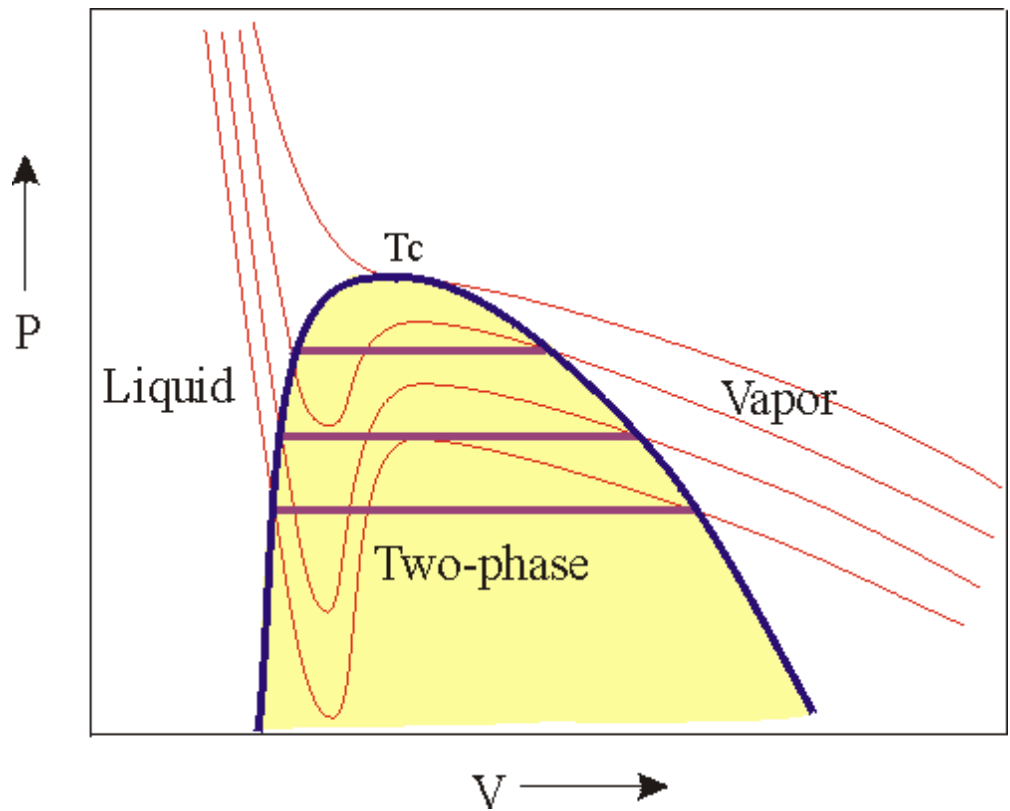

Fig. 2. Cubic isothermal curves for the van der Waals equation of state. For T > T$_c$, the curves are monotonic and contain no extrema. For T < T$_c$, there are two extrema, and phase separation occurs, with the liquid and vapor phases connected by the purple tie lines. The two extrema merge at the critical temperature T$_c$. From R. L. Rowley, Web Module: Van der Waals Equation of State, with author's permission.

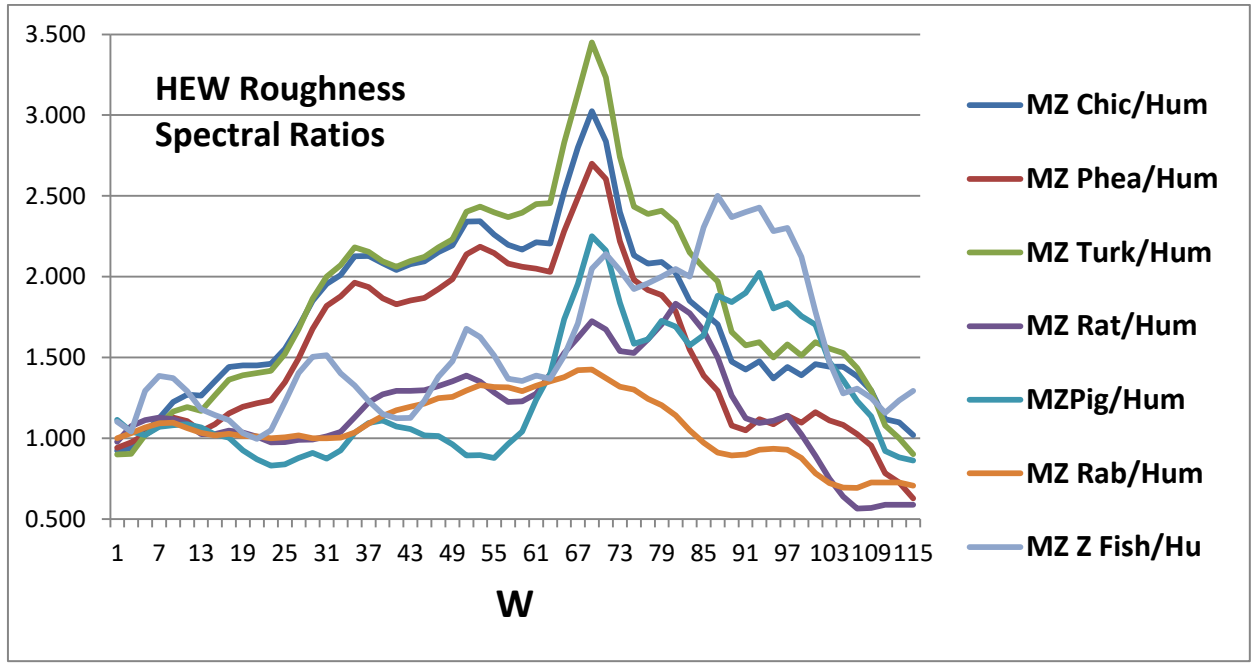

Fig. 3. The roughness $R_S(W)/R_{Human}(W)$ ratios are shown for six terrestrial species [8,50]. These ratios exhibit both a broad and narrow peak, both centered on W = 69, suggesting that the evolution of HEW from chicken to human has been directed towards improving a specific function, avoiding aggregation (amyloidosis).

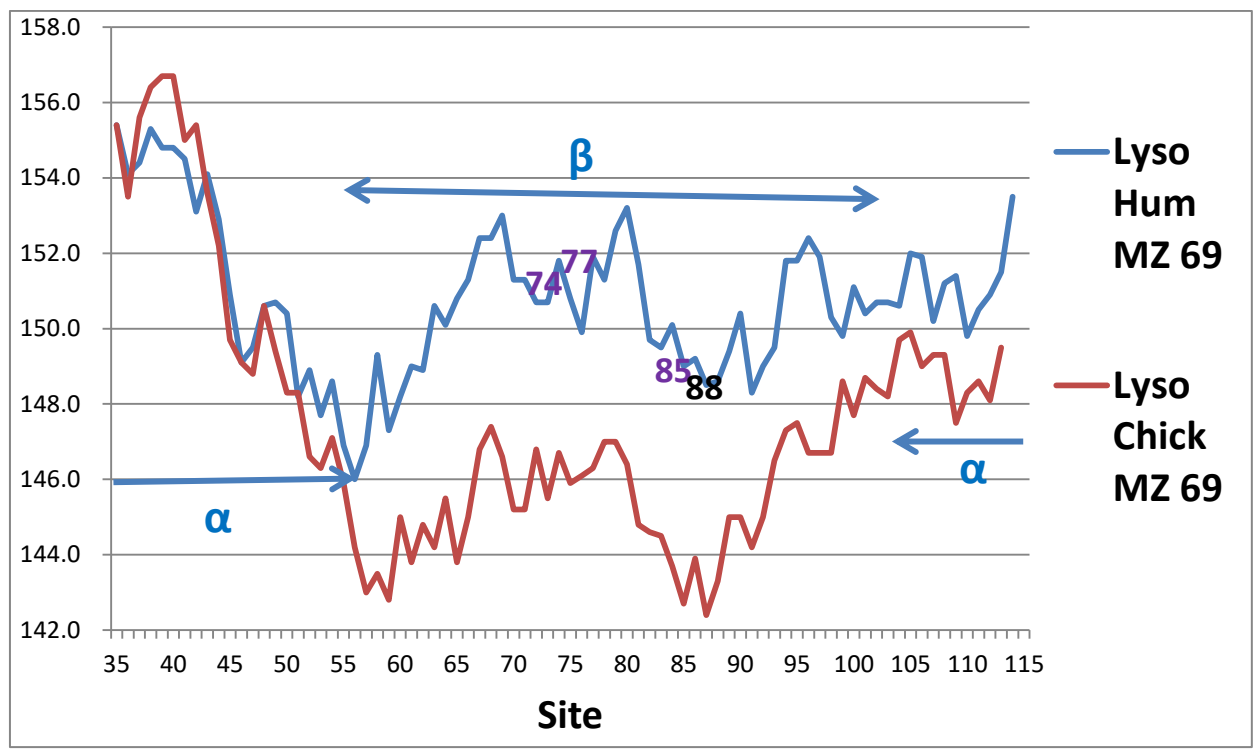

Fig. 4. Hydropathic profiles of human and chicken HEW, using the fractal MZ scale, with W = 69, as suggested by Fig. 3. Note the large stabilization by the human strain in the β region, compared to the small differences in the α regions. Here increasing ordinate corresponds to increasing hydrophobicity and increasing rigidity. The numbered sites near the center show large mutated aggregation rate changes.

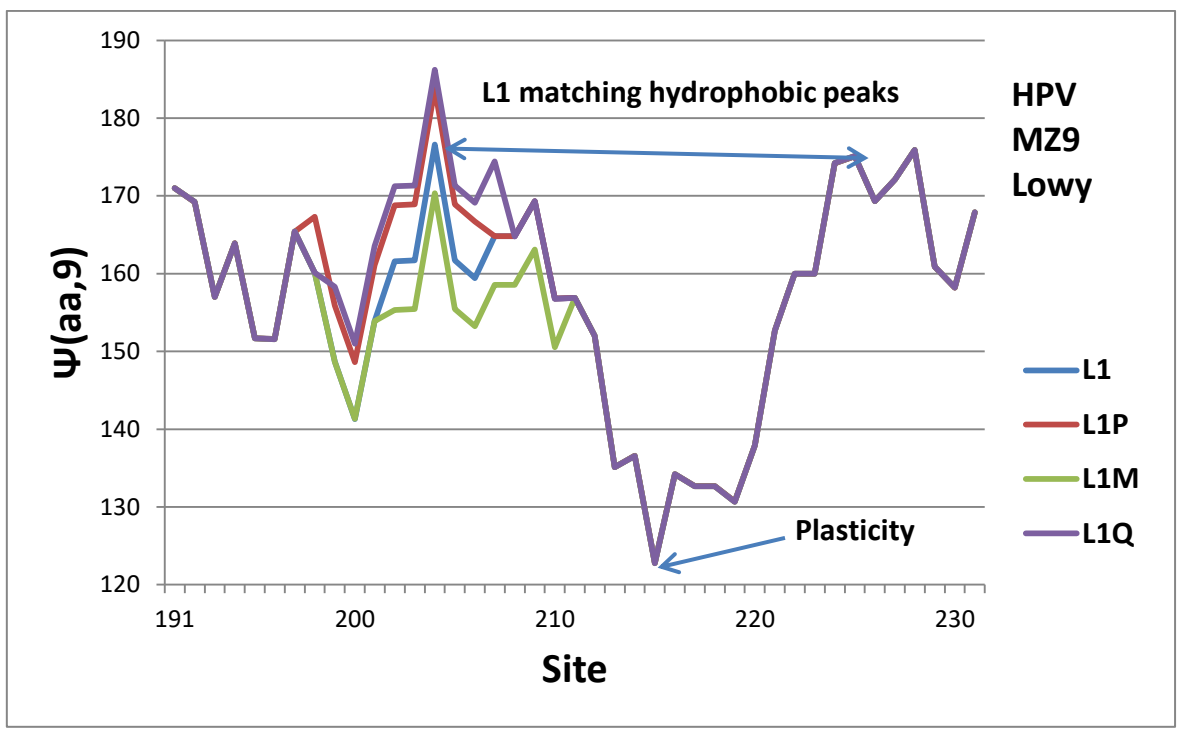

Fig. 5.  Hydroprofile of L1 and several single amino acid mutants, using the fractal scale [25,55].

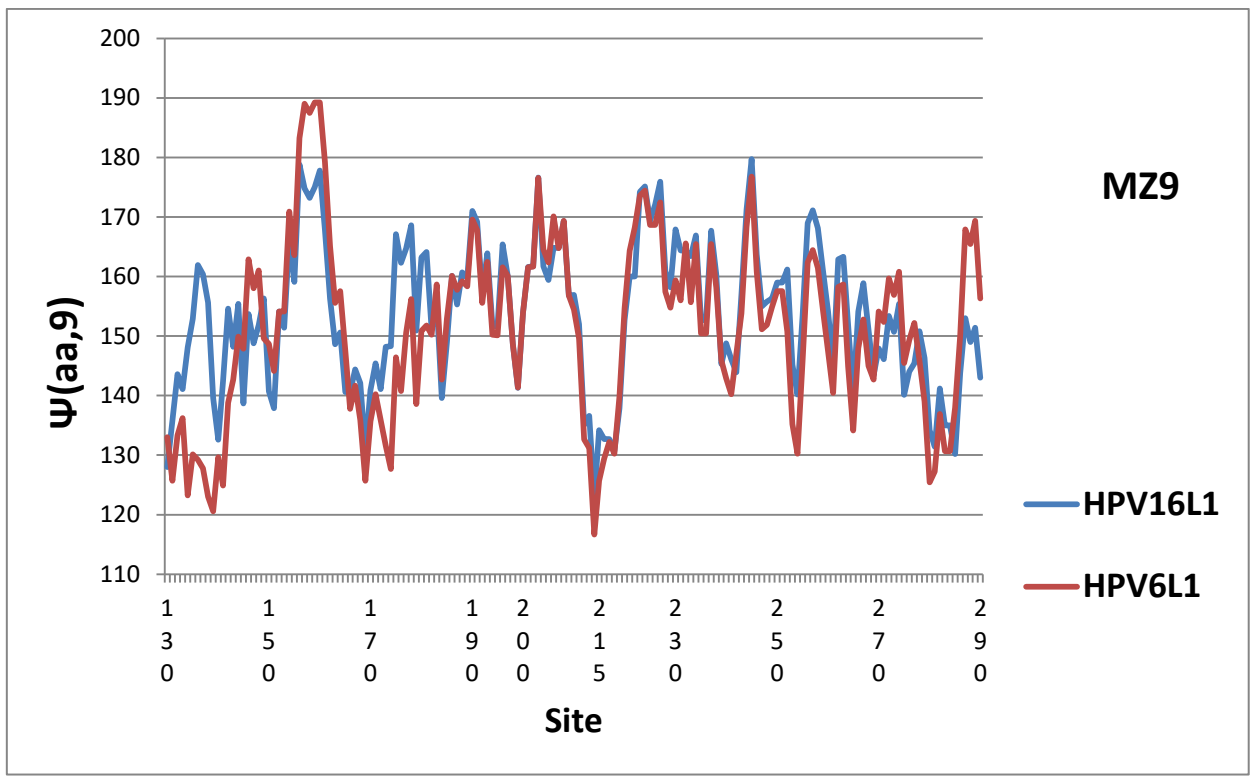

Fig. 6. By aligning the extrema, the MZ9 hydropathic profile reveals strong similarities (r = 0.82) between L1 HPV16 (cervical cancer) and HPV6 (warts). The large differences around 135 and 160 could be the major factor in the functional differences.

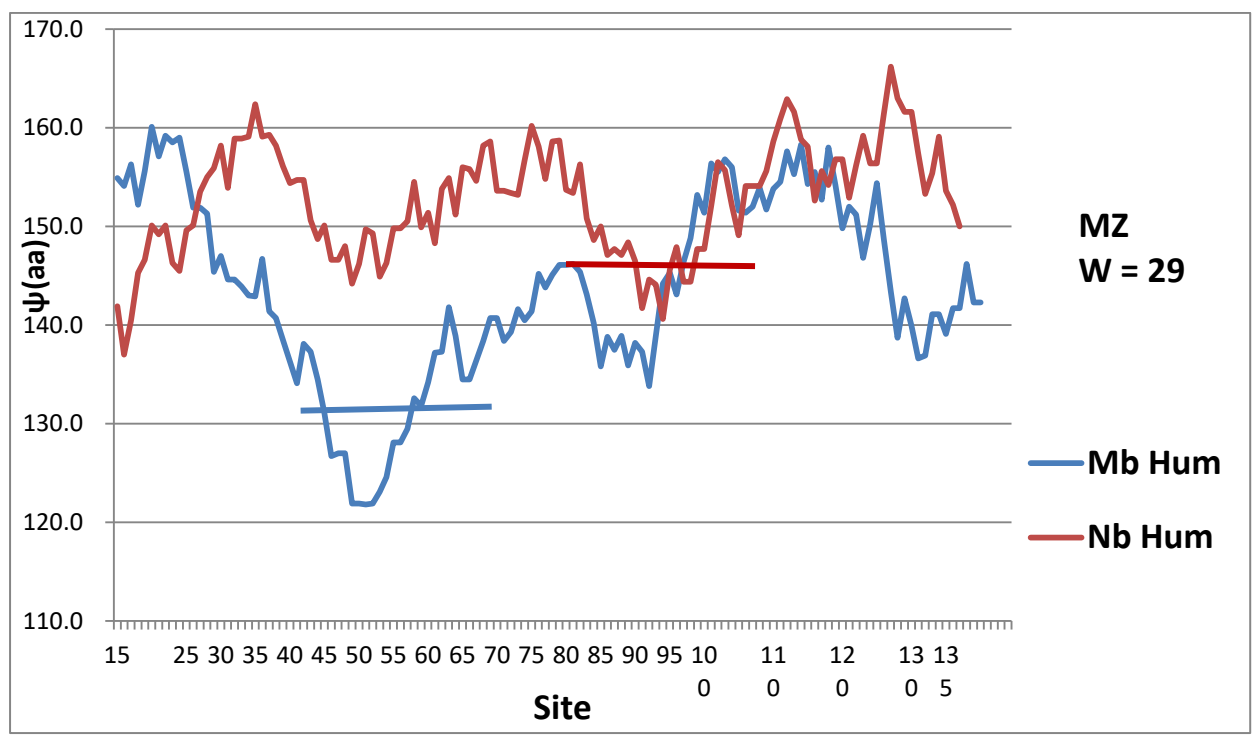

Fig. 7. The Ψ(aa,W= 29) hydropathic profiles for human Myoglobin and Neuroglobin. The ψ(aa) averages for the deepest elastic region minima spanning 29 amino acids are marked by line segments. The much lower elastic blue value for Mb enables the E8 His (64) gate to open and function as a small gas molecule channel. The much larger red elastic value for Ngb stiffens this region, so that the functional channel is switched from Mb. Note that in Mb the deepest hydrophilic minimum is centered near the distal His 64, while in Ngb the deepest minimum is centered near the proximate His 96. This shift is associated with the two different oxidation channels. Note also that the two Ngb minima are nearly level, suggesting a faster synchronized oxygen release.

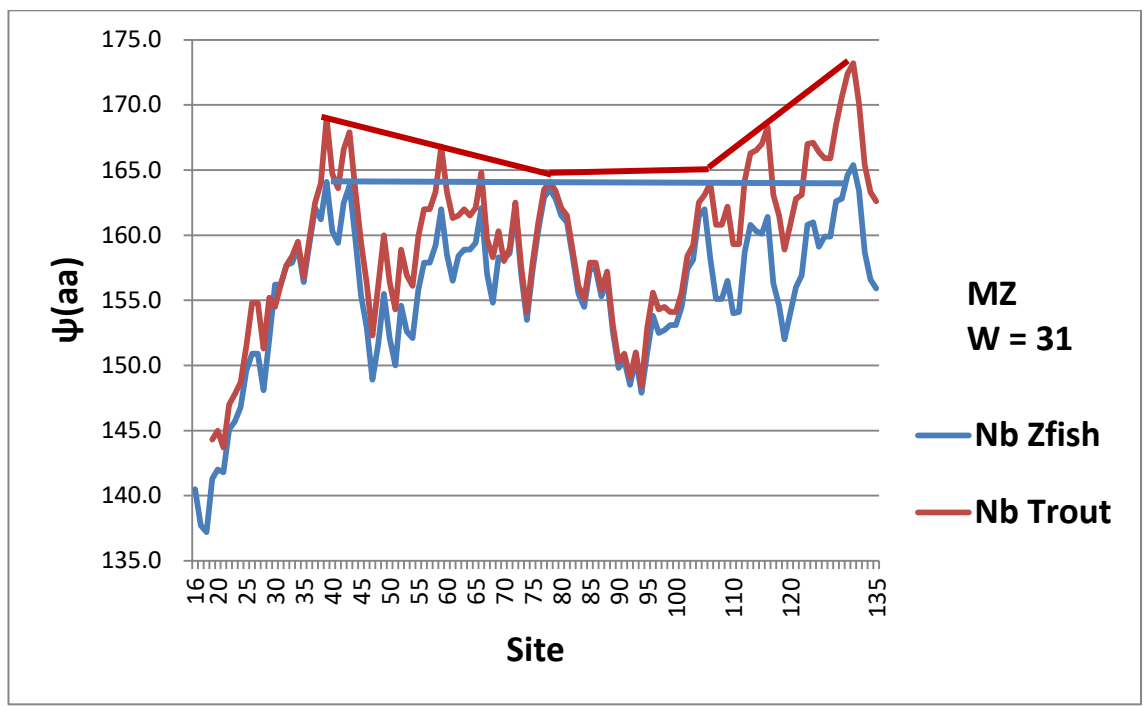

Fig. 8. Comparison of the hydroprofiles of tropical and temperate freshwater fish. The central region involving the heme and the distal His 64 – proximate His 96 channel is conserved, while the remaining regions are parabolically more hydrophobic for temperate trout, and nearly level for tropical Zebrafish. The parabolically more hydrophobic regions are necessary because of larger temperature fluctuations in temperate regions. Note also the unmarked hydrophilic minima at 47 and 94. These also divide the proteins into three nearly equal regions, with the Zebrafish hydrophilic minima again being more nearly level [69].

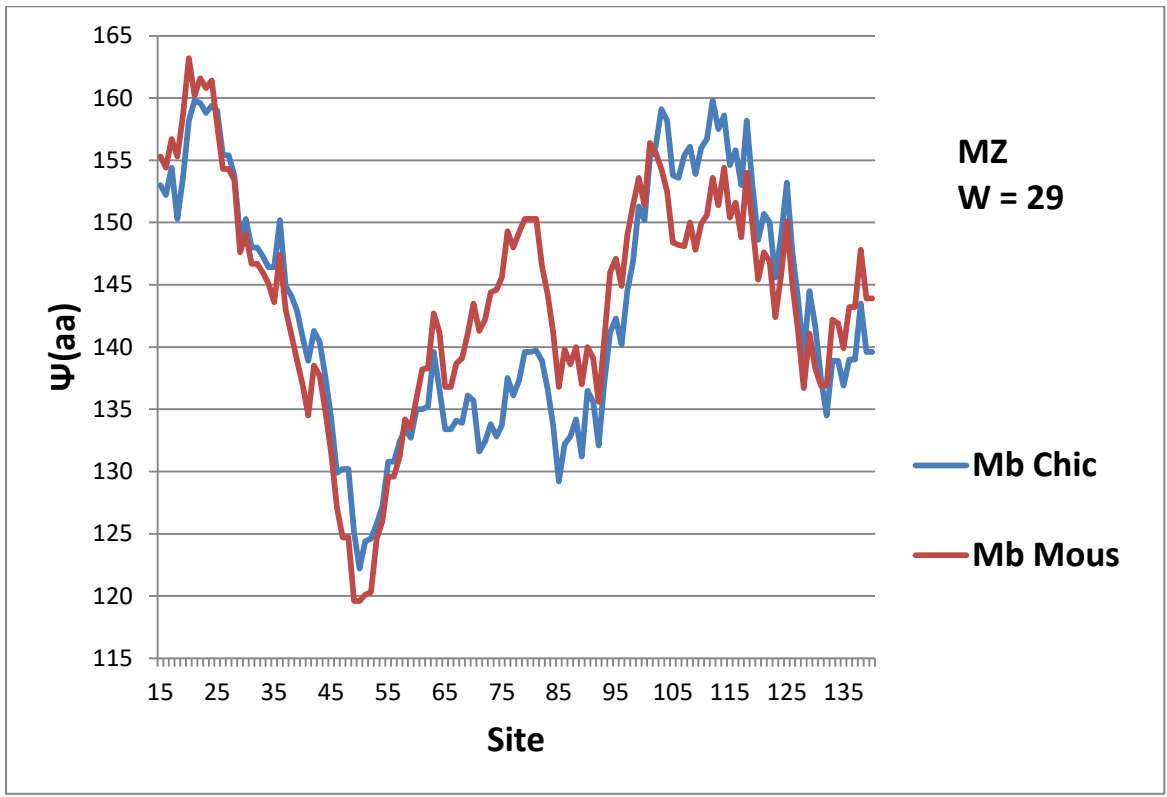

Fig. 9. The largest change here is the mouse apical hydroneutral peak near site 80. It acts as a balanced pivot, which facilitates release of both O and $CO_2$.

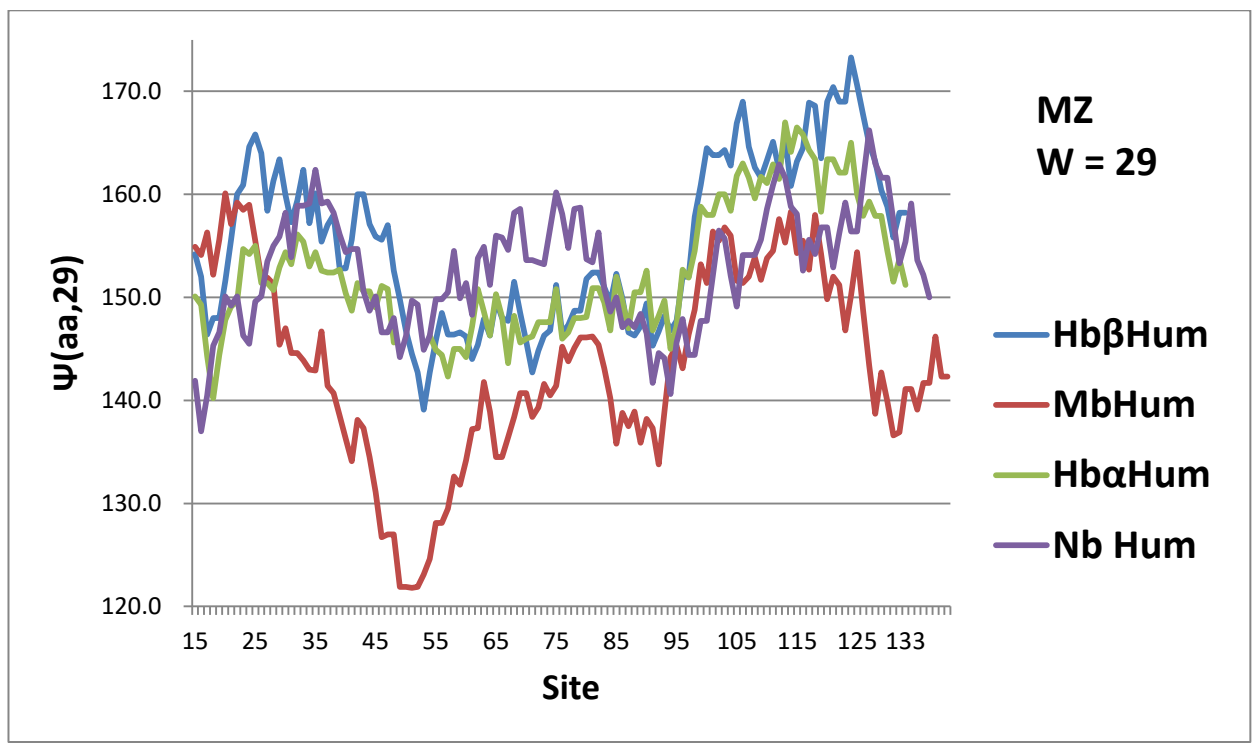

Fig. 10.    A striking feature of these profiles is that Mgb is the "odd globin out", which can be understood as reflecting the function of Mgb as storing oxygen in tissues for long periods.

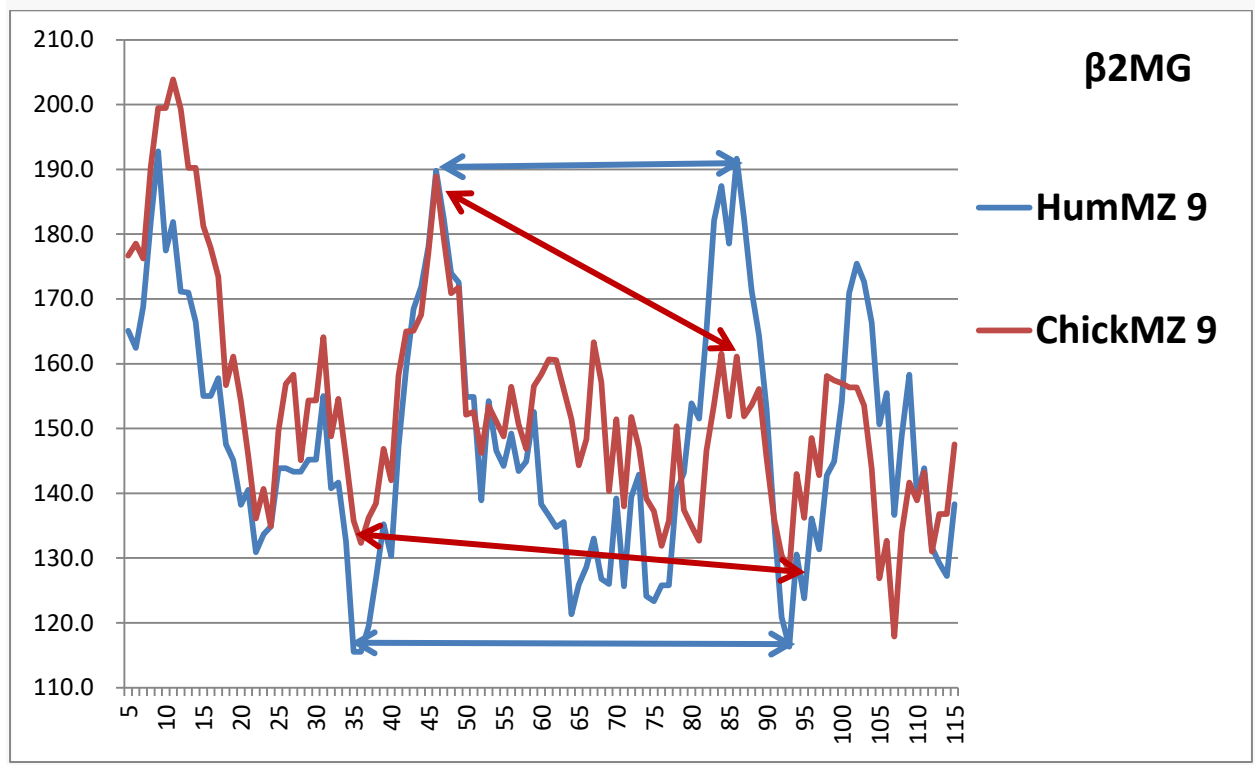

Fig. 11. Profiling the hydropathic shape of β2m with W = 9 reveals striking features, and how these have evolved. The structure is stabilized by two hydrophobic peaks near sites 45 and 85, while its kinetics are driven by the hydropathic amide ends and the hydrophilic carboxyl ends. There are two hinges near 35 and 93, which are nearly level in chicken, and almost exactly level in human. The largest evolutionary change is the leveling of the two hydrophobic peaks near sites 45 and 85 in humans.

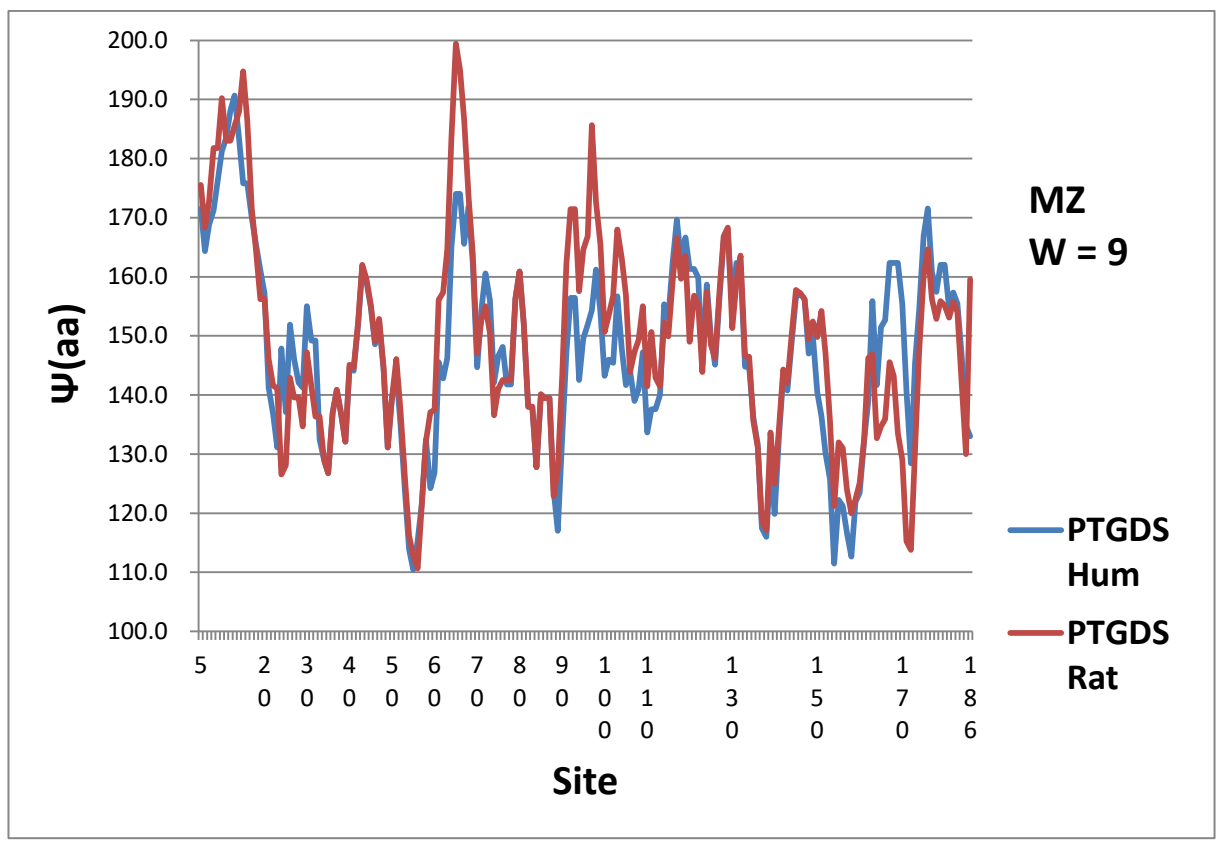

Fig. 12. Comparison of human and rat prostaglandin-H2 D-isomerase MZ W = 9 profiles. The human profile is smoother and flatter across multiple hydrophobic and hydrophilic extrema, which enable this enzyme to function better in multiple contexts [63].

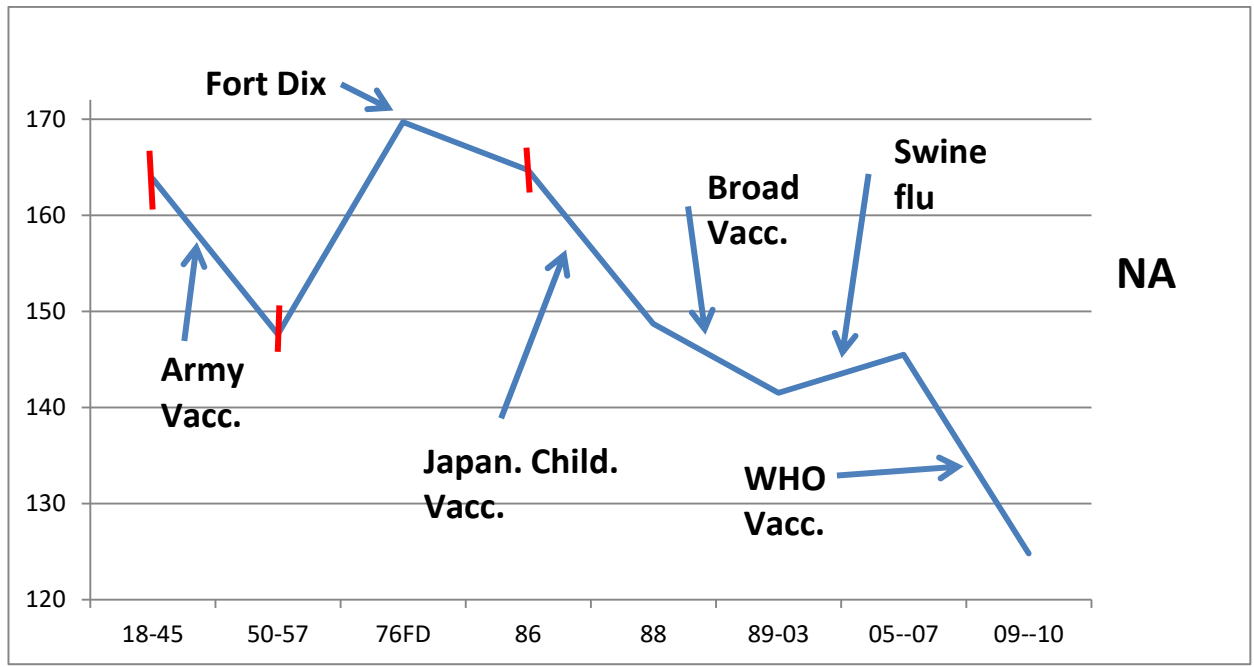

Fig. 13. A panoramic presentation of the opposing effects of migration and vaccination on NA roughness (variance of $\psi$(aa,17), MZ scale) after the first wide-spread vaccination program, begun by the US Army in 1944, as listed in Table I. Flu virulence decreases or increases in tandem with NA roughness. The build-up of swine flu virulence from 2001 on is evident in selected urban areas (New York, Berlin, Houston) with crowded immigrant neighborhoods . A few early error bars are indicated in red, but after 1986 these become too small for this sketch.

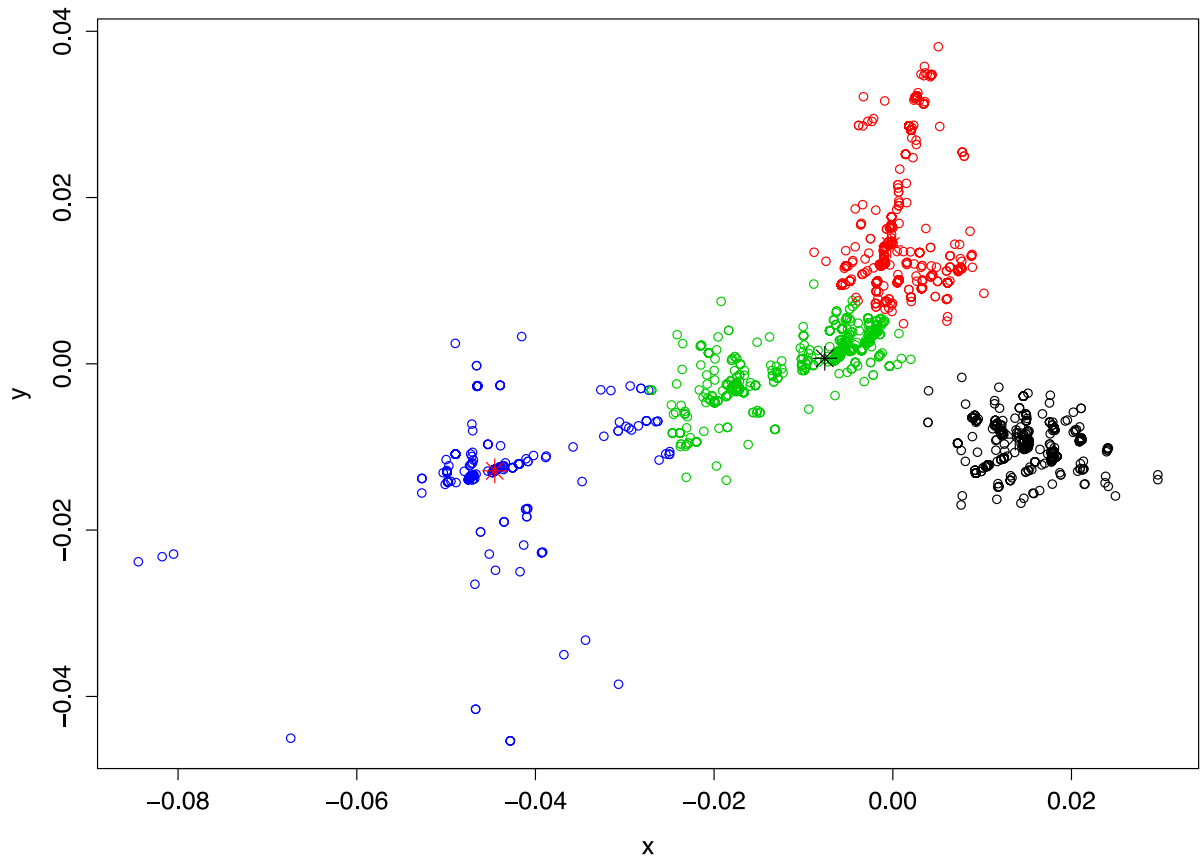

Fig. 14 Voronoi clustering of d = 2 compressed sequence data. Colors represent for 4 different clusters, each centering at the stars. The clusters follow a linear time orderliness from left to right. Black area is the A/Nebraska/4/2014 cluster. A curious and as yet unexplained feature of the three earlier clusters is their linear backbones. The time intervals for these four clusters is altogether 6 months (1.5 months each, 15.7.1-15.12.31), which illustrates how timely predictions based on it can be. Most phylogenic plots use at least 12 months per cluster, and have much weaker resolution.

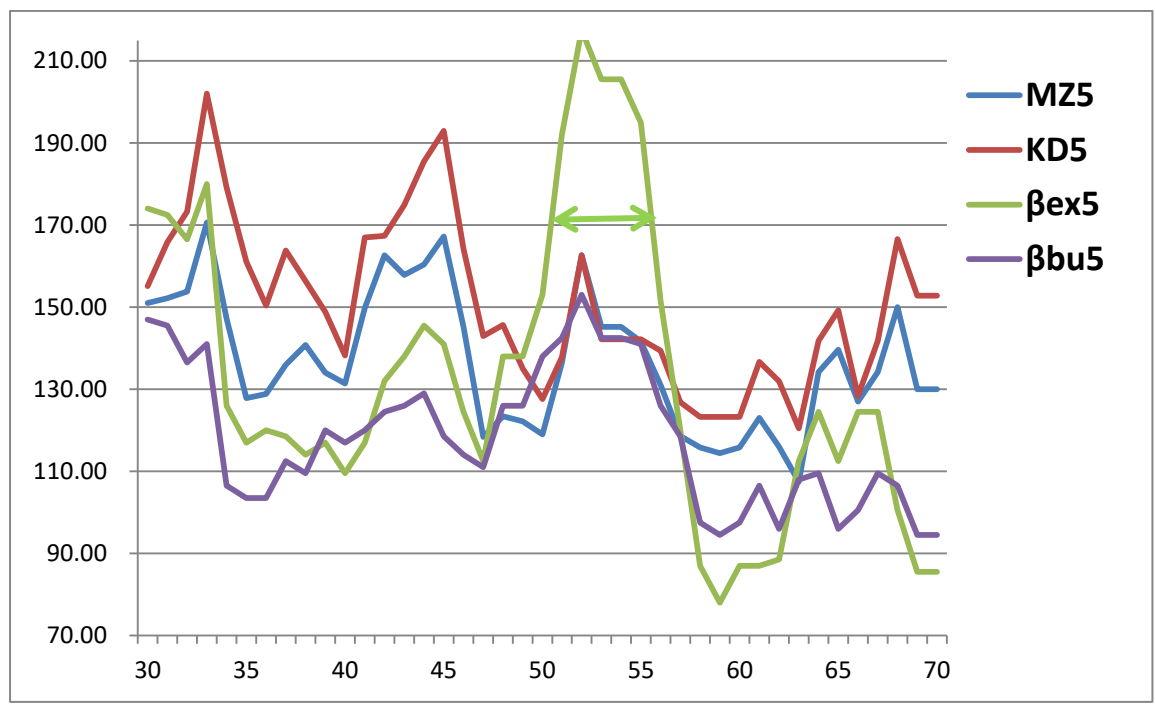

Fig. 15. [110] used four scales (two Ψ, two β) in the 30 – 70 site range where the sensitive epitopes -9-10 (sites 41-60) occur, with W = 5. The choice of a small W produces noisy profiles, but the success of the βex scale compared to the other scales in identifying the central 7-mer 50-56 IEQWFTE epitope is clear (double arrow). Note the Trp53 = W53 at the center of this epitope, and how use of the βex scale lifts the 50-56 compressed epitope signal well above noise level.

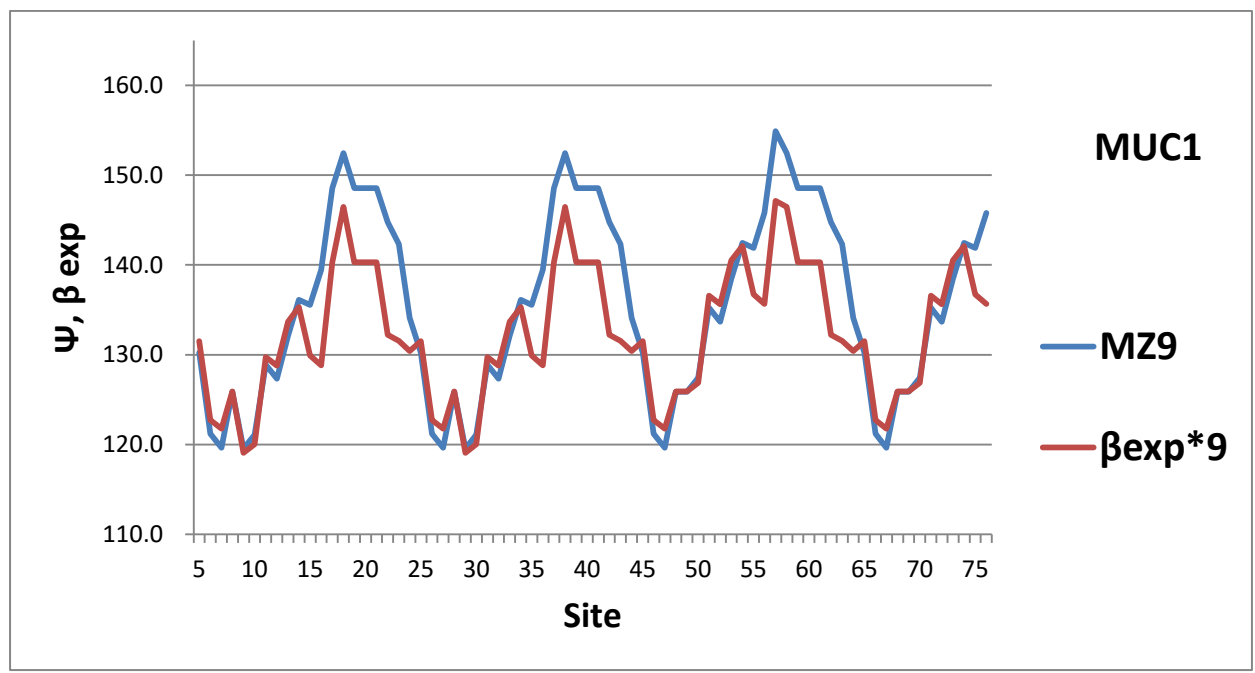

Fig. 16. For the most sensitive MUC1 repeats the βexp* values lie below the ΨMZ values (W = 9), but in a special way: about 10 near the relatively hydrophobic maxima, but near 0 for the extremely hydrophilic minima. Overall the MUC1 ΨMZ values are hydrophilic, and lie below the hydroneutral value for ΨMZ of 155. This is consistent with the overall disordered mucin structures, and is similar to the overlapping 15-mer 40-60 p53 region of Fig. 15.

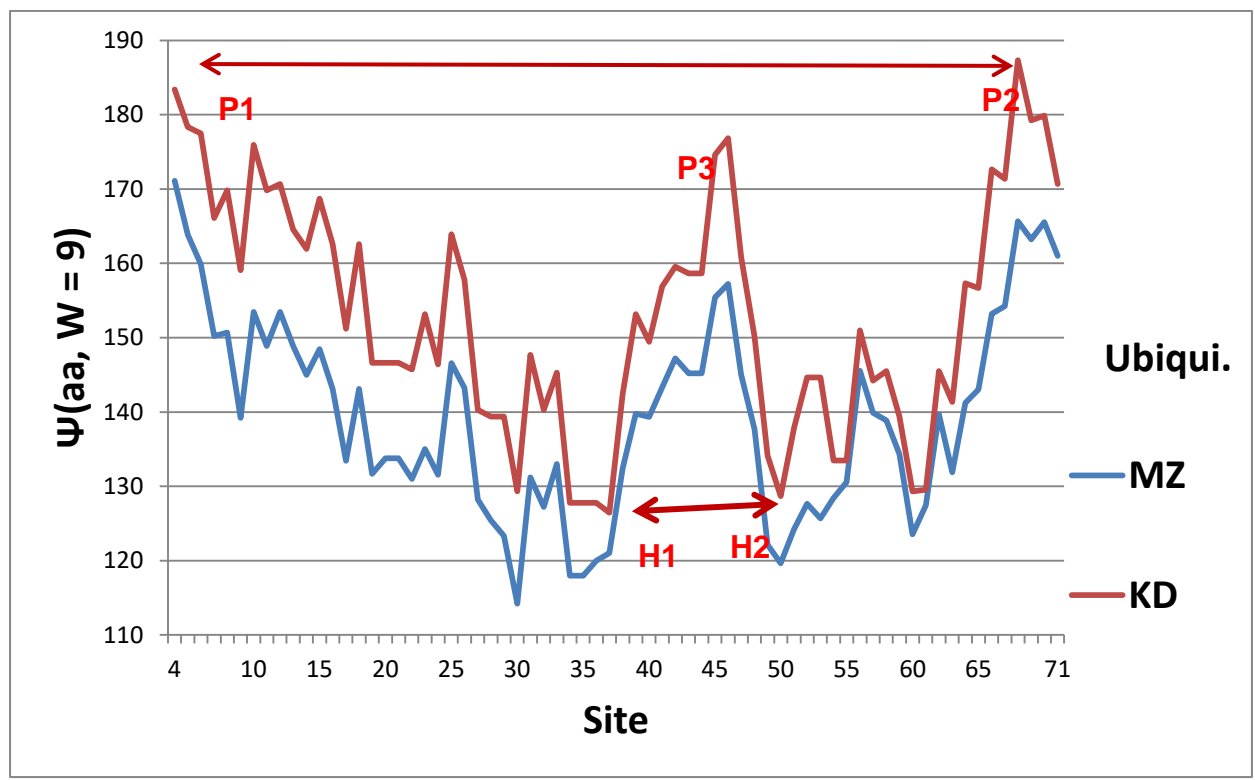

Fig. 17. Hydropathic profiles with sliding window width W = 9 (smallest value associated with MZ scale, and highest resolution) for ubiquitin. Sites are numbered from N terminal to C terminal, and the two scale centers have been slightly offset for clarity, with larger values being more hydrophobic, and smaller values more hydrophilic. Hydroneutral is approximately 155 on both scales. The overall shape is roughly that of a symmetrical parabolic bowl, with a hydrophobic central peak ~ site 45 functioning as an elastic pivot enabling bent tagging. Thus the N and C terminal wings can swing separately, facilitating tagging the opposite sides of a crack in the strain field of a diseased protein. The average hydropathicity of ubiquitin is hydrophilic, which means it is elastically softer than average, which again facilitates tagging diseased cracks near the N and C terminals. The hydrophobic N (P1) and C(P2) terminal peaks are nearly level, which facilitates synchronous tagging, as are the hydrophilic hinges H1 and H2.

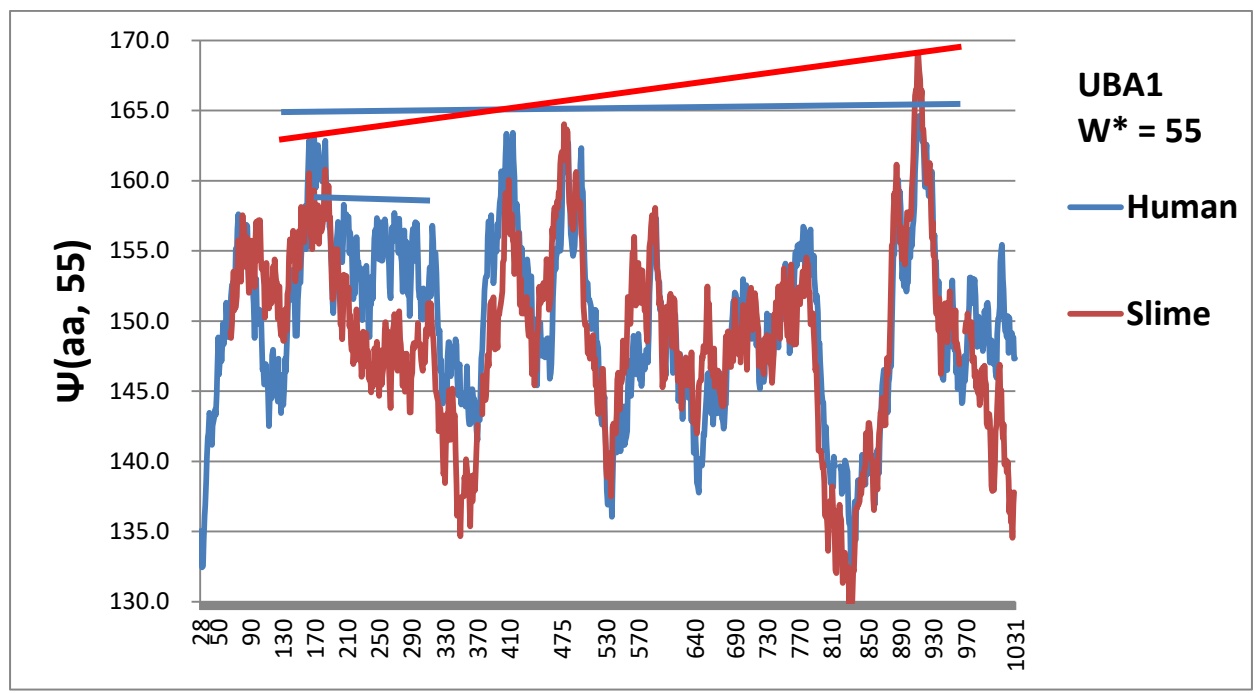

Fig. 18. The $\Psi(aa, W*)$ profiles for human and slime mold Uba1 (E1), using the modern MZ scale, which describes second-order (allosteric) evolutionary approaches to perfection (thermodynamic critical point). A similarly successful value for W* using the standard KD classical (thermodynamically first-order) scale was not found. Note the secondary leveling of the human profile between sites 200 and 330. The site numbering here is that of Uniprot P22314 (UBA_1HUMAN). The four highest human hydrophobic pivots are nearly level, while the slime mold pivots are tilted by ~ 15%. The choice of W* = 55 is consistent with the size of ubiquitin. With > 1000 amino acid sites, such a large value of W* again emphasizes the importance of W as a tuning element for focusing on domain interactions.

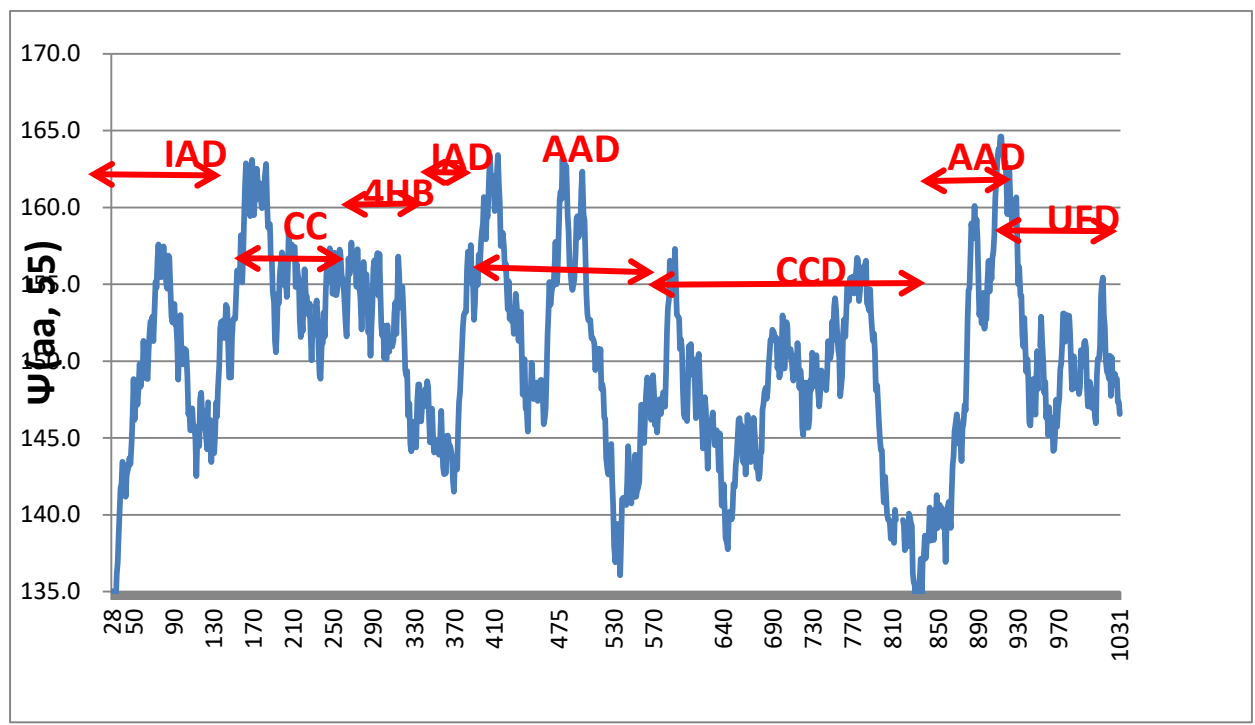

Fig. 19. Labelling of the enzyme Ubal (El) human MZ profile Ψ(aa, 55) according to the structural analysis of reveals many very accurate pivotal set leveling effects. The most obvious one is the four highest pivoting peaks (Fig. 1), associated with the two parts of IAD and AAD, ''inactive'' and ''active'' adenylation domains. Also labelled are the catalytic cysteine half-domains (CCD), which are also level at nearly hydroneutral.

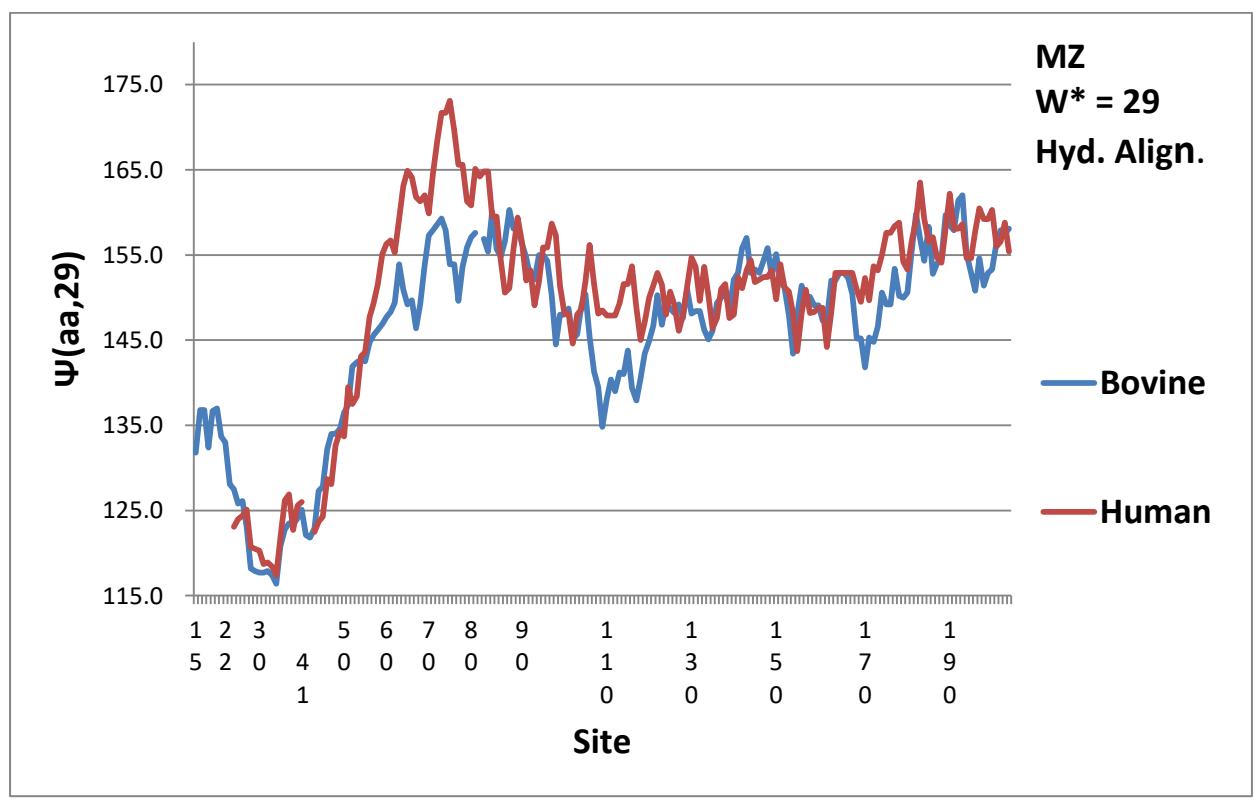

Fig 20. Hydropathic casein beta profiles with the MZ scale. Here the profiles are aligned with a two-site human gap at 41, and a single-site bovine gap at 84, thereby maxinizing their correlation.

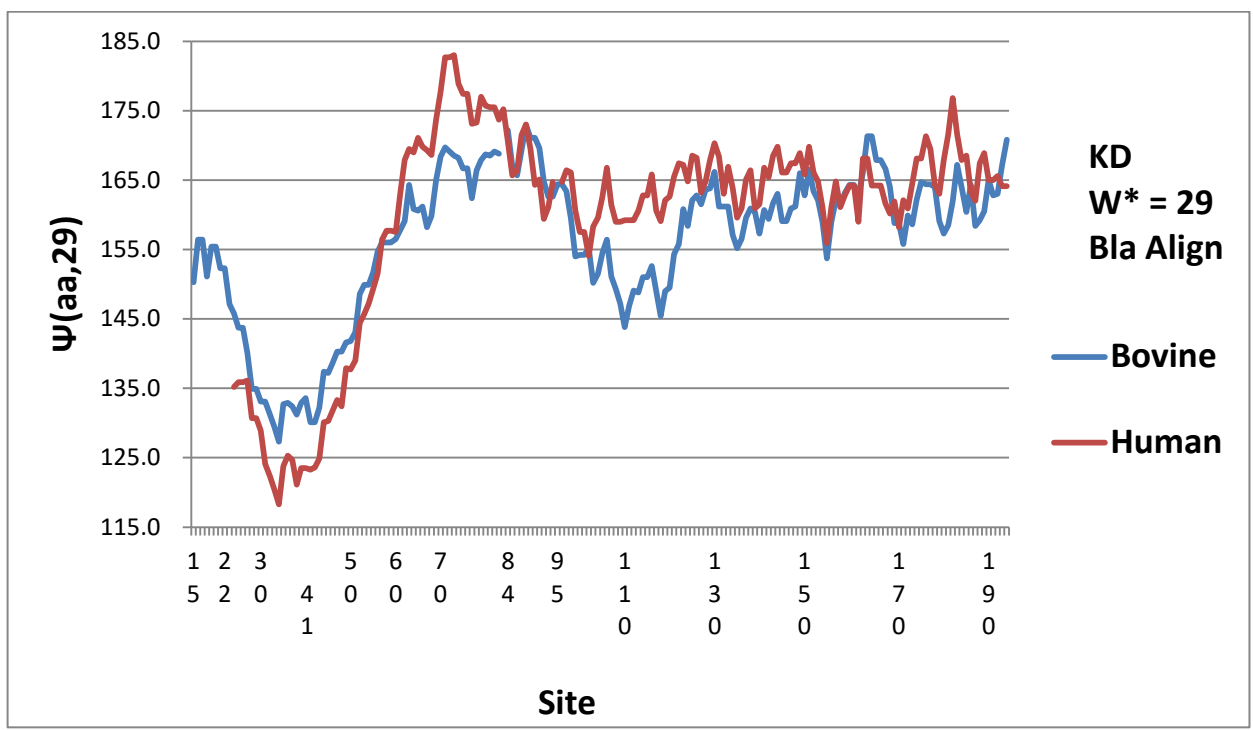

Fig. 21. Hydropathic casein beta profiles with the KD scale. Here the profiles are aligned with the BLAST single-site bovine gap at 84 and single-site human gap at 95.

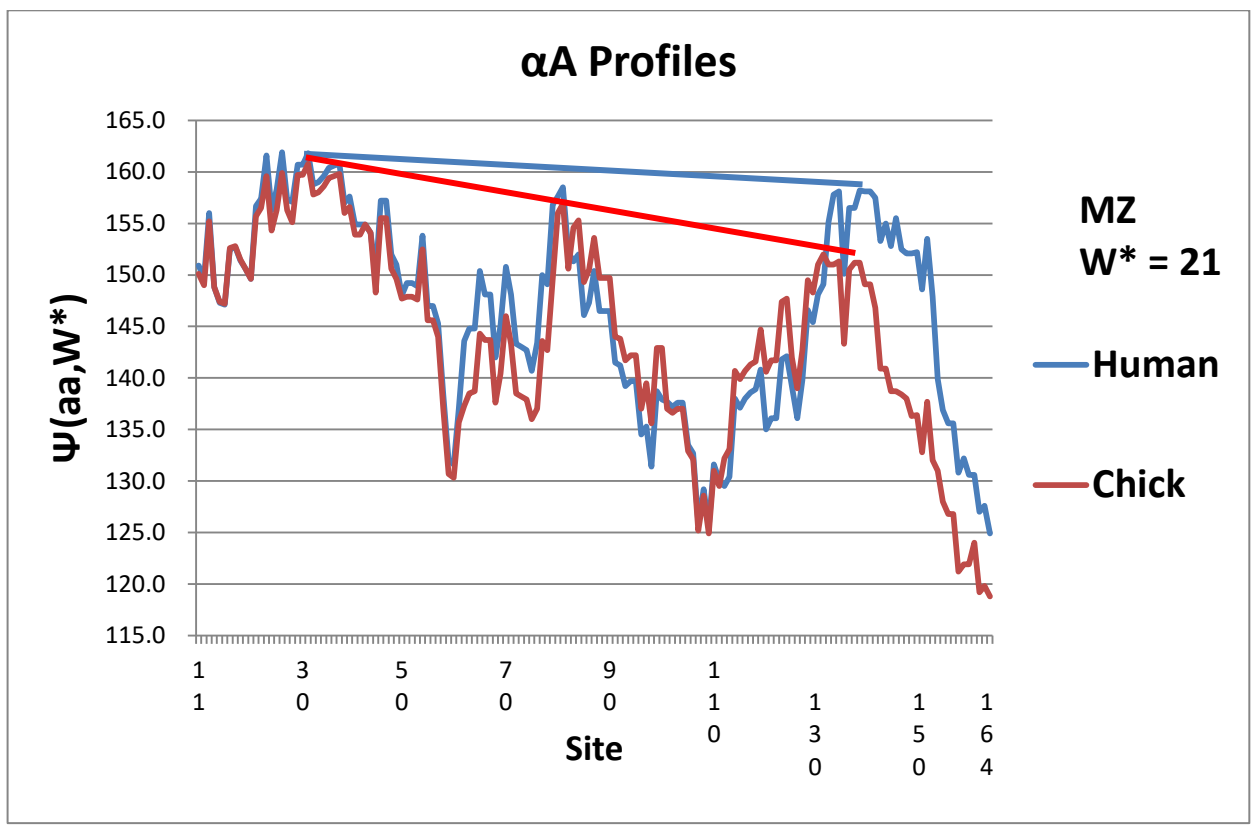

Fig. 22. Hydropathic wave profiles of human and chicken αA crystallins. Note the leveling of the hydrophobic peak at 140 with the peak at 80 in humans. Note also the strong hydrophobic n-terminal conserved peak at 30, and the increase in hydrophobicity of the c-terminal peak in humans. The chicken extrema (both hydrophobic and hydrophilic) show amphiphilic trends, decreasing from n terminal to c terminal. Trend lines are added for the reader's convenience.

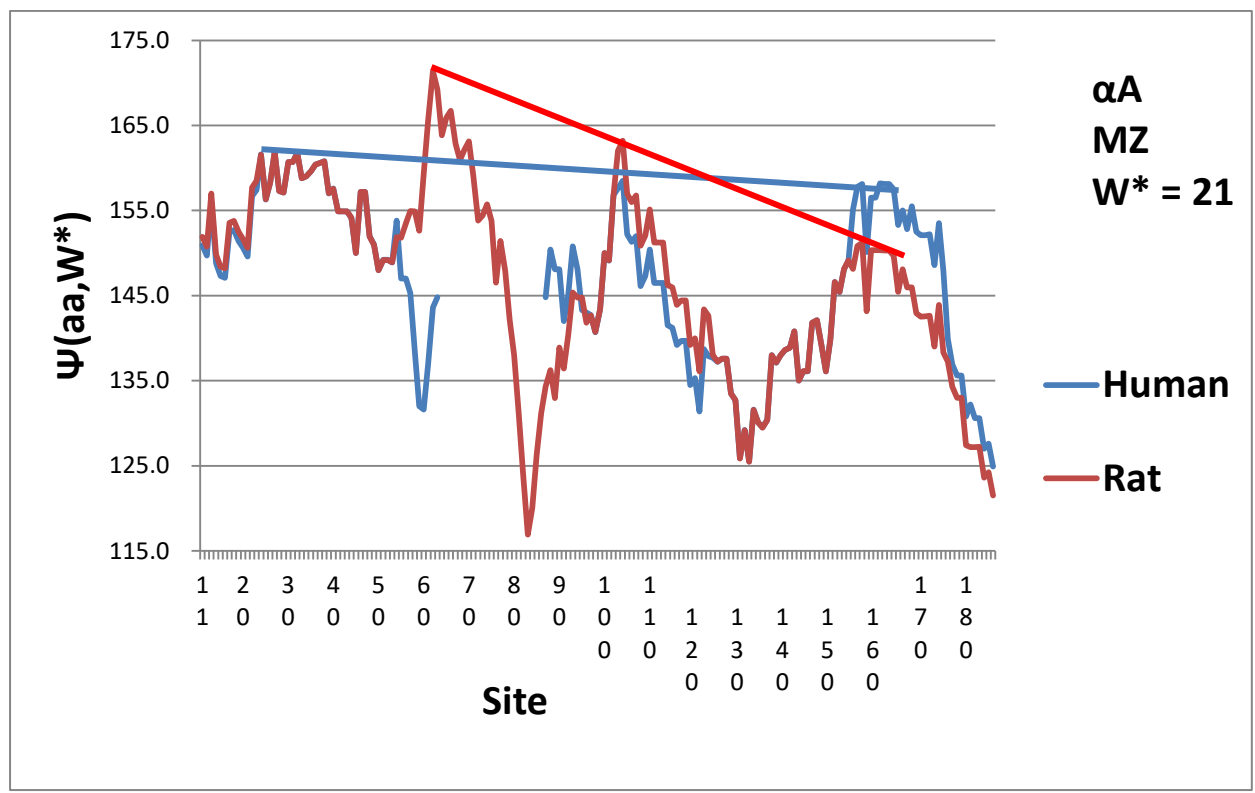

Fig. 23.  Hydropathic wave profiles of human and rat αA crystallins (rat site numbering).  This figure should be compared to Fig. 22.  As described in the text, the differences suggest a possible connection to the formation of cataracts in rats only two years old.  Trend lines are added for the reader's convenience.  The rat αA sequence is from P24623.

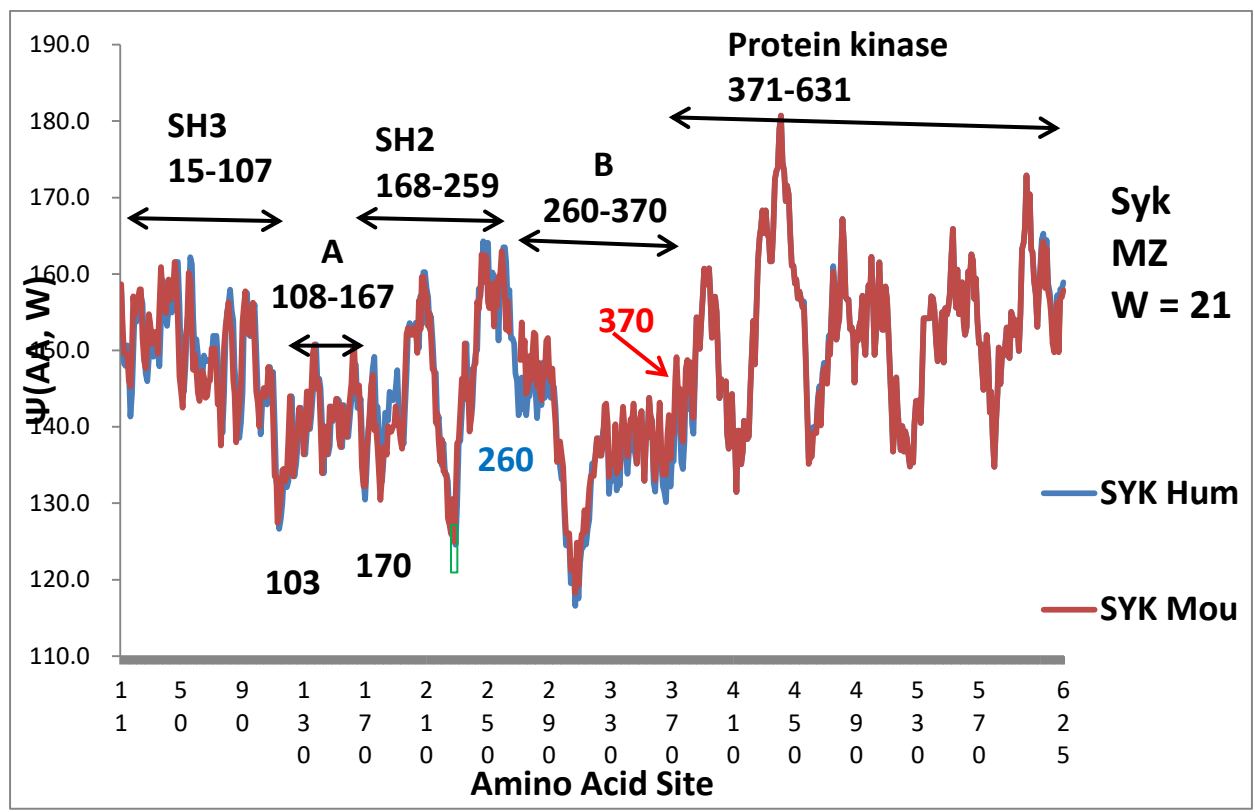

Fig. 24. Hydropathic one-dimensional profiles Ψ(aa,W) for tyrosine kinase Syk for human and mouse using the MZ scale. The annotations are from three+--dimensional structural data in Uniprot P43405. The value W = 21 corresponds to a membrane length, which is appropriate for signaling globular proteins which interact with transmembrane proteins. The catalytic octad cluster 56-63 is located at the center of SH3. The three hydrophobic structural clusters are separated by two hydrophilic regions A and B. Most of the small differences are concentrated in the central functional domain SH2 and its A and B regional wings.

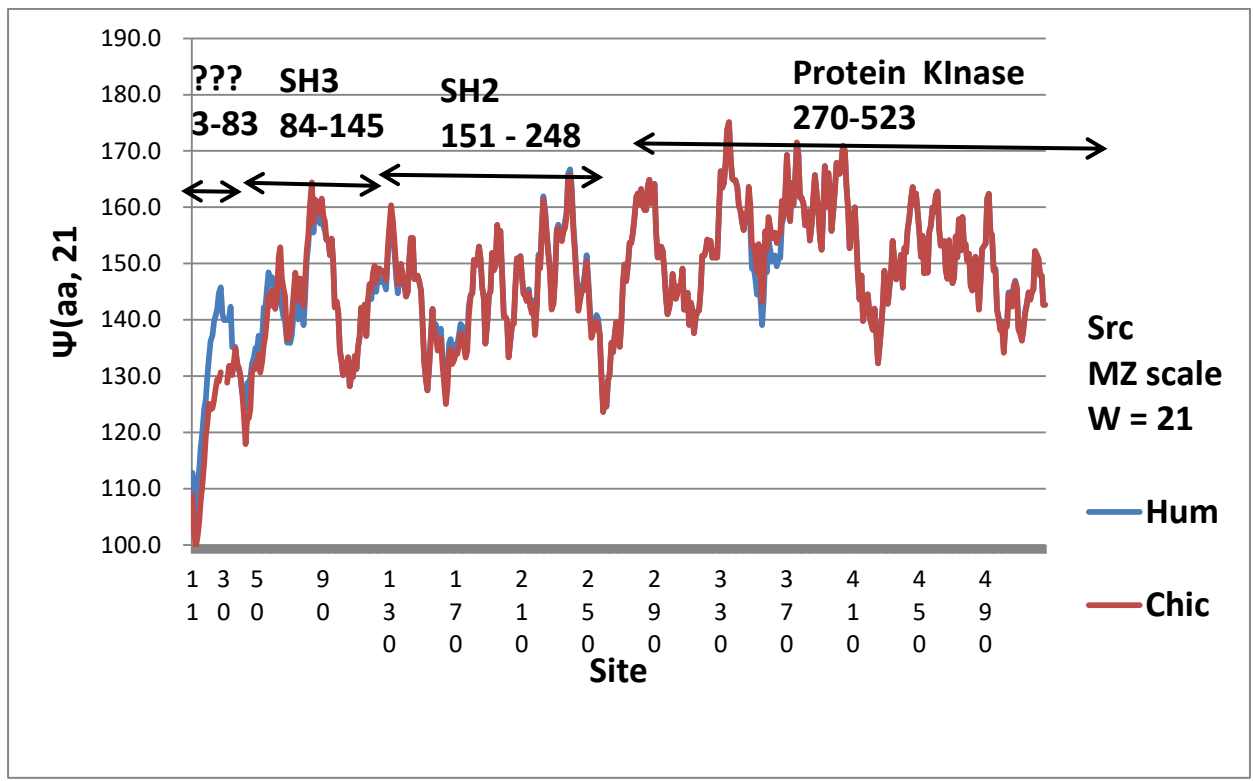

Fig. 25. Hydropathic profiles Ψ(aa,W) for tyrosine kinase Src using the MZ scale. The value W = 21 corresponds to a membrane length, which is appropriate for signaling globular proteins which interact with transmembrane proteins. Note that almost all the evolutionary changes are centered near 30 in the mysterious U region near the N terminal. They are associated with the hydroneutral human septad 25-31 discussed in the text. The SH3 target recognition domain and its companion SH2 domain are formed at an early stage of evolution.

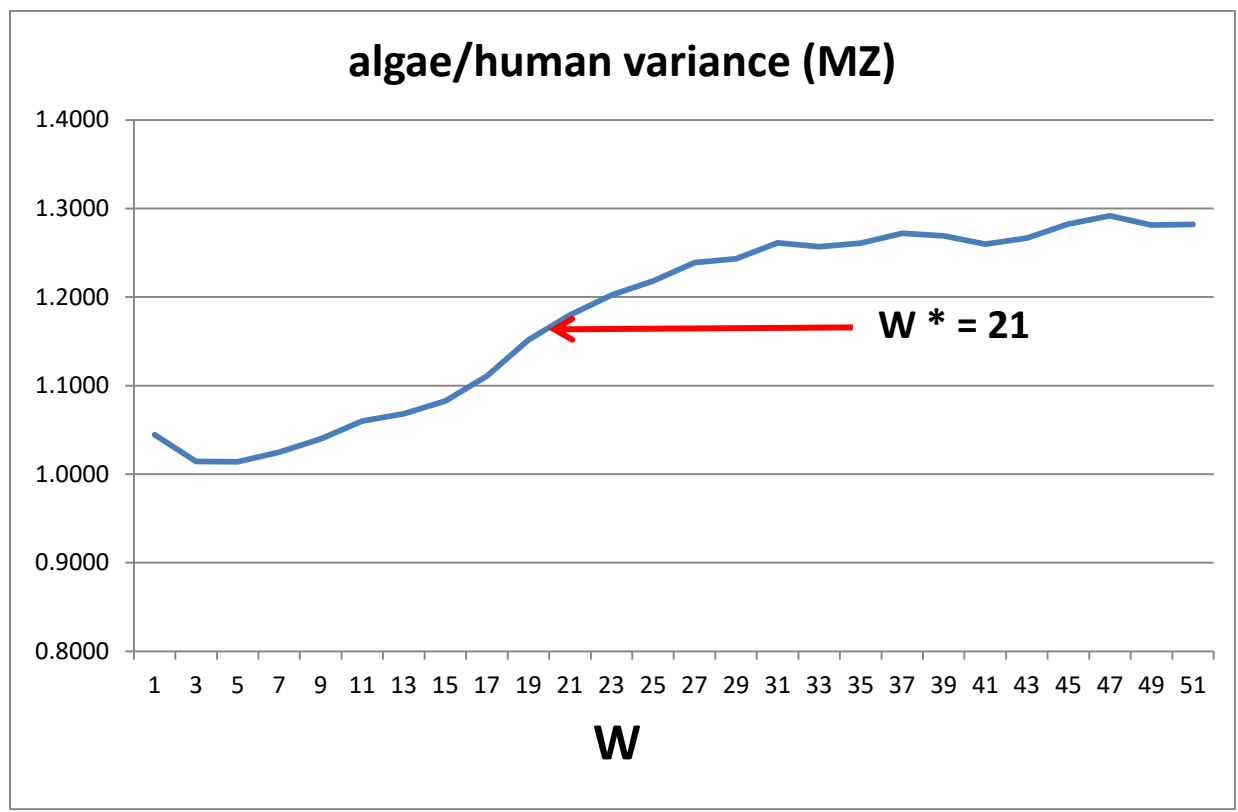

Fig 26. Actin variance ratios as a function of sliding window averaging width W. The maximum in $dV_r/dW$ occurs at W* = 21. Human sequence P68133, algae sequence P53500.

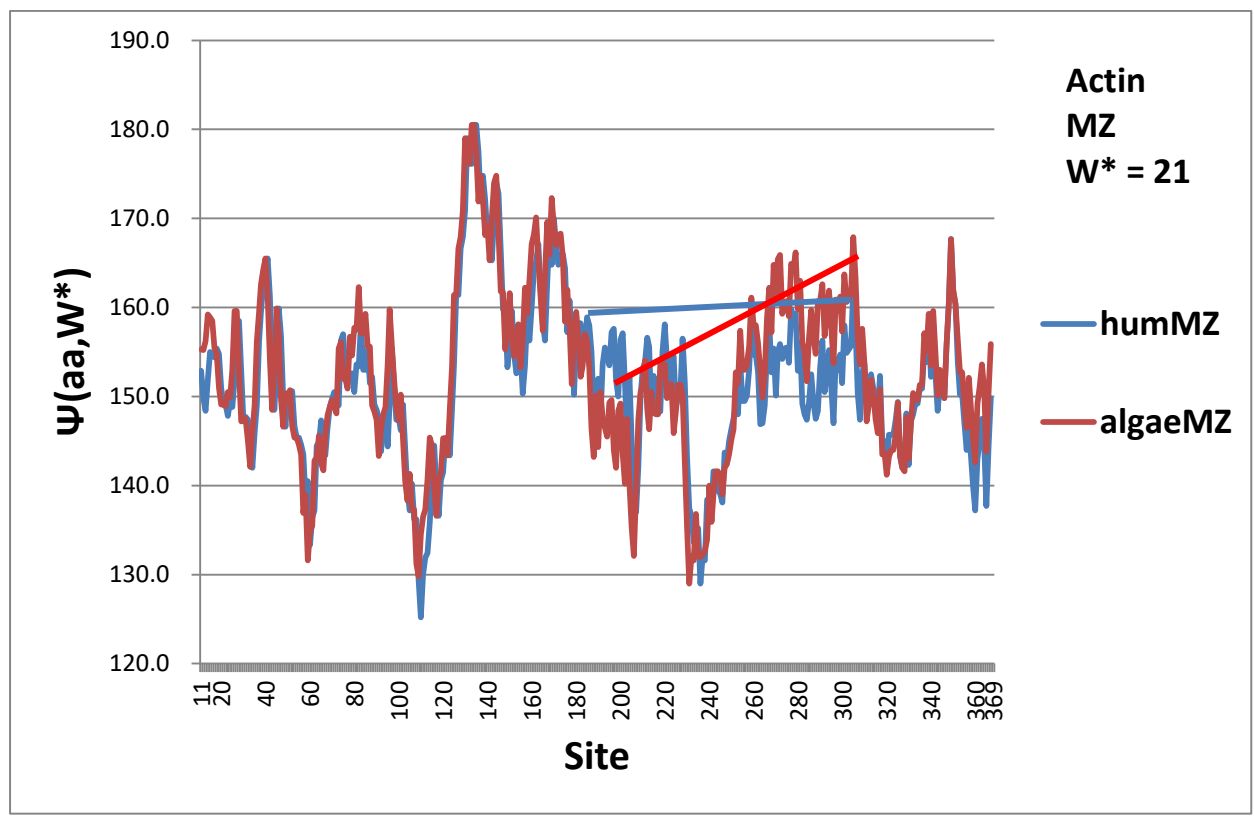

Fig. 27. Because of strong evolutionary conservation, actin algae and human profiles $\Psi$(aa,W*) differ only slightly, but the nature of these differences still show important Darwinian (positive evolutionary) improvements (see text). Overall human actin hydropathic extrema are more level (beter synchronized) than algae extrema.

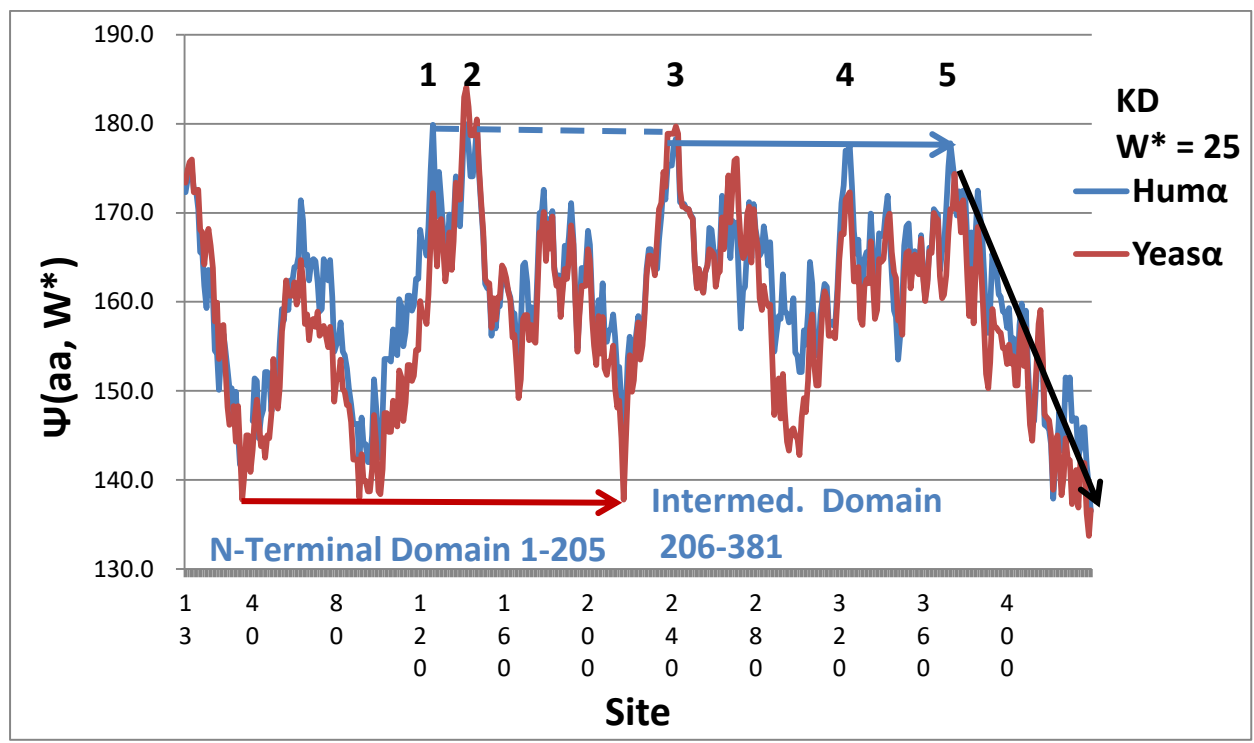

Fig. 28. The α tubulin W* = 25 profiles for human and yeast are quite similar (KD scale). Note the match between profile extrema and the tripartite domain edges identified structurally [131,132]. Also impressive are the three extremely level (Ψ ave. deviation. 0.1) human hydrophobic extrema (3-5) in the intermediate domain. The dashed line shows that two human peaks (1,2) in the N-terminal domain are also nearly level with these three (Ψ higher by only 1). The amphiphilic linearity of the 382-440 C-terminal domain is emphasized by the black arrow. Note also the water-driven flexibility of the N-terminal domain associated with its three deep and nearly level hydrophilic hinges. Note the upward profile shifts of human above yeast around site 210 by about 10% of the profile range, and around site 300 by about 35% of the profile range. These shifts stabilize mechanically active human cells, which are subjected to greater stresses than passive yeast cells.

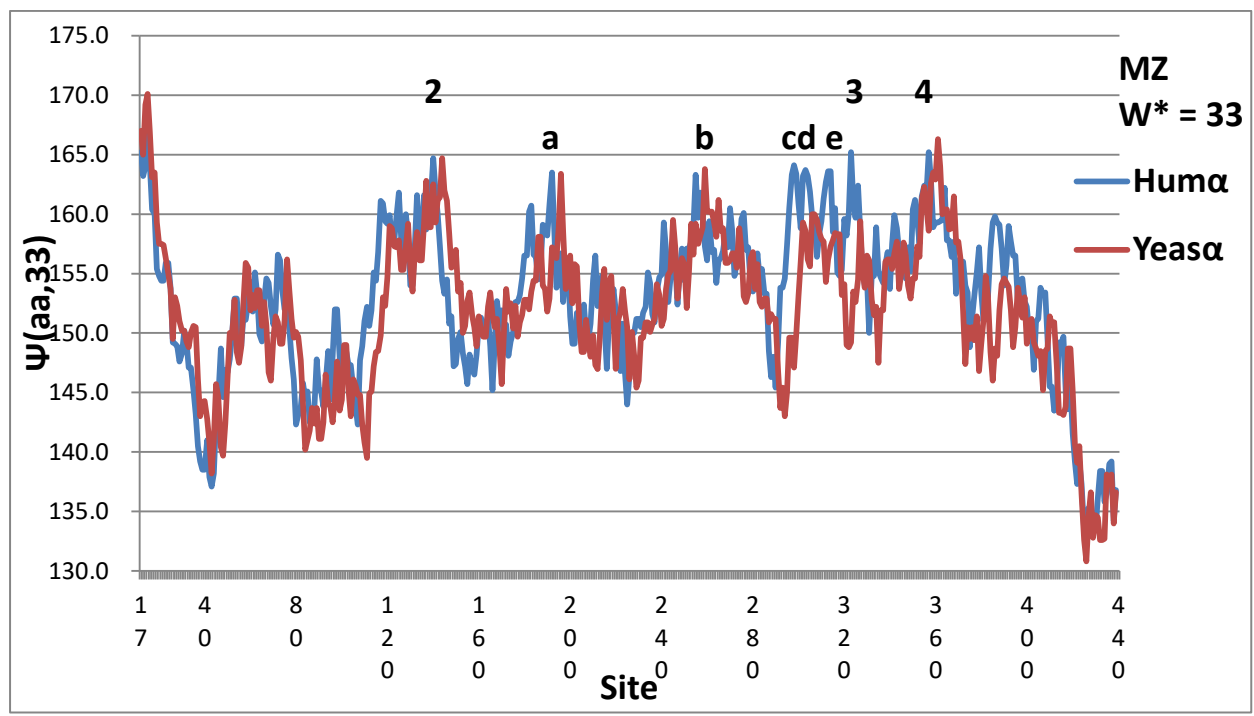

Fig. 29. The MZ scale with W* = 33 shows eight level peaks in sets of 3 and 5 for Human α tubulin. The sets are well separated at 165.0 (2) and 163.4 (1) respectively. Therefore the choice W* = 33 for the MZ scale is at least as effective in defining level sets as W* = 25 was for the KD scale in Fig. 28. The BLAST alignment of the Human and Yeast α sequences has a gap of 5 amino acids at human 40, which produces the profile offsets. The offsets facilitate comparison of Human and Yeast extrema. The differences are usually small, but between 300 and 330 they are large, with Humα much more hydrophobic. This is similar to the KD W* = 25 Human and Yeast differences, while here the Human profile is refined and includes extra peaks c, d,e and 4. These extra hydrophobic peaks both stabilize (because more hydrophobic) and increase flexiblity (because the additional level peaks facilitate synchronized motion) of the Intermediate domain, and couple it (through 2 and a) to the N-terminal domain. At the same time, the larger value of W* = 33 and use of the MZ scale does not give the successful domain separation shown in Fig. 1 for W* = 25 with the KD scale.

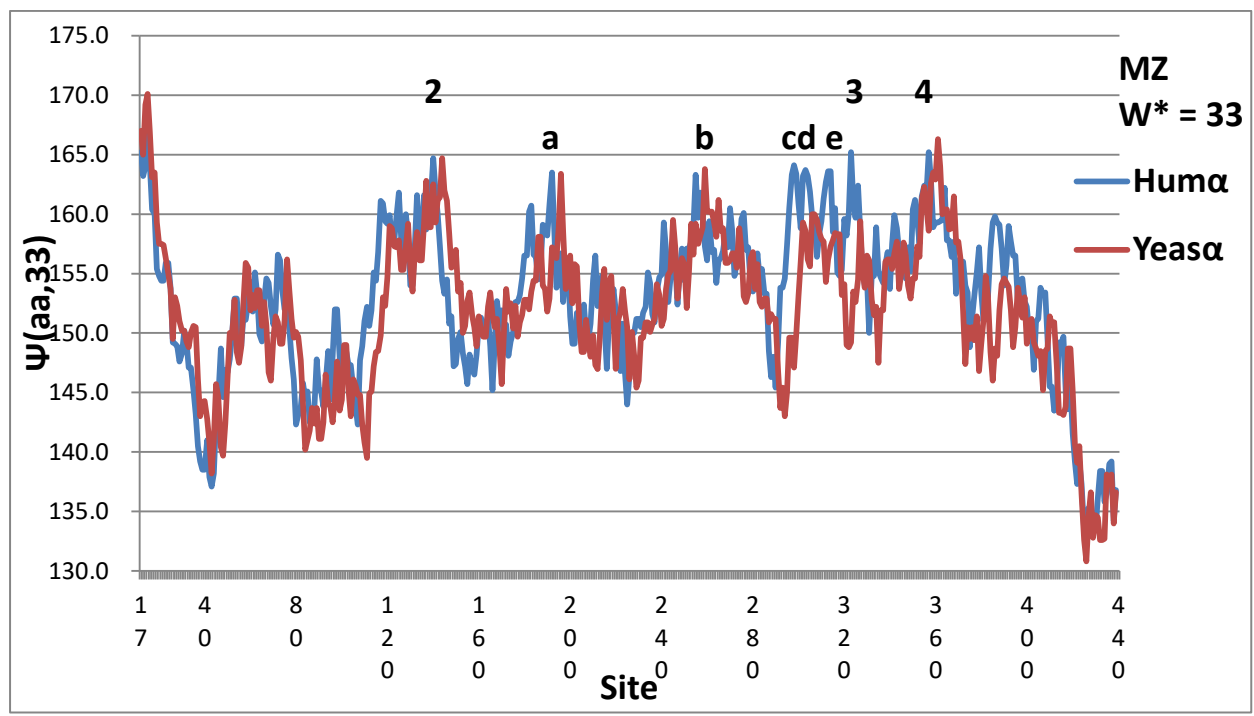

Fig. 30. The MZ scale with W* = 33 shows eight level peaks in sets of 3 and 5 for Human α tubulin. The sets are well separated at 165.0 (2) and 163.4 (1) respectively. Therefore the choice W* = 33 for the MZ scale is at least as effective in defining level sets as W* = 25 was for the KD scale in Fig. 28. The BLAST alignment of the Human and Yeast α sequences has a gap of 5 amino acids at human 40, which produces the profile offsets. The offsets facilitate comparison of Human and Yeast extrema. The differences are usually small, but between 300 and 330 they are large, with Humα much more hydrophobic. This is similar to the KD W* = 25 Human and Yeast differences, while here the Human profile is refined and includes extra peaks c, d,e and 4. These extra hydrophobic peaks both stabilize (because more hydrophobic) and increase flexiblity (because the additional level peaks facilitate synchronized motion) of the Intermediate domain, and couple it (through 2 and a) to the N-terminal domain. At the same time, the larger value of W* = 33 and use of the MZ scale does not give the successful domain separation shown in Fig. 28 for W* = 25 with the KD scale.

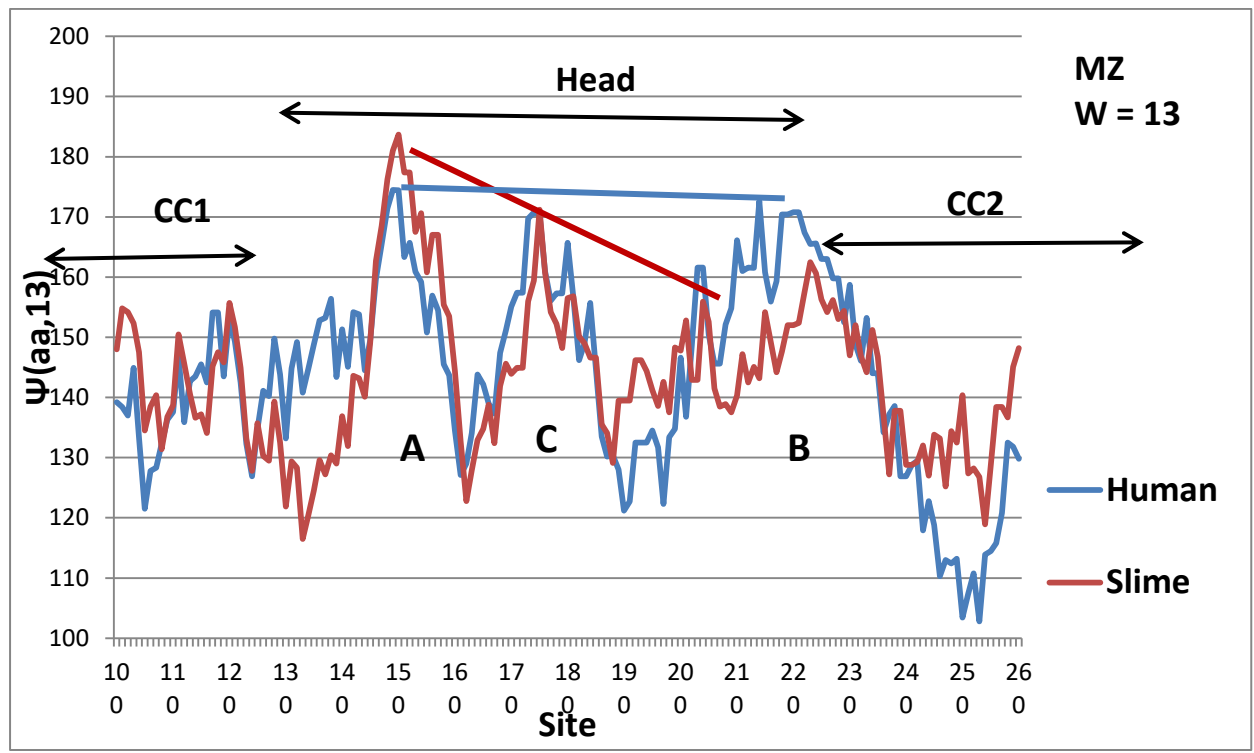

Fig. 31. The three hydrophobic maxima A-C are tilted in slime mold, but are level in mouse and human dynein.

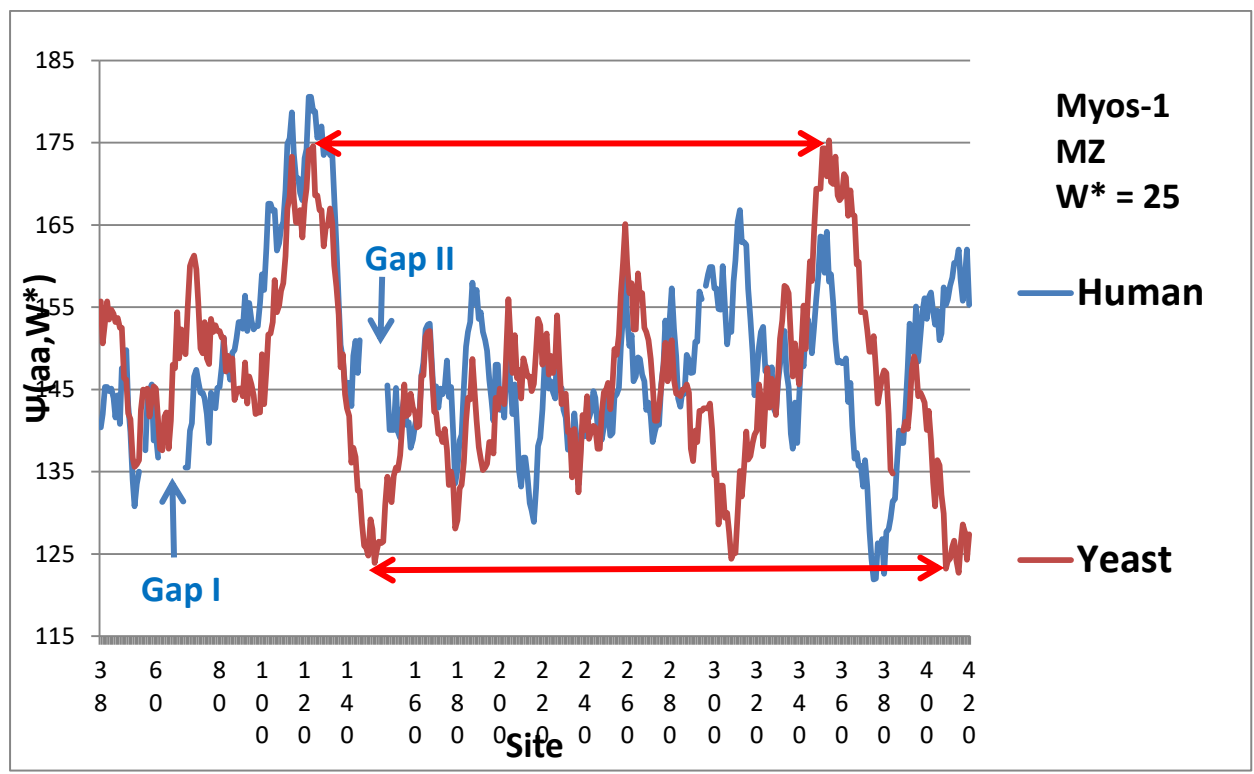

Fig. 32. Comparison of N - terminal Ψ(aa,W*) profiles (MZ scale) of Human and Yeast Myosin-1. BLAST alignment shows two 12 aa gaps in the Human sequence P12882 compared to the Yeast sequence P08964, as marked. Two major hydrophobic peaks and three major hydrophic valleys are also level in the Yeast profile, but not in the Human profile. Human site numbering.

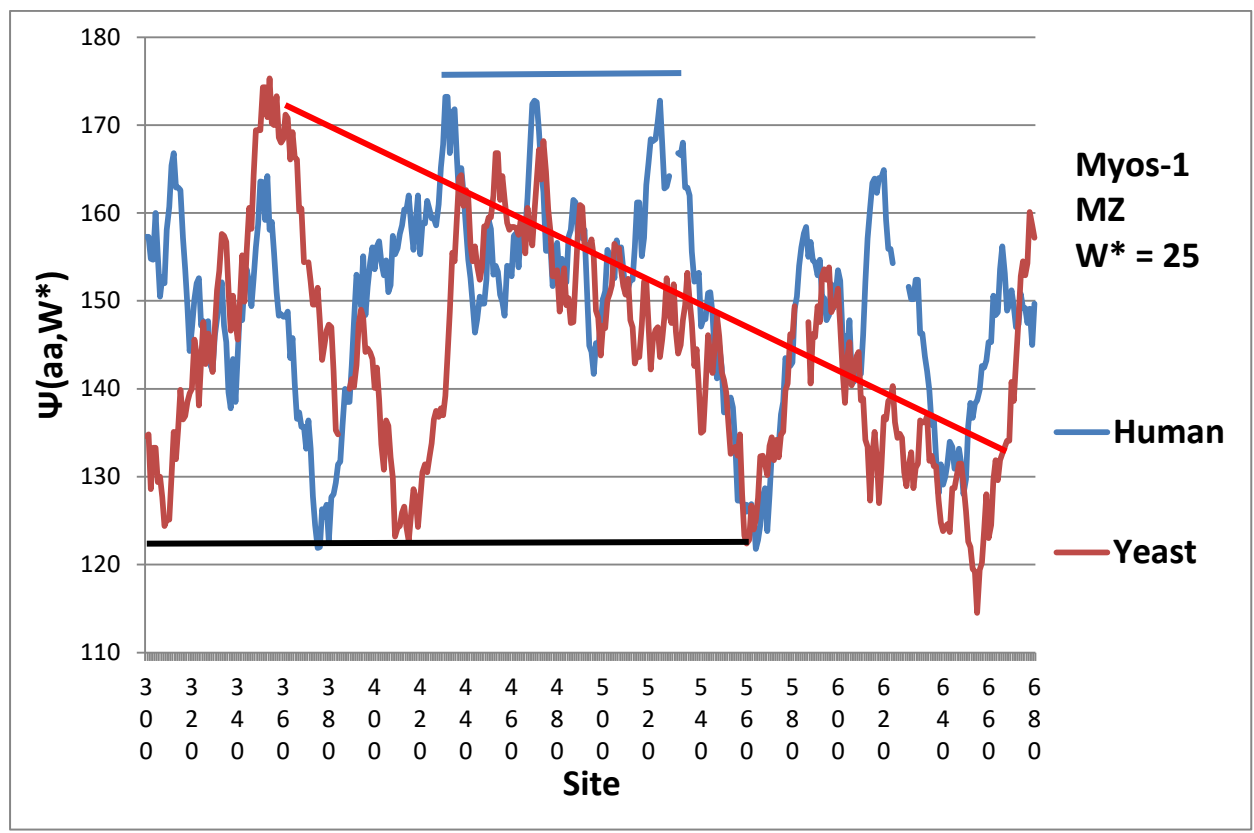

Fig. 33. Whereas in Fig. 32, in a wide Myosin-1 N-terminal region, Yeast showed more level sets, in the central region 300-680 shown here the Human hydrophobic extrema are more level. The Yeast extrema show a strongly amphiphilic (cascade) trend, while the Human hydrophobic extrema are flatter, with three level large hydrophobic extrema dominating the 430-550 region. Note that the hydrophilic floor remains for both cases at the same level as in Fig. 32. The larger actin-binding region 659-681 (Uniprot) lies at the end of the Yeast amphiphilic region (red line online).

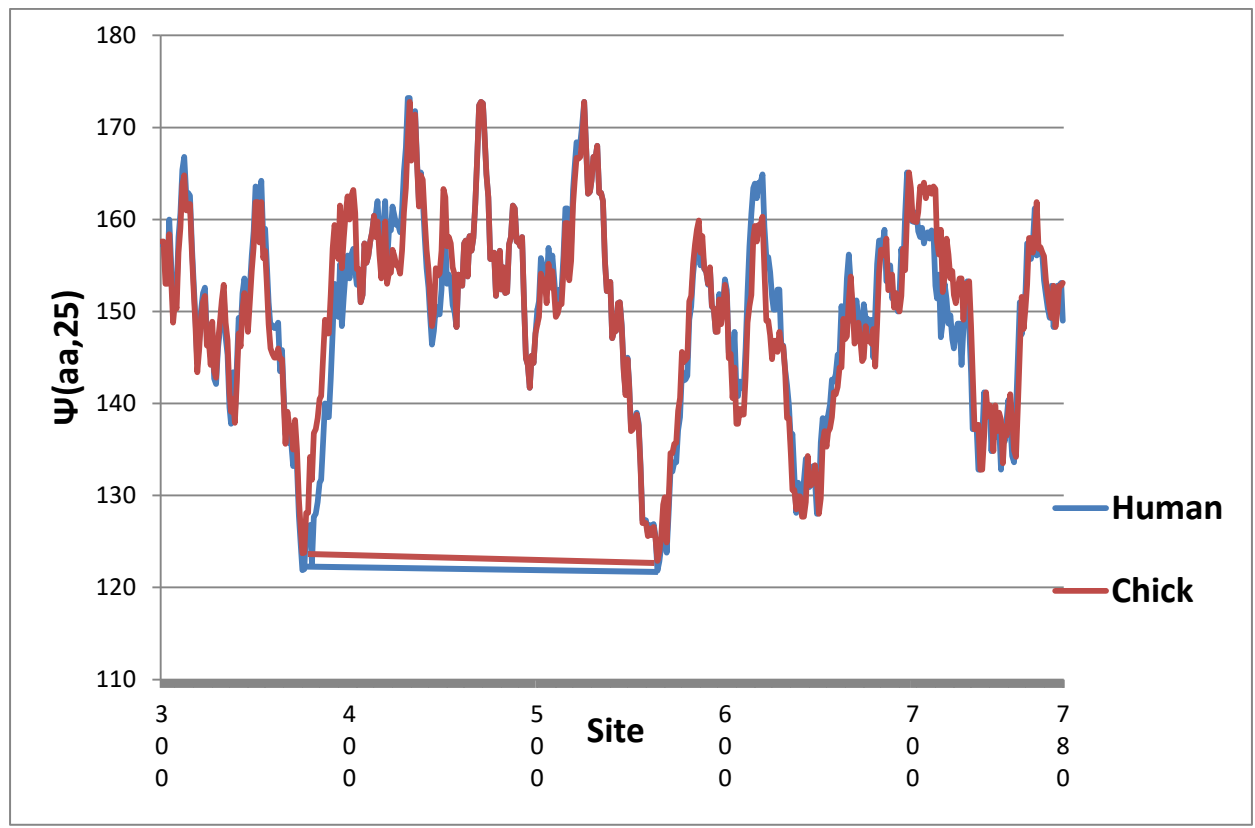

Fig. 34. Because chicken and human Myosin-1 are so similar, their sequences can be used to test the accuracy of the MZ scale. The two hydrophilic extrema are both level, and the human extrema are both 1.5 (0.5)% more hydrophilic than the chicken extrema. This accuracy of 0.5% can be compared to the BLAST positives difference of 4%. Myosin-2 is similar to Myosin-1 chicken, and thus is less evolved than Myosin-1.

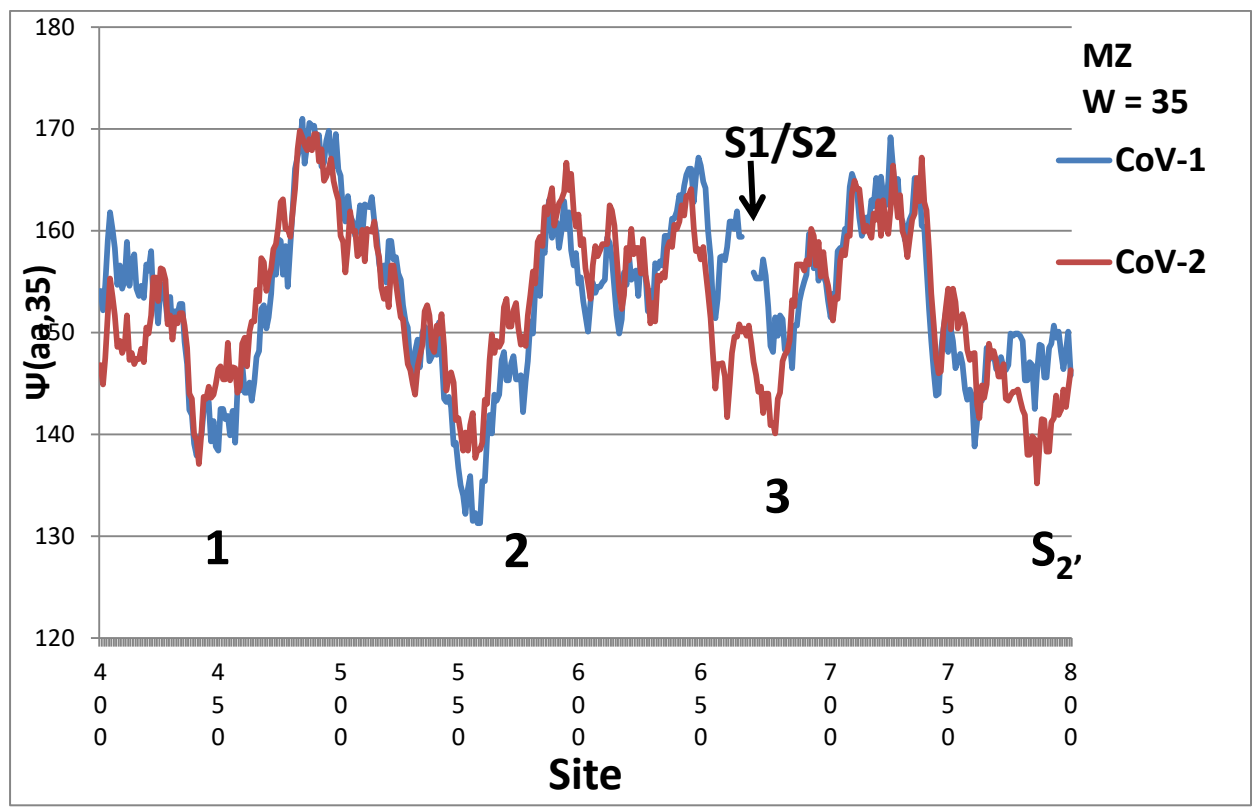

Fig. 35. The hydropathic profiles of CoV-1 and CoV-2 reveal a hidden symmetry when plotted using the MZ scale (second-order phase transitions). The three hydrophilic minima of CoV-2, labeled 1-3, are nearly equal at 140, whereas the similar minima of CoV-1 range from 131 to 147. Note especially the very deep minimum of CoV-1 at 559. The new sequence PRRA in CoV-2, inserted at 681 in CoV-1 (the S1/S2 cleavage interface) [2], has an average MZ hydropathicity 108.5. This lowers $\Psi(aa,35)$ to 140.9 at the 3 minimum, aligning it with ~140 minima 1 and 2. The three minima span ~ 250 amino acids sites, which makes their water-driven synchronization for CoV-2, but not CoV-1, outside the range of most simulation or modeling methods.

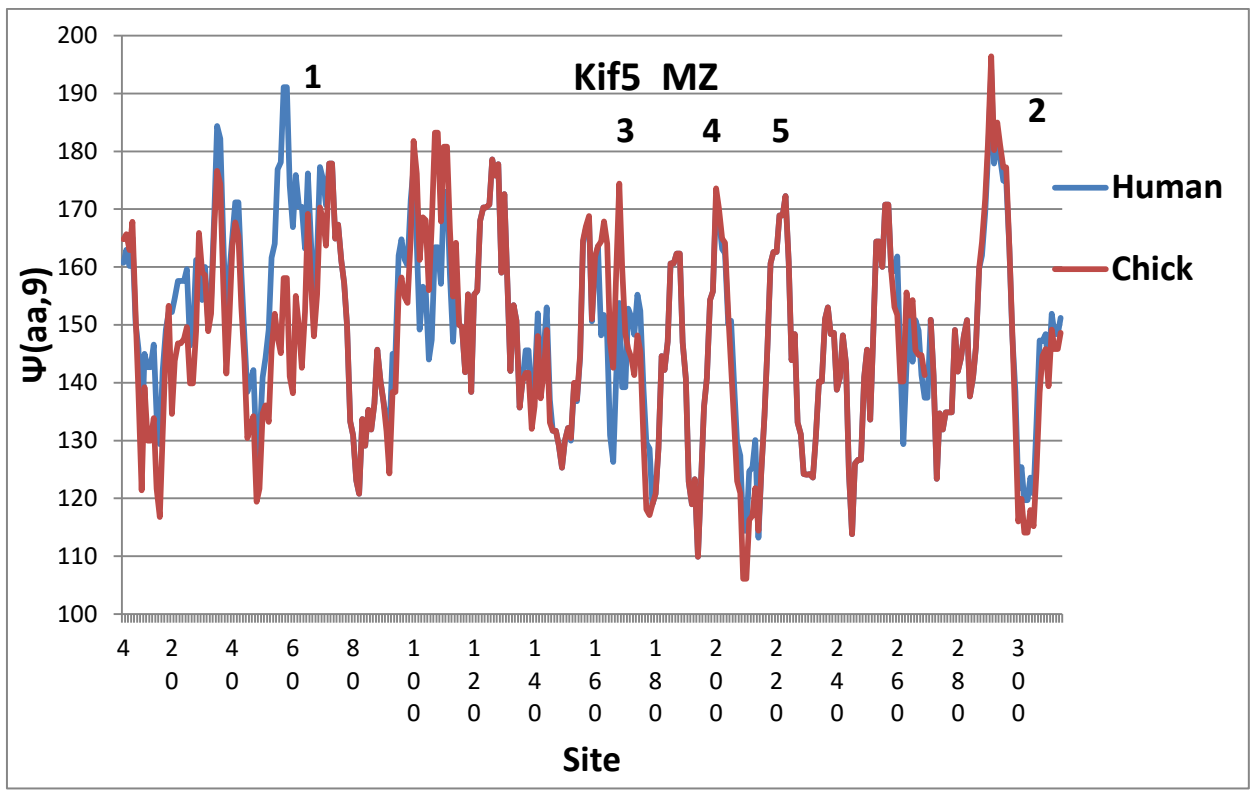

Fig. 36. When Kif5 evolves from Chicken to Human, it develops a second large hydrophobic peak 1 that is nearly level with the single hydrophobic peak 2 found already in Chicken. A similarly good match is found for W = 1 and 11. Peaks 3-5 are a secondary level set near 172, and peak 5 is dropped to 163 in Rat by a single mutation.

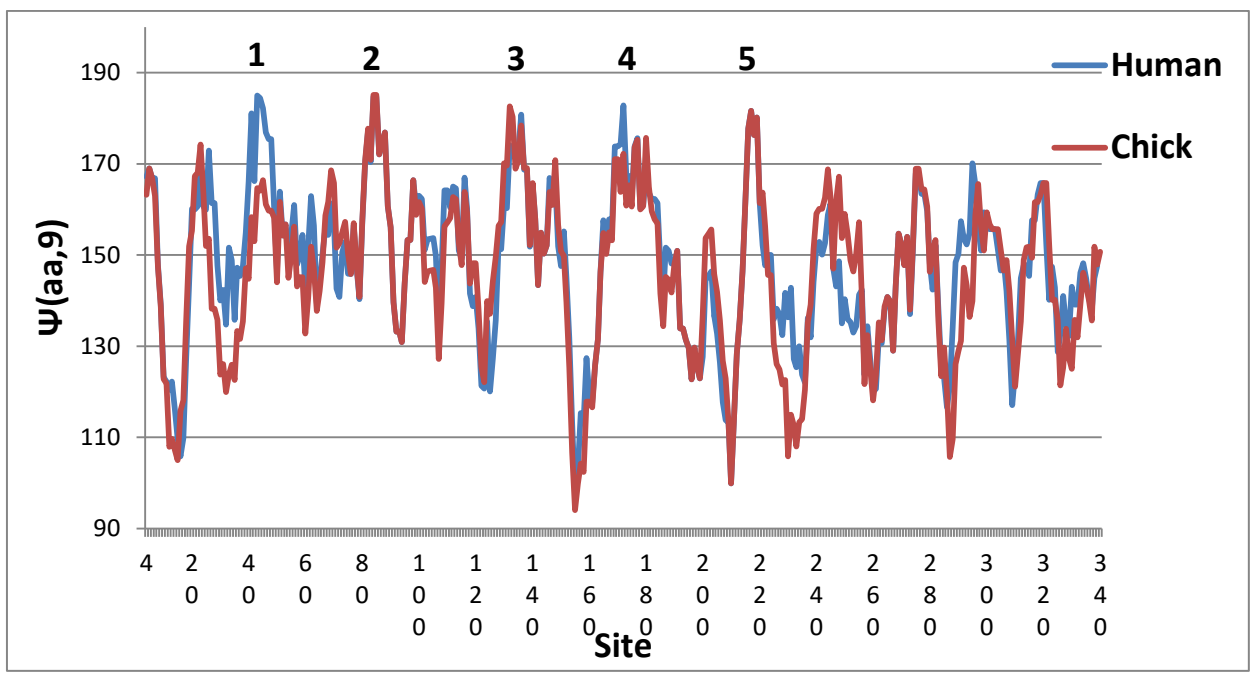

Fig. 37. Evolution leveled hydrophobic peaks 2, 3 and 5 in chicken, and further evolution added level peaks 1 and 4 in human Kif14. The reader may also observe progressive partial leveling in chickens and humans of the five lowest hydrophilic minima near 10, 150, 210, 230 and 290.

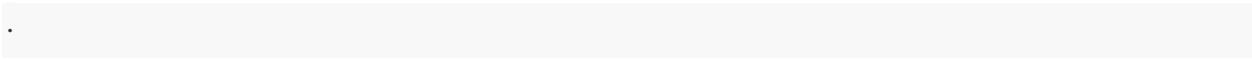